\documentclass{jpp}
\usepackage{color}

\usepackage{graphicx}
\usepackage{natbib}

\ifCUPmtlplainloaded \else
  \checkfont{eurm10}
  \iffontfound
    \IfFileExists{upmath.sty}
      {\typeout{^^JFound AMS Euler Roman fonts on the system,
                   using the 'upmath' package.^^J}%
       \usepackage{upmath}}
      {\typeout{^^JFound AMS Euler Roman fonts on the system, but you
                   dont seem to have the}%
       \typeout{'upmath' package installed. JPP.cls can take advantage
                 of these fonts, if you use 'upmath' package.^^J}%
       \providecommand\upi{\pi}%
      }
  \else
    \providecommand\upi{\pi}%
  \fi
\fi

\ifCUPmtlplainloaded \else
  \checkfont{msam10}
  \iffontfound
    \IfFileExists{amssymb.sty}
      {\typeout{^^JFound AMS Symbol fonts on the system, using the
                'amssymb' package.^^J}%
       \usepackage{amssymb}%
       \let\le=\leqslant  \let\leq=\leqslant
       \let\ge=\geqslant  \let\geq=\geqslant
      }{}
  \fi
\fi

\ifCUPmtlplainloaded \else
  \IfFileExists{amsbsy.sty}
    {\typeout{^^JFound the 'amsbsy' package on the system, using it.^^J}%
     \usepackage{amsbsy}}
    {\providecommand\boldsymbol[1]{\mbox{\boldmath $##1$}}}
\fi

\providecommand\bcdot{\boldsymbol{\cdot}}

\newsavebox{\astrutbox}
\sbox{\astrutbox}{\rule[-5pt]{0pt}{20pt}}

\def\Alfven{Alfv$\acute {\mathrm{e}}$n}
\def\Alfvenic{Alfv$\acute {\mathrm{e}}$nic}
\newcommand{\diff}{\mathrm{d}}
\def\const{\mathrm{const.}}

\newcommand{\jump}[1]{\ensuremath{[\![#1]\!]} }
\newcommand{\Ljump}[1]{\ensuremath{\left[\!\!\left[#1\right]\!\!\right]} }
\newcommand{\average}[1]{\ensuremath{\langle#1\rangle} }

\def\vh{\hat{v}}
\def\ph{\hat{p}}
\def\vecBth{\hat{\boldsymbol{B}_t}}
\def\Bth{\hat{B}_t}
\def\gM{\gamma M_0^2}
\newcommand{\vsp}{\vspace{.3em}}

\newcommand\bm[1]{\boldsymbol{#1}}

\title[Exact MHD Riemann solver]{Exact Riemann solver for ideal magnetohydrodynamics that can handle all types of intermediate shocks and switch-on/off waves}

\author[K. Takahashi and S. Yamada]%
{K.\ns T\ls A\ls K\ls A\ls H\ls A\ls S\ls H\ls I$^1$%
  \thanks{Email address for correspondence: ktakahashi@heap.phys.waseda.ac.jp}\ns
\ and S.\ns Y\ls A\ls M\ls A\ls D\ls A$^{2,3}$}

\affiliation{$^1$Department of Physics, Waseda University,
3-4-1 Okubo, Shinjuku, 169-8555, Japan (ktakahashi@heap.phys.waseda.ac.jp)\\[\affilskip]
$^2$Science \& Engineering, Waseda University, 3-4-1 Okubo, Shinjuku, 169-8555, Japan\\[\affilskip]
$^3$Advanced Research Institute for Science and Engineering, Waseda University, 3-4-1 Okubo, Shinjuku, 169-8555, Japan}

\pubyear{?}
\volume{?}
\pagerange{?--?}
\date{?; revised ?; accepted ?. - To be entered by editorial office}
\begin{document}

\maketitle

\begin{abstract}
We have built a code to obtain the exact solutions of Riemann problems in ideal magnetohydrodynamics (MHD) for an arbitrary initial condition.
The code can handle not only regular waves but also switch-\textcolor{black}{on/}off rarefactions and all types of non-regular shocks: intermediate shocks and switch-on/off shocks. 
Furthermore, the initial conditions with vanishing normal or transverse magnetic fields can be handled 
although the code is partly based on the algorithm proposed by \citet{T02} (Torrilhon 2002 Exact Solver and Uniqueness Condition for Riemann problems of Ideal Magnetohydrodynamics.
Research report 2002-06, Seminar for Applied Mathematics, ETH, Zurich), which cannot deal with all types of non-regular waves nor the initial conditions without normal or transverse magnetic fields. 
Our solver can find all the solutions for a given Riemann problem and hence, as demonstrated in this paper, one can investigate the structure of the solution space in detail. 
Therefore the solver is a powerful instrument to solve the outstanding problem of the existence and uniqueness of solutions of 
MHD Riemann problems. Moreover, the solver may be applied to numerical MHD schemes like the Godunov scheme in the future.

\end{abstract}


\section{Introduction}
The Riemann problem is a kind of initial value problems for hyperbolic systems such as the system of equations of ideal hydrodynamics or ideal magnetohydrodynamics (MHD), 
in which the initial condition is given by two constant states separated by a discontinuity. Not only do the solutions of Riemann problems have mathematical interest,
but also solving Riemann problems is one of main tasks in numerical schemes for fluid dynamics because 
the solutions are used to obtain numerical fluxes. Although the theory of partial differential equations underlies that of Riemann problems, solving Riemann problems in one-dimensional space 
turns to be equivalent to solving the algebraic equations and therefore the solution is obtained by the Newton-Raphson method with a good initial guess. 
This facilitation, however, does not necessarily mean that Riemann problems can be easily solved. In ideal MHD, for example, the system of  algebraic equations is highly non-linear and complex in 
addition to the five-dimension parameter space, reflecting the non-linearity and largeness of the original system of partial differential equations. 

Moreover, there exists an outstanding problem in MHD Riemann problems that there is no convincing criterion for the physically relevant solution.
It is well-known that the solution of Riemann problems is generally not unique in the sense of the weak solution and other conditions should be imposed to single out the physically 
relevant one \citep{JT64}. A famous and obviously acceptable condition is the so-called entropy condition, which admits only the shocks across which the entropy increases. 
The entropy condition discards manifestly unphysical solutions such as those including expanding shocks, across which the entropy is decreased, and the condition works well indeed in ordinary 
hydrodynamics to uniquely choose a solution. In ideal MHD, however, the entropy condition is insufficient to uniquely choose a solution. In fact, some initial conditions have more than one 
solutions that satisfy the entropy condition \citep{T02, T03, T03b, TY12a}. Therefore the so-called evolutionary conditions are introduced, which 
\textcolor{black}{require that physically relevant shocks should be structurally stable. Here, we must note that structural stability is totally different from the more familiar stability that discusses the exponentially growth. Structurally stable shocks just remain close to the initial discontinuity when they are perturbed, while structurally unstable ones will instantaneously split into other waves \citep{Landau}.} 
The evolutionary conditions discard the so-called intermediate shocks, across which the transverse magnetic field is reversed (the definition and detailed 
classification of the shocks are given in Sec. \ref{sec.MHD}), and the uniqueness of the solution seems to be recovered. Indeed, the intermediate shocks had been considered to be unphysical 
in the literatures \citep[e.g.][]{JT64,KP66}.

However, the relevance of the intermediate shocks is still under debate. In fact, in spite of the evolutionary conditions, the intermediate shocks are commonly observed as stable shocks 
in numerical simulations \citep{Wu87, Wu88a, Wu88b, Wu90, BW88, Wu&Kennel92}. 
The evolutionary conditions are also reconsidered in the context of dissipative MHD \citep{Hada94,Markovskii98,I&I07}. They found that the new modes that do not exist in ideal MHD are responsible for 
the evolutionary conditions and the intermediate shocks become evolutionary in the dissipative system. Furthermore, some interplanetary experiments have reported the detection of the 
intermediate shocks \citep{Chao95, FW08,FWC09}. These results cast doubt on the classical theory of MHD and support the relevance of the intermediate shocks. 
On the other hand, there also exist completely opposite arguments, defending the classical theory. \citet{FK97,FK01} pointed out that the intermediate shocks are observed 
in numerical simulations only because the initial conditions have a special symmetry, where the initial transverse magnetic fields and velocities are confined in a plane. Since there is no reason to 
break the symmetry, \Alfven \ waves, which rotate the fields, do not emerge and the absence affects the evolutionary conditions. In fact, some authors demonstrated numerically that some 
intermediate shocks break into other waves if one breaks the symmetry by adding another component of the field \citep{BKP96, FK97,FK01}. Although \citet{FK01} agreed with \citet{Wu90} 
that the temporal survival of some intermediate shocks in their interaction with \Alfven \ waves is due to the non-unique dissipative structures, they claimed that the shocks should be regarded 
as transients. \textcolor{black}{\citet{Nikolai} also comes to the same conclusion.}

Due to the lack of the understanding of the intermediate shocks, it is desirable for MHD Riemann solvers to treat these shocks.
Therefore we will present a Riemann solver that can handle all types of intermediate shocks. Furthermore, our solver can also treat the switch-on/off shocks and switch-\textcolor{black}{on/}off rarefactions 
(see Sec. \ref{sec.MHD} for the details of these waves), which have never been considered in previous studies. This feature can become critical because it happens that the initial condition 
has only a solution in which these non-regular shocks and switch-off rarefactions exist and does not have any solution without them \citep{TY12a}.
Our solver can handle any initial condition; It does not matter whether normal or transverse magnetic field is absent. 
Although \citet{ATJweb} released an exact MHD Riemann solver online, their solver does not consider either of these waves and requires the initial conditions where both normal and transverse 
magnetic field exist. 
We also note that \citet{T02} proposed an idea of treating the intermediate shocks although they neglected some types. While our solver is partly based on the idea of \citet{T02},  
we drastically modified it to handle all types of intermediate shocks. Furthermore, the details of the main techniques are released for the first time since \citet{T02} did not 
show the details of their method, which must be rather complicated as described in Sec. \ref{sec.method} to Sec. \ref{sec.method2}.

Our solver has potential to solve an outstanding problem associated with the uniqueness and existence of the solution of MHD Riemann problems. In fact,
even the local existence and uniqueness are no longer guaranteed by the Lax's theorem \citep{Lax1957, JT64, Serre} because the system of ideal MHD is not 
strictly hyperbolic and the characteristic fields are neither linear nor genuinely non-linear \citep{FK01}. 
\textcolor{black}{There are some analytical studies on the existence and uniqueness of solutions of ideal MHD Riemann problems. \citet{Gogosov1961,Gogosov1962} investigated the wave-pattern of the solution in MHD Riemann problems, considering only the evolutionary waves and switch-off waves.
Considering the intermediate shocks, \citet{T03} investigated the uniqueness of the solution.
However, they assumed that a particular type of intermediate shock emerges on only one side and therefore it is not complete.} 
On the other hand, our solver can find all the solutions for a given Riemann problem
and hence it can investigate the structure of the solution space without any restriction. Therefore the solver is a powerful instrument to examine the non-uniqueness and existence of the solution.
The solver may be applicable to numerical MHD, on the other hand, where the Riemann solver provides numerical fluxes. \textcolor{black}{Actually, there are several works on the numerical MHD codes with Riemann solvers \citep[e.g.][]{DW,Sano,II}.}
Furthermore, since our solver gives the exact solution of the Riemann problems, 
one can know which solution the other approximate MHD Riemann solvers approximate, 
which has not been investigated ever and turns to be one of the essential criteria for the appropriate scheme when the physically relevant conditions for the solutions are clarified in the future.

The paper is organized as follows. 
In Sec. \ref{sec.MHD}, we give a brief review of the shock waves and simple waves in ideal MHD, which are constituents of the solution of the Riemann problems. 
In Sec. \ref{sec.method} and Sec. \ref{sec.method1.5}, the main procedure to solve the Riemann problems is given. 
Other technical details are given in Sec. \ref{sec.method2}. 
In Sec. \ref{sec.results}, we demonstrate our solver by showing the solutions of some MHD Riemann problems.
We summarize the features of our solver in Sec. \ref{sec.summary}.

\section{Ideal MHD} \label{sec.MHD}
We here review the simple waves and discontinuities in ideal MHD, which are the constituents of the solutions of Riemann problems.
Note that this section provides essential knowledge to construct the MHD Riemann solver in the succeeding sections although the section is largely quoted from Sec. 3 in \citet{TY12a}.
For the general theory of Riemann problems, see other text books or our previous paper \citep[e.g.,][]{JT64, TY12a}.

In plane symmetry, the ideal MHD equations are given by
\begin{eqnarray}
\label{mass}
\frac{\partial \rho}{\partial t} +\frac{\partial }{\partial x}(\rho v_n) = 0, \\
\frac{\partial }{\partial t}(\rho v_n) +\frac{\partial }{\partial x} \left( \rho v_n^2 +p +\frac{\bm{B}_t^2}{2}\right) =0, \\
\frac{\partial }{\partial t}(\rho \bm{v}_t) +\frac{\partial }{\partial x} ( \rho v_n\bm{v}_t -B_n\bm{B}_t) =\bm{0}, \\
\frac{\partial \bm{B}_t}{\partial t} +\frac{\partial }{\partial x}(v_n \bm{B}_t -B_n \bm{v}_t) = \bm{0}, \\
\label{energy}
\frac{\partial e}{\partial t} +\frac{\partial }{\partial x}\left[ \left( e +p +\frac{\bm{B}^2}{2}\right) v_n -B_n\bm{B}\bcdot \bm{v} \right] = 0,
\end{eqnarray}
where $\rho$, $p$, $\bm{v}$ and $\bm{B}$ are density, pressure, flow velocity and  magnetic field respectively \citep{Landau}. 
The subscripts $n$ and $t$ indicate the normal component, i.e., $x$-component, and transverse component, i.e., $y$ or $z$-component respectively.
The total energy density is denoted by $e = p/(\gamma -1) +\rho \bm{v}^2/2 +\bm{B} ^2/2 $, where the equation of state for ideal gas is
assumed and $\gamma$ is the ratio of specific heats. 
\textcolor{black}{In this expression, we have used units so that factors such as $4 \upi$ and $c$ (the speed of light) do not appear for no special reason.}
The normal component of magnetic field, $B_n$, is constant owing to the divergence-free 
condition.

\subsection{Simple waves in ideal MHD\label{smp}}
Simple waves are defined as waves in which the conservative variables are all functions of one variable or, equivalently, defined by the $N-1$ generalized Riemann invariants, where $N$ stands for 
the number of the system equations \citep{JT64, TY12a}.
These waves make one-parameter families by definition. The loci in phase space, which connect the states of the head and tail of the waves, are constructed by 
the right eigenvectors of the Jacobian matrix for the system equations. The eigenvalues of the Jacobian matrix for the system (\ref{mass})-(\ref{energy}) are
\begin{eqnarray}
\label{eigenvalues}
v_n \mp c_f, \quad
v_n \mp c_A, \quad
v_n \mp c_s, \quad
v_n,
\end{eqnarray}
where $c_f$, $c_A$ and $c_s$ are called the fast, \Alfven \ and slow speeds respectively, and are expressed as 
\begin{eqnarray}
\label{fast}
c_{f,s} := \left[ \frac{1}{2}\left(\frac{\bm{B}^2}{\rho} +a^2\right) \pm \sqrt{\frac{1}{4}\left( \frac{\bm{B}^2}{\rho} +a^2\right)^2 -a^2\frac{B_n^2}{\rho} } \right]^{1/2}, \\
\label{Alfven}
c_A := \sqrt{\frac{B_n^2}{\rho}}.
\end{eqnarray}
In the above expressions, $a = \sqrt{\gamma p/\rho}$ is the acoustic speed. In (\ref{eigenvalues}), the minus (plus) sign is applied to 
the left-going (right-going) waves. The simple waves corresponding to these eigenvalues are referred to as the fast, \Alfven , slow and entropy waves
respectively. The right eigenvectors for fast and slow waves are given as
\begin{eqnarray}
\label{fseigenvec}
\bm{r}_{f,s}^\mp =  \xi _{f,s} \left[
\begin{array}{ccccc}
-\rho \\
-\gamma p \\
\pm c_{f,s}   \vsp \\
\displaystyle
\pm \frac{c_{f,s}}{1-(c_{f,s}/c_A)^2}\frac{\bm{B}_t}{B_n} \vsp \\
\displaystyle
\frac{\bm{B}_t}{(c_A/c_{f,s})^2 -1} \\
\end{array} \right], 
\end{eqnarray}
in which the new variables, $\xi _f$ and $\xi _s$, are introduced as follows:
\begin{eqnarray}
\xi _f := \sqrt{\frac{a^2 -c_s^2}{c_f^2 -c_s^2}}, \quad
\xi _s := \sqrt{\frac{c_f^2 -a^2}{c_f^2 -c_s^2}}.
\end{eqnarray}
These factors are necessary to ensure that the eigenvectors do not vanish for any $\bm{u}$, where $\bm{u}$ is the vector of the primitive variables
\textcolor{black}{\citep{BW88}}.
In deriving the above expressions of right eigenvectors we assume that $\bm{u} = (\rho , p, v_n, \bm{v}_t, \bm{B}_t)$. 
Since the density is decreased across the fast and slow simple waves, these waves are also referred to as fast and slow rarefaction waves respectively.
We do not give the explicit forms of eigenvectors for the \Alfven \ and entropy waves here because the corresponding waves turn to be discontinuous in MHD Riemann problems 
and can be expressed by the Rankine-Hugoniot relations (see the next sub-section).

The eigenvalues are degenerate in the following two cases:
\begin{eqnarray}
\label{degenerate_1}
B_n  = 0 & : & \quad c_f = \sqrt{a^2+\frac{\bm{B}_t^2}{\rho}}, \quad c_s = c_A = 0 , \\
\label{degenerate_2}
B_n \ne 0,\ \bm{B}_t = \bm{0} & : & \quad c_f = \mathrm{max}(a, c_A), \quad c_s = \mathrm{min}(a, c_A).
\end{eqnarray}
In the former case, the limits of the right eigenvectors for fast waves are given as
\begin{equation}
\bm{r}_f^\mp \rightarrow \sqrt{\frac{a^2}{c_f^2}}\left[
\begin{array}{ccccc}
-\rho \\
-\gamma p \\
\pm c_f \\
\bm{0} \\
-\bm{B}_t
\end{array} \right].
\end{equation}
The slow waves form now a discontinuous wave called tangential discontinuity (see the next sub-section).

In the latter case ($B_n \ne 0,\ \bm{B}_t = \bm{0}$), the limits of the right eigenvectors for fast and slow waves as $\bm{B}_t \rightarrow \bm{0}$ depend on 
the magnitudes of the acoustic and \Alfven \ speeds. For $a > c_A$, we obtain  
\begin{eqnarray}
\bm{r}_f^\mp \rightarrow \left[
\begin{array}{ccccc}
-\rho \\
-\gamma p \\
\pm a \\
\bm{0} \\
\bm{0}
\end{array} \right]
, \quad
\bm{r}_s^\mp \rightarrow a \left[
\begin{array}{ccccc}
0 \\
0 \\
0 \\
\pm \mathrm{sgn}(B_n) \, \bm{e}_t \\
\sqrt{\rho} \, \bm{e}_t
\end{array} \right],
\end{eqnarray}
where $\bm{e}_t$ is a unit vector that has the same direction as the transverse magnetic field.
Note that $\bm{r}_f^\mp$ is reduced to the eigenvectors for the rarefaction waves in the ordinary hydrodynamics.
In the opposite case, i.e., $a < c_A$, we get 
\begin{eqnarray}
\label{sw-off eigenvector}
\bm{r}_f^\mp \rightarrow a \left[
\begin{array}{ccccc}
0 \\
0 \\
0 \\
\mp \mathrm{sgn}(B_n) \bm{e}_t \\
-\sqrt{\rho} \bm{e}_t
\end{array}\right]
,\quad
\bm{r}_s^\mp \rightarrow \left[
\begin{array}{ccccc}
-\rho \\
-\gamma p \\
\pm a \\
\bm{0} \\
\bm{0}
\end{array}\right],
\end{eqnarray}
in which $\bm{r}_s^\mp$ is reduced to the eigenvectors for the ordinary rarefaction waves in hydrodynamics.
Finally, in the case of $a = c_A$, we find 
\begin{eqnarray}
\bm{r}_f^\mp \rightarrow \frac{1}{\sqrt{2}} \left[
\begin{array}{ccccc}
-\rho \\
-\gamma p \\
\pm a \\
\mp a\bm{e}_t \\
-a\sqrt{\rho} \bm{e}_t 
\end{array}\right]
,\quad
\bm{r}_s^\mp \rightarrow \frac{1}{\sqrt{2}} \left[
\begin{array}{ccccc}
-\rho \\
-\gamma p \\
\pm a \\
\pm a\bm{e}_t \\
a\sqrt{\rho} \bm{e}_t 
\end{array}\right].
\end{eqnarray}
As mentioned earlier, the right eigenvectors are chosen in our code so that these degenerate cases could be 
properly handled as the limits of non-degenerate cases.

In the fast rarefaction waves, the magnitude of transverse magnetic field is decreased and, as a limiting case, 
the field vanishes behind the so-called switch-off rarefaction waves. Since the fast rarefaction waves cannot reverse the direction 
of the transverse magnetic field, the switch-off rarefactions lie at the end point of the fast rarefaction loci.
\textcolor{black}{
We also note that in the slow rarefaction waves there is a family across which the transverse magnetic field is produced while it does not exist on the head, which are called the switch-on rarefaction waves.
They can emerge when $a \ge c_A$ is satisfied. 
}

\subsection{Discontinuities in ideal MHD}
As mentioned earlier, discontinuities are another important element in the solutions of Riemann problems. The quantities on
both sides of a discontinuity satisfy the Rankine-Hugoniot relations, which in ideal MHD are expressed as
\begin{eqnarray}
\label{mass flux}
m = \const , \\
m^2\jump{v} +\Ljump{ p+\frac{\bm{B}_t^2}{2} } = 0, \\
m\jump{ \bm{v}_t } -B_n\jump{ \bm{B}_t } = \bm{0}, \\
m\jump{ v\bm{B}_t } -B_n\jump{ \bm{v}_t } = \bm{0}, \\
\label{energy jump}
m\left( \Ljump{\frac{pv}{\gamma -1} } +\average{p}\jump{v} +\frac{1}{4}\jump{ v }\jump{ \bm{B}_t }^2 \right) = 0,
\end{eqnarray}
in the rest frame of the discontinuity. In the above expressions, $m := \rho _0 {v_n}_0 = \rho _1 {v_n}_1$ is the mass flux, 
$v := 1/\rho$ is the specific volume. $\jump{X} := X_0 -X_1$ denotes the jump in a quantity, $X$, across the discontinuity, where $X_0$ and $X_1$ are the value of ahead of and behind 
the discontinuity respectively.
$\average{X} := (X_0 +X_1)/2$ stands for the arithmetic mean of upstream and downstream quantities. 
In what follows, we summarize those features that are needed for later discussions.

Following \citet{T02,T03}, we normalize all quantities with those upstream as
\begin{eqnarray}
\label{vh}
\hat{v} := \frac{v_1}{v_0}, \quad
\hat{p} := \frac{p_1}{p_0}, \quad
\hat{\bm{B}}_t := \frac{ {\bm{B}_t}_1}{\sqrt{p_0}}, \\
\bm{A} := \frac{ {\bm{B}_t}_0}{\sqrt{p_0}}, \quad
B := \frac{B_n}{\sqrt{p_0}}, \quad
\label{M_0}
M_0 := \frac{{v_n}_0}{a_0},
\end{eqnarray}
and employ in the following the dimensionless MHD Rankine-Hugoniot relations, which are obtained by substituting 
(\ref{vh})-(\ref{M_0}) and eliminating $\jump{\bm{v}_t}$ in (\ref{mass flux})-(\ref{energy jump}): 
\begin{eqnarray}
\label{Euler_normal_hat}
\ph -1 +\gM (\vh -1) +\frac{1}{2}(\vecBth ^2-\bm{A}^2) = 0, \\
\label{Euler_trans_hat}
\gM (\vh \vecBth -\bm{A}) -B^2(\vecBth -\bm{A}) = 0, \\
\label{energy_hat}
M_0\left[ \frac{1}{\gamma -1}(\ph \vh -1) +\frac{1}{2}(\vh -1)(\ph +1) +\frac{1}{4}(\vh -1)(\vecBth -\bm{A})^2 \right]= 0.
\end{eqnarray}
Fixing the upstream quantities, $A$, $B$ and $M_0$, 
we solve (\ref{Euler_normal_hat})-(\ref{energy_hat}) and use 
(\ref{vh})-(\ref{M_0}) to obtain $v_1$, $p_1$ and $\bm{B}_{t1}$.
The other downstream quantities can be calculated as
\begin{eqnarray}
\label{vnormal}
{v_n}_1 = \vh {v_n}_0, \\
\label{vtrans}
{\bm{v}_t}_1 = {\bm{v}_t}_0 \pm \frac{a_0 B}{\gamma M_0}\jump{\vecBth}.
\end{eqnarray}
In (\ref{vtrans}), the plus and minus signs correspond to the left- and right-going discontinuities respectively.

\subsubsection{Contact, tangential and rotational discontinuities}
The solutions of (\ref{Euler_normal_hat})-(\ref{energy_hat}) that have a vanishing mass flux, i.e., $M_0=0$, but a non-vanishing 
normal component of magnetic field, i.e., $B \ne 0$, are called the contact discontinuity and satisfy the following relations:
\begin{eqnarray}
\vh = \mathrm{arbitrary}, \quad \ph = 1, \quad \vecBth = \bm{A}, \\
\jump{\bm{v}} = \bm{0}.
\end{eqnarray}
That is, only the density is discontinuous at the contact discontinuity and other quantities, pressure, magnetic field and velocity field, are continuous.

The solutions with $M_0 = 0$ and $B = 0$, on the other hand, are named the tangential discontinuity, for which the following 
relations hold:
\begin{eqnarray}
\label{tolpre}
\vh = \mathrm{arbitrary}, \quad  \ph -1 +\frac{1}{2}(\vecBth^2 -\bm{A}^2) = 0, \\
\jump{v_n} = 0,\quad \jump{\bm{v}_t} = \mathrm{arbitrary}.
\end{eqnarray}
At the tangential discontinuity, the total pressure and normal velocity are continuous while other quantities can be discontinuous.

The solutions with $M_0 \ne 0$ and $B \ne 0$ are either a linear wave ($\vh = 1$) or a shock wave ($\vh > 1$).
The former is referred to as the rotational discontinuity, since the transverse component of magnetic field rotates, not 
varying its magnitude during its passage. The rotational discontinuities meet the following conditions:
\begin{eqnarray}
\label{rot}
\vh = 1, \quad \ph = 1, \quad \vecBth ^2= \bm{A}^2, \quad M_0^2 = \frac{B^2}{\gamma}, \\
\jump{v_n} = 0, \quad \jump{\bm{v}_t} = \pm \frac{1}{\sqrt{\rho}}\jump{\bm{B}_t},
\end{eqnarray}
where the plus and minus signs correspond to the left- and right-going waves respectively. 
The upstream and downstream Mach number turn to be equal to the ratio of the \Alfven \ velocity to the acoustic speed from the above relations.

All the above discontinuities satisfy the evolutionary conditions except for the rotational discontinuity in which the transverse 
magnetic field rotates by $180^{\circ}$. The latter is sometimes called weakly evolutionary in the literature \citep{JT64}, since 
the neighboring rotational discontinuities are all evolutionary.

\subsubsection{Shock waves}
The solutions of (\ref{Euler_normal_hat})-(\ref{energy_hat}) for which $M_0 \ne 0$ and $\vh > 1$, i.e., matter is compressed as it passes through the discontinuities, 
are called shock waves. Their notable feature is that magnetic fields are either planar or coplanar. This is apparent from (\ref{Euler_trans_hat}). Indeed, recalling $\vh > 1$ and ${v_n}_1 = \vh{v_n}_0$, we obtain
\begin{equation}
\vecBth = \frac{{v_n}_0^2 -{c_A}_0^2}{{v_n}_1^2 -{c_A}_1^2} \bm{A},
\end{equation}
which shows immediately that transverse magnetic fields are coplanar if and only if the upstream flow velocity is super-\Alfvenic \ whereas
the downstream speed is sub-\Alfvenic.
The shocks with planar transverse magnetic fields are either fast or slow shocks, the former of which amplifies the magnitude of transverse
magnetic fields whereas the latter reduces it. The shocks that change the direction of transverse magnetic fields are referred to as  
intermediate shocks. 

Recalling $c_f \geq c_A \geq c_s$, we assign $1$ to the states with super-fast velocities, $2$ to those with sub-fast and 
super-\Alfvenic \ velocities, $3$ to those with sub-\Alfvenic \ and super-slow velocities and $4$ to those with sub-slow velocities
in the shock-rest frame. With this allocation, the fast shocks are denoted by $1 \rightarrow 2$ shocks, since the upstream velocity is
super-fast (state $1$) whereas the downstream speed is sub-fast and super-\Alfvenic \ (state $2$)
\textcolor{black}{\citep[e.g.][]{Hans}}.
Similarly the slow shocks are designated 
as $3 \rightarrow 4$ shocks. The intermediate shocks normally belong to one of the following four types: $1 \rightarrow 3$, 
$1 \rightarrow 4$, $2 \rightarrow 3$ and $2 \rightarrow 4$ shocks. The $1 \rightarrow 3$ and $2 \rightarrow 4$ intermediate shocks are 
called over-compressive shocks and $9$ out of $14 (=7 \times 2)$ characteristics run into these shock waves. To the $1 \rightarrow 3$ 
and $2 \rightarrow 4$ shocks converge the fast and \Alfven \ characteristics and the \Alfven \ and slow characteristics respectively.
The $1 \rightarrow 4$ intermediate shock is doubly over-compressive and $10$ characteristics  of all types go into the shock.
In the case of $2\rightarrow 3$ shock, only the \Alfvenic \ characteristic converges to the shock wave and the number of in- and out-going 
waves are right. 
In some cases, the flow velocity coincides with one of characteristic velocities. We employ a pair of numbers to specify those states;
'$1,2$', '$2,3$' and '$3,4$' represent those states whose flow speed are equal to the fast, \Alfven \ and slow speeds respectively.
The shock wave with the upstream velocity being super-\Alfvenic \ and the downstream speed being equal to the slow velocity, for example, 
is designated as the $2 \rightarrow 3,4$ shock. Of our special concern among these intermediate shocks are the so-called 
switch-on ($1 \rightarrow 2,3$) and switch-off ($2,3 \rightarrow 4$) shocks, the details of which will be given shortly.
Note that all the intermediate shocks and switch-on/off shocks do not satisfy the evolutionary conditions and referred to as non-regular waves.

\subsubsection{Fast and slow loci} \label{sec.Fast and slow loci}
We regard the shock solutions of (\ref{Euler_normal_hat})-(\ref{energy_hat}) as functions of the upstream Mach number ($M_0$), 
normal ($B$) and transverse ($\vecBth$ or $\bm{A}$) components of magnetic field, divide them into two families and look into 
their loci in some detail. Since magnetic fields in shock waves are 
either planar or coplanar as pointed out earlier, in the following we assume without loss of generality that magnetic fields are confined in 
the $(x,y)$-plane and treat $\Bth$ and $A \ (\geq 0)$ as scalar variables.

Eliminating $\ph$ and $\Bth$ from (\ref{Euler_normal_hat})-(\ref{energy_hat}) we obtain the following cubic equation for 
the specific volume, $\vh$
\begin{eqnarray}
&&(\gM \vh) ^3 -\left[ \frac{2}{\gamma +1} +\frac{\gamma -1}{\gamma +1} \gM +2B ^2+\frac{\gamma A^2}{\gamma +1}\right] ( \gM \vh )^2 \nonumber \\
\label{fast branch}
&&+\left[2B^2 \left( \frac{2\gamma}{\gamma +1} +\frac{\gamma -1}{\gamma +1}\gM \right) +B^4 -\frac{2 -\gamma}{\gamma +1}\gM A^2 +A^2B^2\right] \gM \vh \\ 
&&-\left[ B^4 \left( \frac{2\gamma}{\gamma +1}  +\frac{\gamma -1}{\gamma +1}\gM \right) +\frac{\gamma -1}{\gamma  +1}\gM A^2B^2 \right] = 0, \nonumber
\end{eqnarray}
where we used the assumption of $\vh \ne 1$ in deriving the equation. 
Or, alternatively, we obtain the following quadratic equation for the specific volume by eliminating $M_0$:
\begin{eqnarray}
&&\left\{\frac{\Bth }{2}\left[ \frac{4\gamma}{\gamma -1}+(\Bth -A)^2 +\frac{\gamma +1}{\gamma -1}(A^2 -\Bth^2 )\right] -\frac{\gamma +1}{\gamma -1}B^2(\Bth -A)\right\} \vh ^2\nonumber \\
\label{quadra}
&& \ +\frac{2\gamma}{\gamma -1}\left[ \frac{A}{2}(\Bth ^2 -A^2) +B^2(\Bth -A) -(\Bth +A)\right] \vh \\
&& \ +\frac{2\gamma A}{\gamma -1} -(A^2 +B^2)(\Bth -A) = 0. \nonumber
\end{eqnarray}

\begin{figure}
\begin{tabular}{cc}
\begin{minipage}{0.45\hsize}
\begin{center}
\includegraphics[scale=0.26]{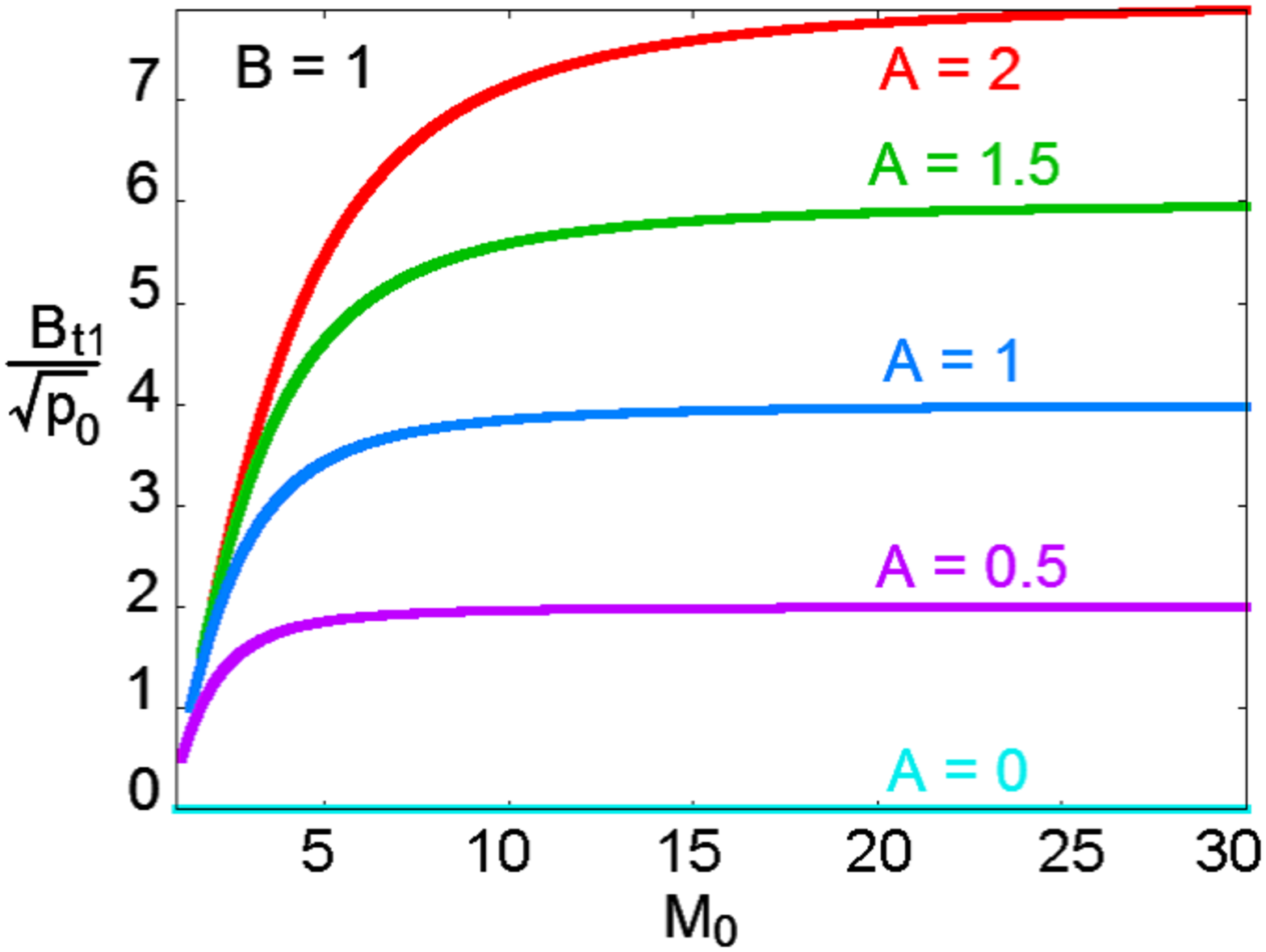}
\end{center}
\end{minipage} &
\begin{minipage}{0.45\hsize}
\begin{center}
\includegraphics[scale=0.26]{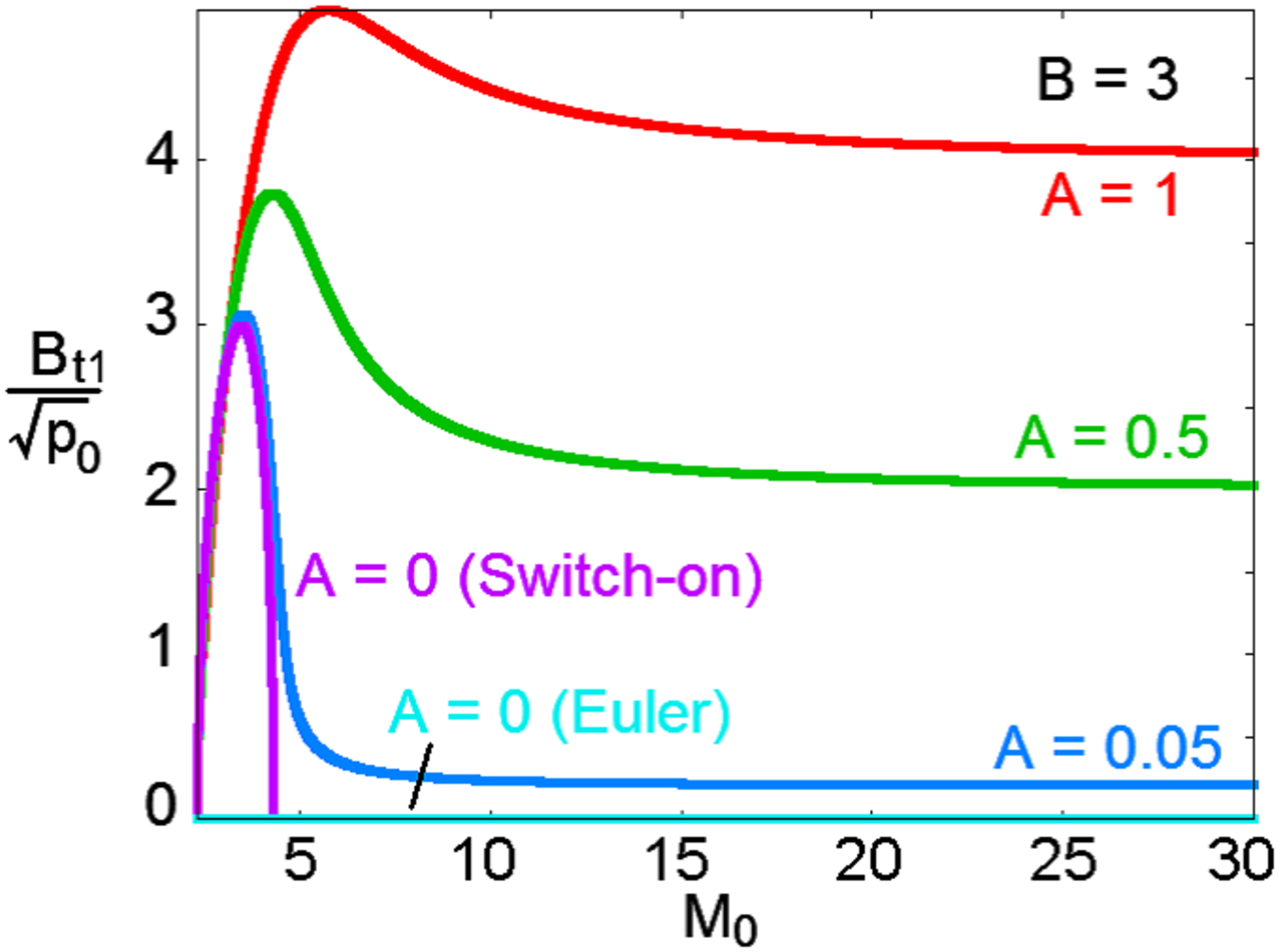}
\end{center}
\end{minipage} \\
\begin{minipage}{0.45\hsize}
\begin{center}
\includegraphics[scale=0.26]{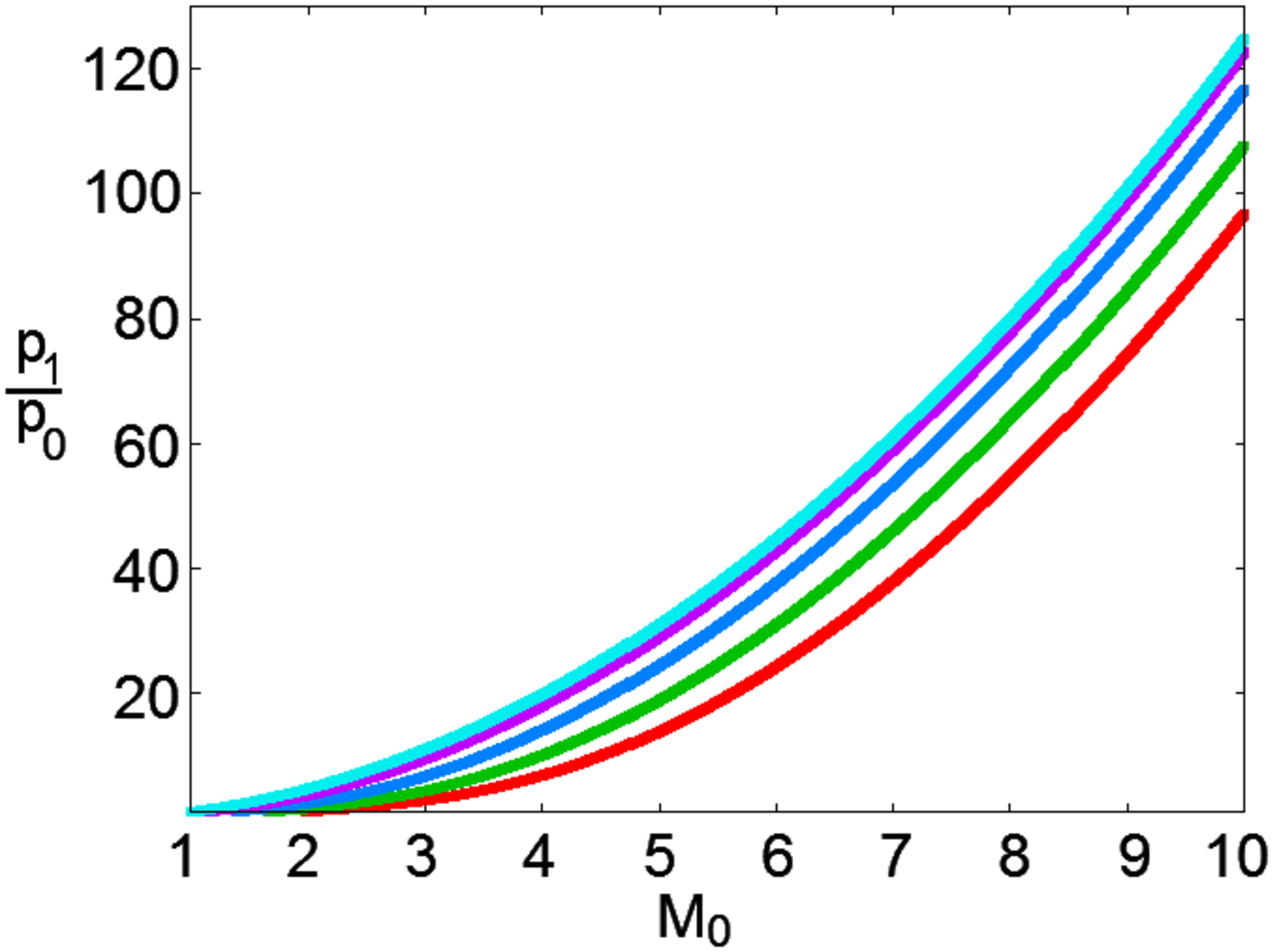}
\end{center}
\end{minipage} &
\begin{minipage}{0.45\hsize}
\begin{center}
\includegraphics[scale=0.26]{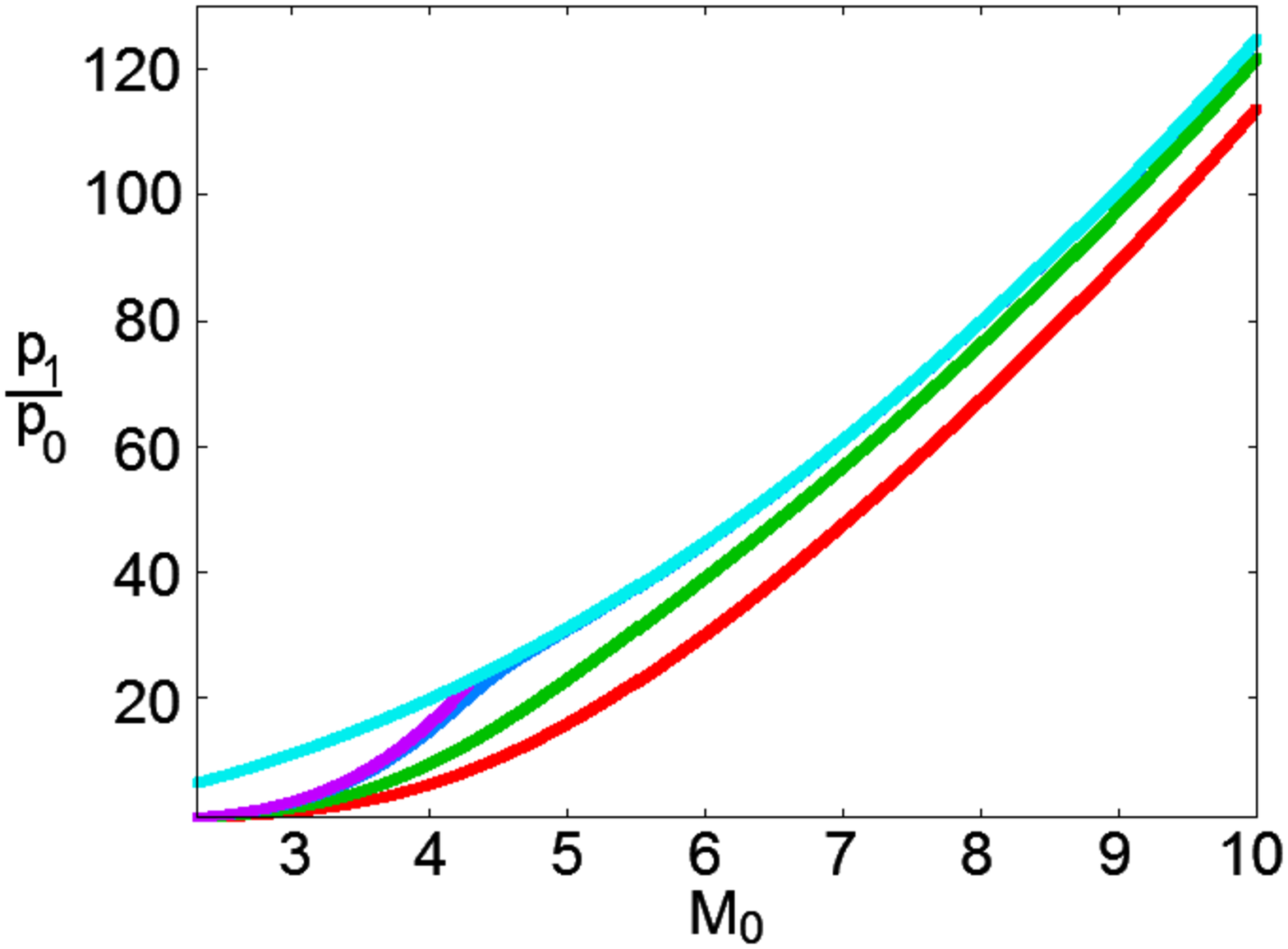}
\end{center}
\end{minipage} \\
\begin{minipage}{0.45\hsize}
\begin{center}
\includegraphics[scale=0.26]{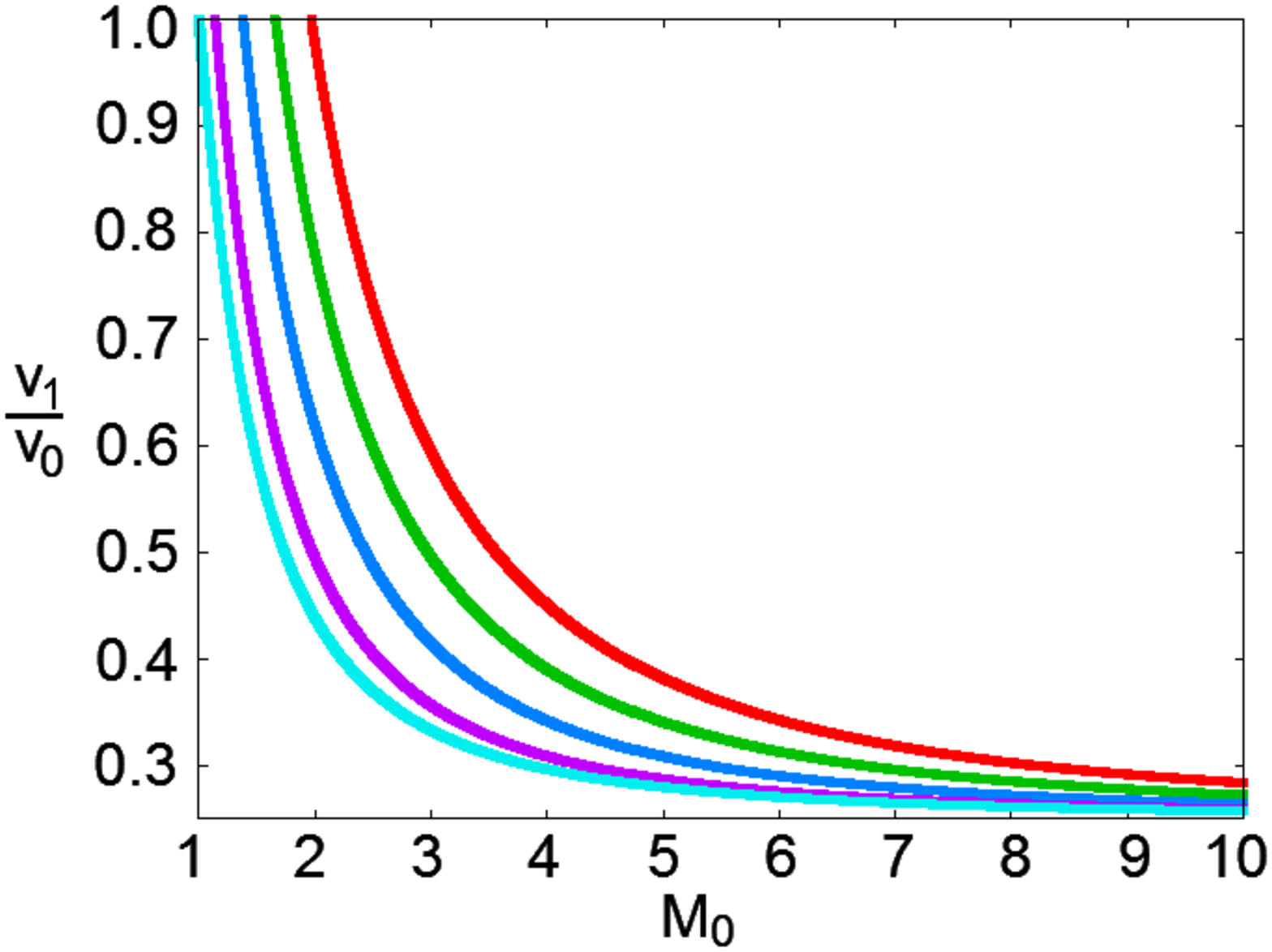}
\end{center}
\end{minipage} &
\begin{minipage}{0.45\hsize}
\begin{center}
\includegraphics[scale=0.26]{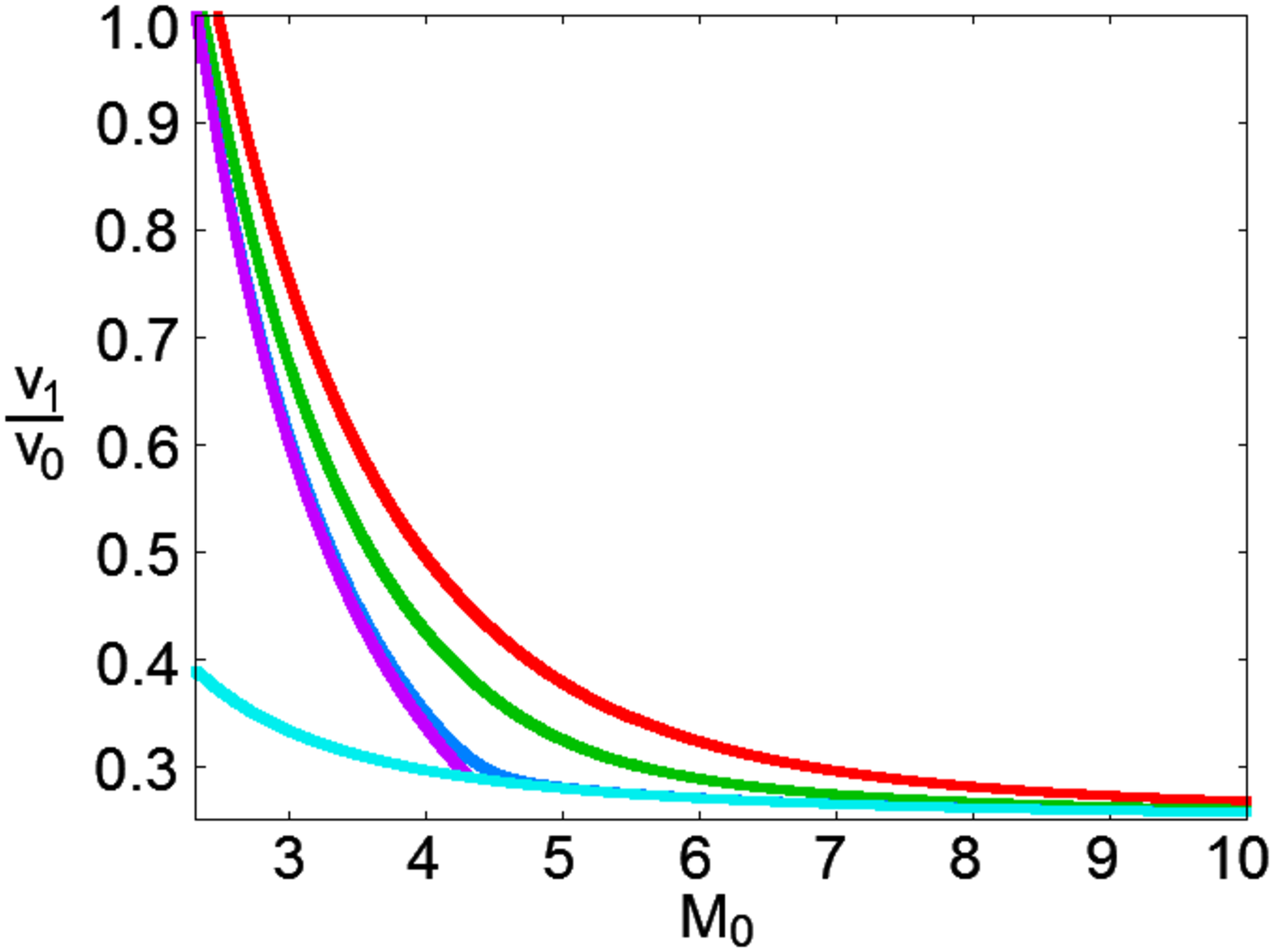}
\end{center}
\end{minipage} 
\end{tabular}
\caption{The fast loci for \textcolor{black}{$\gamma =5/3$ and} different combinations of the normal ($B$) and transverse ($A$) component of magnetic field.
The left panels: $B=1$ and $A=2$ (red), $1.5$ (green), $1$ (blue), $0.5$ (purple) and $0$ (light blue). 
The right panels: $B=3$ and $A = 1$ (red), $0.5$ (green), $0.05$ (blue), $0$ (purple, switch-on shock) and $0$ 
(light blue, Euler shock). The switch-on shock does not exist for $B=1$. See the text for details. The figure is quoted from \citet{TY12a}.}
\label{fig.locus1}
\end{figure}

The family of the fast shock is the solutions characterized 
by the feature that matter is compressed and the transverse magnetic field is amplified by the passage of shock wave. It is then found
that this branch of solutions satisfies the inequality $\vh _\mathrm{min} < \vh < 1$, where the minimum is given by 
\begin{equation}
\vh _\mathrm{min} = \mathrm{max} \left( \frac{B^2}{\gM}, \frac{\gamma -1}{\gamma +1} \right).
\end{equation}
The loci of the solutions are shown in Fig.~\ref{fig.locus1}, taken from \citet{TY12a}, as a function of the upstream Mach number ($M_0$) for some combinations of 
the upstream normal ($B$) and transverse ($A$) components of magnetic field. The fast shock can be parameterized by the upstream Mach number 
if one fixes the other parameters ($A$ and $B$), which are determined by upstream variables, in almost case as seen in Fig.~\ref{fig.locus1}. 
The only exception in which some special treatment is required is seen in the right panel of Fig.~\ref{fig.locus1}, where the fast locus is divided into two branches for the special case of $A = 0$, i.e.,
the vanishing upstream transverse magnetic field, with Mach numbers satisfying the following inequalities:
\begin{equation}
\label{sw-on_ineq}
\hat{c}_{f0} < M_0 < \sqrt{\frac{\gamma +1}{\gamma -1}\frac{B^2}{\gamma} -\frac{2}{\gamma -1}},
\end{equation}
where $\hat{c}_{f0} := c_{f0}/a_0$ is the normalized fast velocity. One of the branches that generates non-vanishing transverse magnetic 
fields by the shock passage is called the switch-on shock branch and the other is referred to as the Euler shock branch, in which
the transverse component of magnetic field remains zero. The post-shock specific volume and transverse magnetic field are given by
\begin{eqnarray}
\label{bifurcate1}
\vh & = & \frac{2+(\gamma -1)M_0^2}{(\gamma +1)M_0^2}, \quad\frac{B^2}{\gM }, \\
\label{bifurcate2}
\Bth & = & 0, \quad \sqrt{\frac{\gM - B^2}{B^2}\left[ (\gamma -1) \left( \frac{\gamma +1}{\gamma -1}B^2 -\gM \right) -2\gamma \right]},
\end{eqnarray}
respectively. In the above expressions, the first options correspond to the Euler shock and the second ones to the switch-on shock.
The requirement that the quantity in the square root be non-negative gives the inequality (\ref{sw-on_ineq}).
It is noted that the flow speed behind the switch-on shock is equal to the \Alfven \ speed. The switch-on shock is hence designated as
the $1 \rightarrow 2,3$ shock and is non-regular. It is also noteworthy that the Euler shock is essentially a hydrodynamical shock
wave and its locus is extended to $M_0 < \hat{c}_{f0}$, where it is smoothly connected to the slow-shock counterpart.
The Euler shock is evolutionary except the range within which the switch-on shock branch appears, i.e., the range satisfying the inequality (\ref{sw-on_ineq}).

\begin{figure}
\begin{tabular}{cc}
\begin{minipage}{0.45\hsize}
\begin{center}
\includegraphics[scale=0.26]{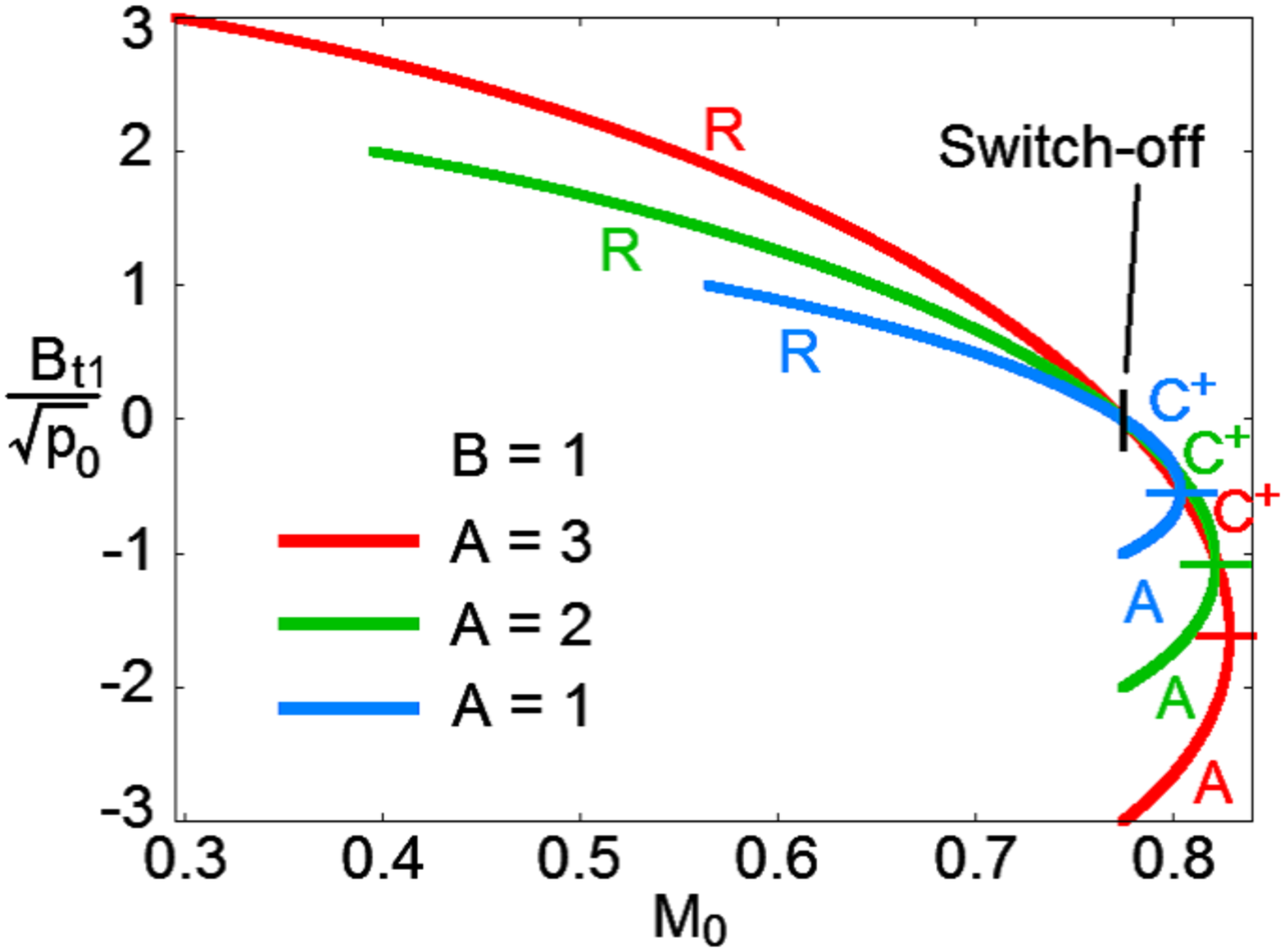}
\end{center}
\end{minipage} &
\begin{minipage}{0.45\hsize}
\begin{center}
\includegraphics[scale=0.26]{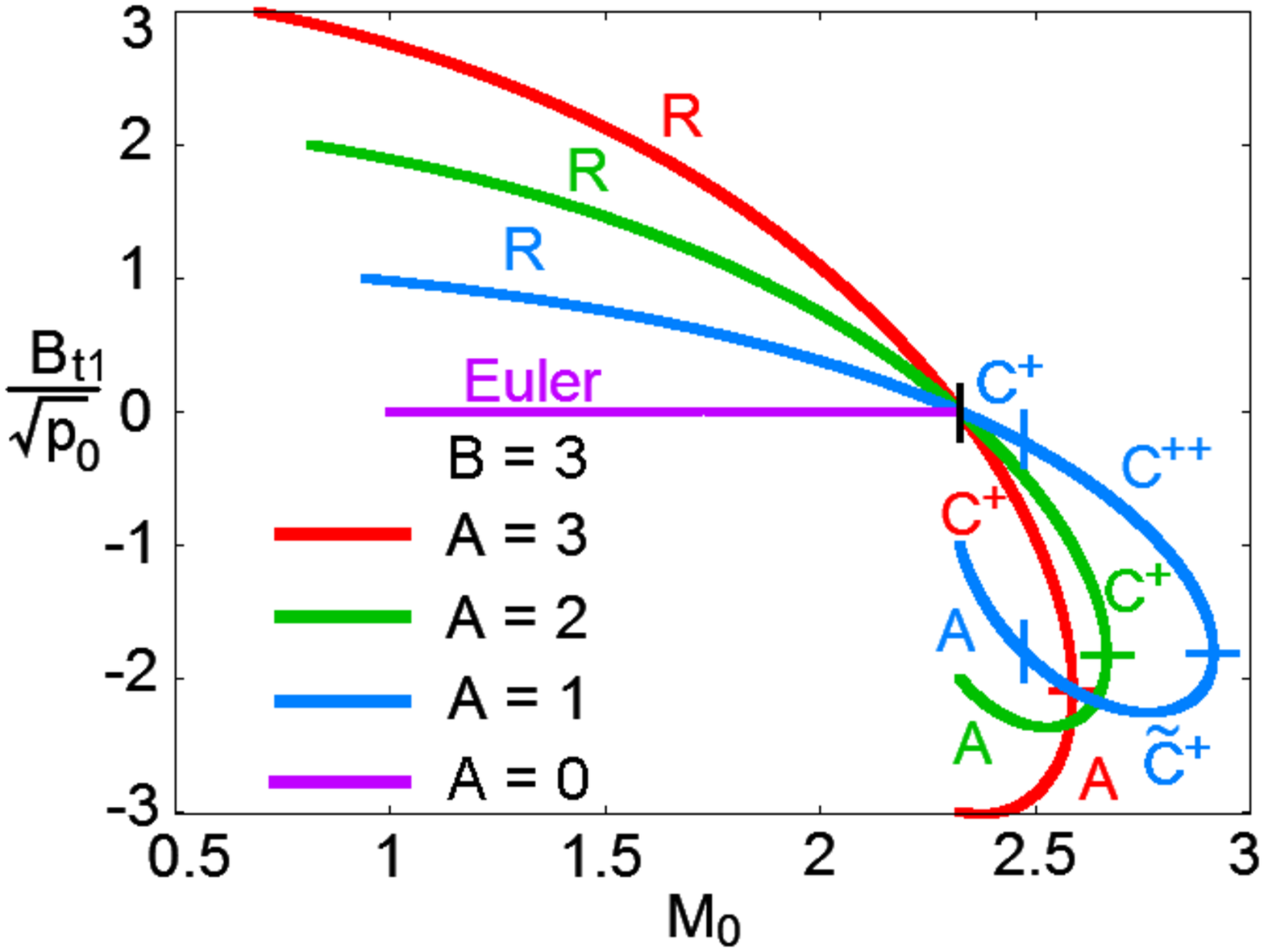}
\end{center}
\end{minipage} \\
\begin{minipage}{0.45\hsize}
\begin{center}
\includegraphics[scale=0.26]{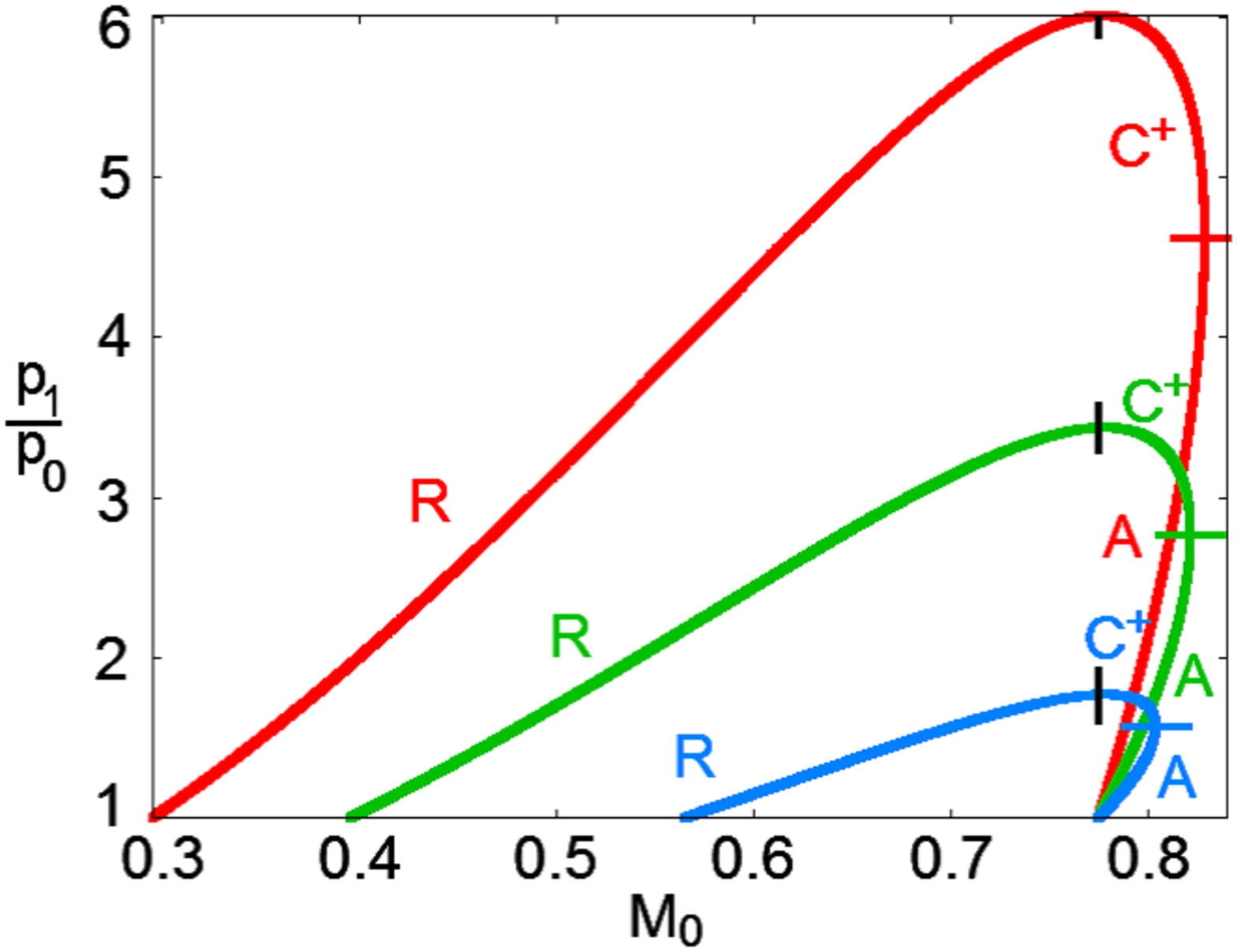}
\end{center}
\end{minipage} &
\begin{minipage}{0.45\hsize}
\begin{center}
\includegraphics[scale=0.26]{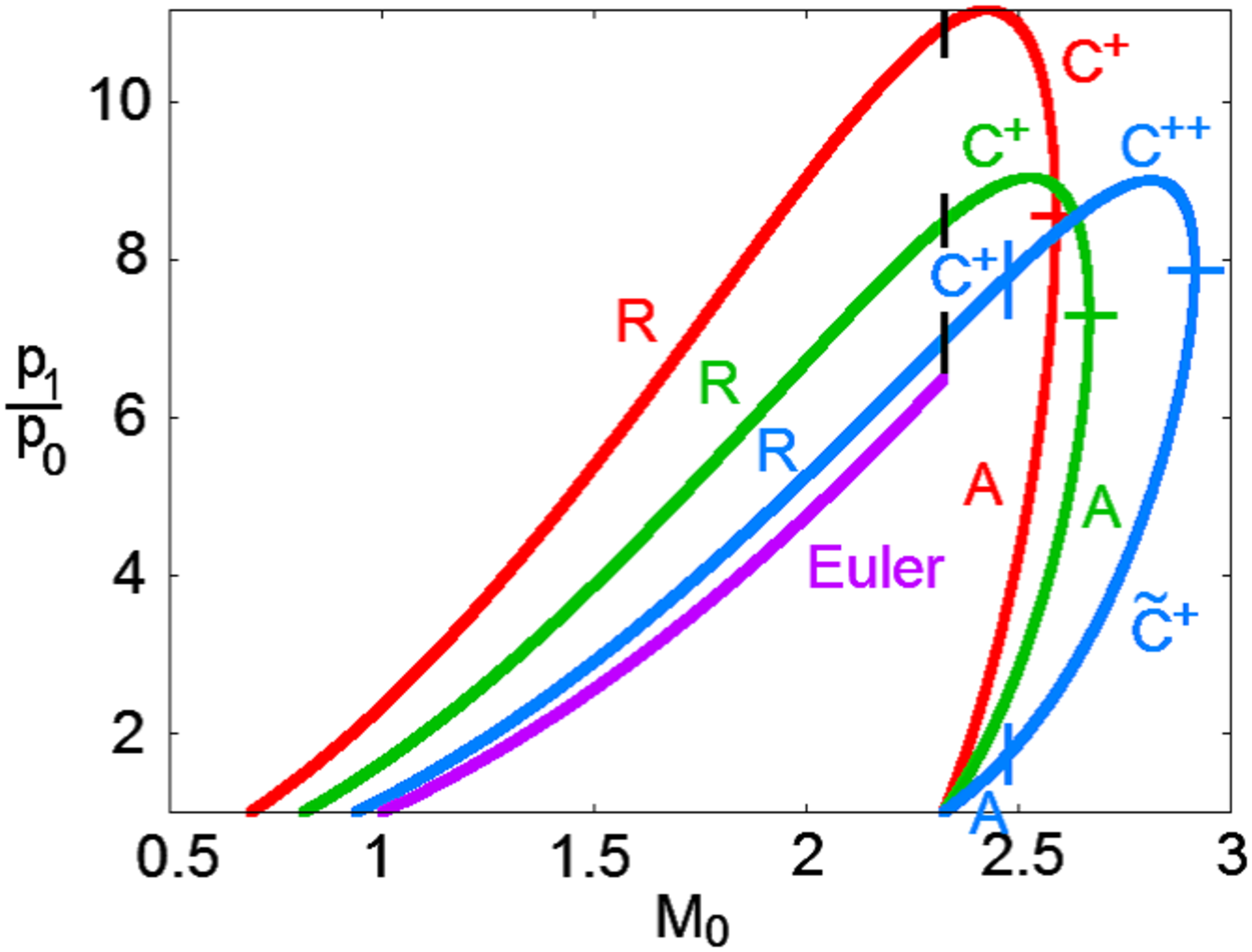}
\end{center}
\end{minipage} \\
\begin{minipage}{0.45\hsize}
\begin{center}
\includegraphics[scale=0.26]{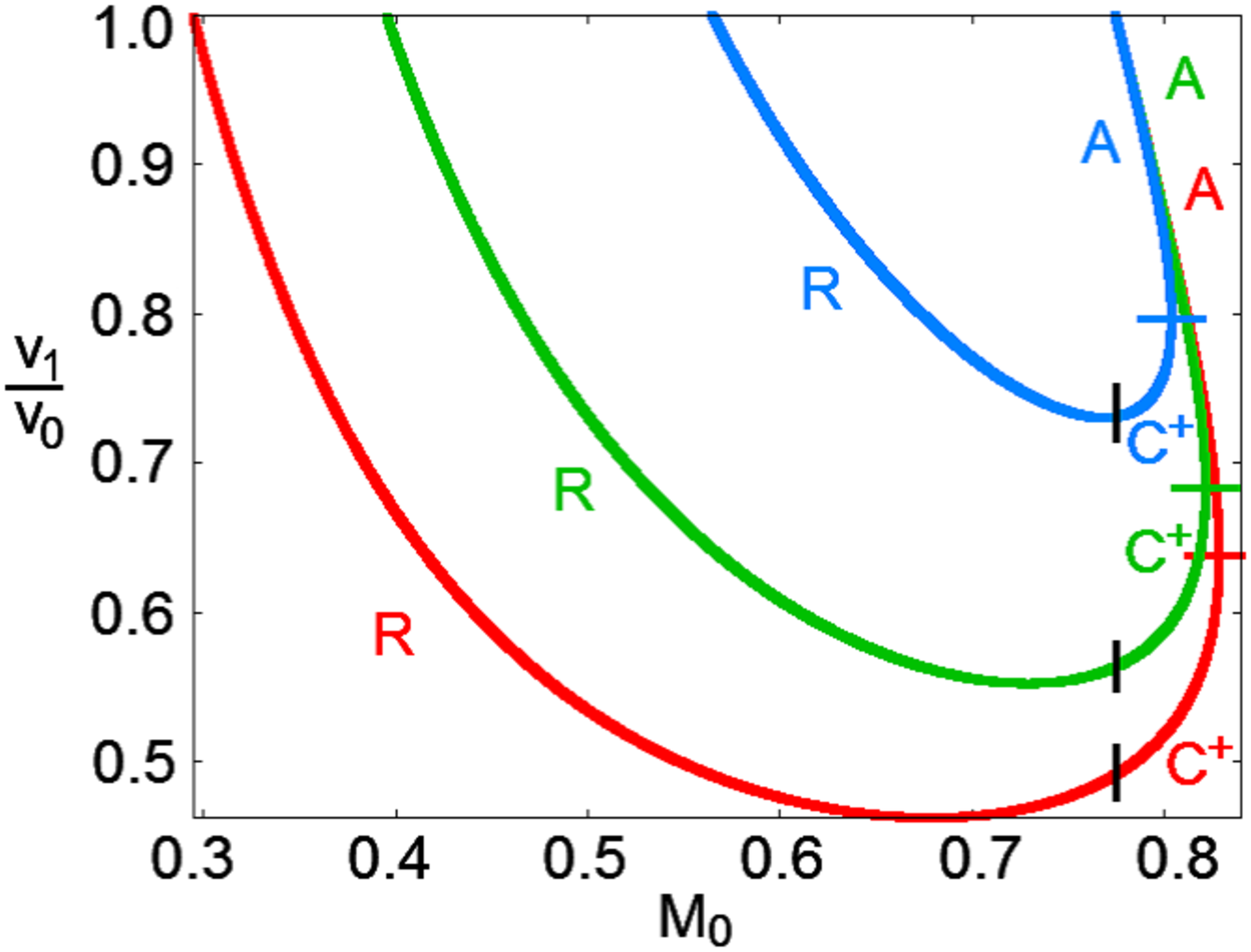}
\end{center}
\end{minipage} &
\begin{minipage}{0.45\hsize}
\begin{center}
\includegraphics[scale=0.26]{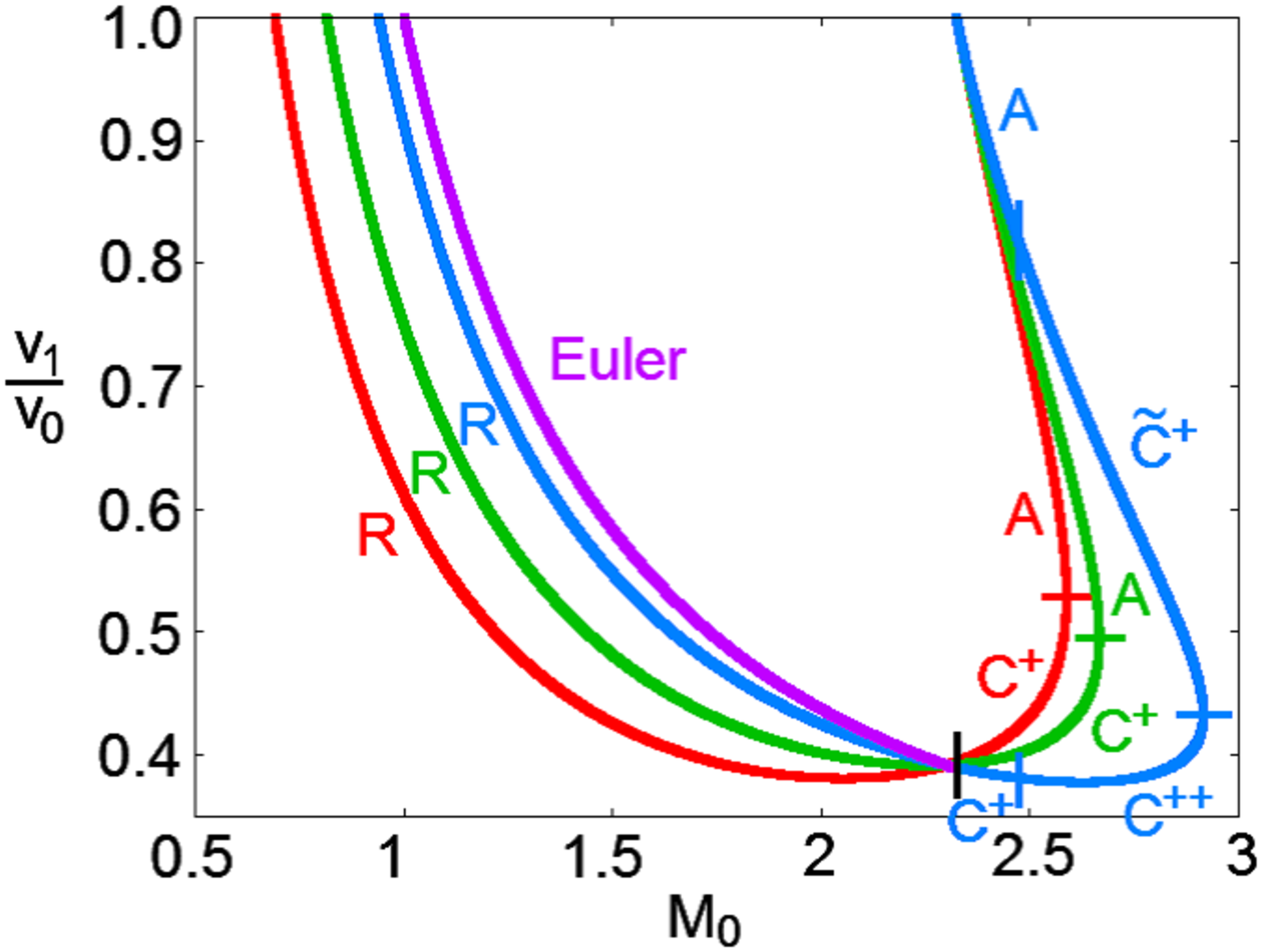}
\end{center}
\end{minipage} 
\end{tabular}
\caption{The slow loci for \textcolor{black}{$\gamma =5/3$ and} different combinations of the normal ($B$) and transverse ($A$) component of magnetic field.
The left panels: $B=1$ and $A = 3$ (red), $2$ (green) and $1$ (blue). The vertical black dashes indicate the points, at which 
$M_0 = \hat{c}_{A0}$ and switch-off shocks ($2,3 \rightarrow 4$ shocks) occur. Since $\hat{c}_{A0}$ is independent of $A$, the 
Mach numbers, $M_0$'s, at the points for all loci coincide with one another. These points mark the boundary between the regular slow 
shocks and non-regular intermediate shocks. The characters, $R$, $C^+$ and $A$, attached to each locus stand for the regular 
slow ($3 \rightarrow 4$), $2 \rightarrow 4$ intermediate and $2 \rightarrow 3$ intermediate shocks, respectively. The horizontal 
dash on each locus shows the point, at which the Mach number reaches its maximum on the locus and a $2 \rightarrow 3,4$ shock occurs.
This is the boundary between the $2 \rightarrow 4$ ($C^+$) and $2 \rightarrow 3$ ($A$) intermediate shocks. 
The right panels: $B=3$ and $A = 3$ (red), $2$ (green), $1$ (blue) and $0$ (purple, Euler shocks).
The vertical black dashes again give the boundary between the regular and non-regular shocks, at which switch-off shocks 
($2,3 \rightarrow 4$ shocks) occur. The characters, $R$, $C^+$ and $A$, have the same meaning as in the left panels whereas 
$C^{++}$ and $\tilde{C}^+$, which emerge only for small $A$'s, represent the $1\rightarrow 4$ and $1\rightarrow 3$ intermediate 
shocks respectively. The horizontal dash on each locus marks again the point, at which the maximum Mach number is reached. 
On the other hand, the two vertical blue dashes on each blue locus indicate the points, at which $M_0 = \hat{c}_{f0}$. 
A $1,2\rightarrow 4$ shock occurs at the point closer to the vertical black dashes whereas a $1,2 \rightarrow 3$ shock emerges at 
the other point. A $1 \rightarrow 3,4$ shock occurs at the point indicated by the horizontal blue dash.
Note in passing that the locus vanishes at $B=0$. The figure is quoted from \citet{TY12a}.
}
\label{fig.locus2}
\end{figure}

The slow family is characterized by the feature that matter is compressed but the transverse magnetic field is reduced and in some cases 
reversed by the shock passage. 
The slow loci are shown in Fig.~\ref{fig.locus2}, taken from \citet{TY12a}, as a function of the upstream Mach number ($M_0$) for a number of combinations of 
the upstream normal ($B$) and transverse ($A$) components of magnetic field. It is evident that some loci are two-valued as a function of $M_0$.

The intermediate shocks are those that give negative downstream transverse magnetic fields. The shocks that nullify the transverse magnetic field are called 
the switch-off shock ($2,3 \rightarrow 4$ shock). Switch-off shocks are located at the boundary between the regular slow shocks and the intermediate shocks. 
Each locus is terminated at the point that corresponds to a rotational discontinuity, which is incompressible and rotates magnetic field by $180^\circ$, i.e., $\Bth = -A$.
The minimum value of the downstream transverse magnetic field, $\hat{B}_{t,\mathrm{min}}$, is given as 
\begin{eqnarray}
\label{Bthmin}
\hat{B}_{t,\mathrm{min}} = -\frac{4(B^2 -\gamma )^2 +\gamma ^2A^4 +4A^2(B^2 +\gamma ^2)}{2\gamma AB^2 +\gamma (2 -\gamma )A(2 +A^2) +4B\sqrt{(\gamma -1)\Delta (A,B)}},\\
\Delta (A,B):= (B^2 -\gamma )^2 +A^2 \frac{\gamma (\gamma ^2 -2\gamma +2)}{\gamma -1} +A^2 (2B^2 +A^2),
\end{eqnarray}
which satisfies the condition that the discriminant of (\ref{quadra}) becomes zero \citep{T02}. Note that $\hat{B}_{t,\mathrm{min}} \le -A$ as seen in the figure.
We shall divide the locus by this minimum point. The part from the maximum in $(M_0,\Bth )$-plane, i.e., corresponding to $\Bth = A$, to the minimum is called the 'plus-branch', named after
a fact that the branch gives a larger $\vh$ in (\ref{quadra}) \citep{T02}. The other part, from the minimum to the end point, is called the 'minus-branch', which gives a smaller $\vh$.
Note that the minus-branch does not always exist as seen in the figure.
In the limit of $B \rightarrow 0$ or $A \rightarrow 0$, the whole slow branch vanishes, i.e., there is no solution that satisfies (\ref{fast branch}) and 
the inequalities: $\hat{c}_{s0} <M_0 < \hat{c}_{f0}$ and $0 < \vh < 1$, except when $A =0$ and the upstream \Alfven \ speed 
is larger than the acoustic speed, i.e., $\hat{c}_{A0}( := c_{A0}/a_0) > 1$, in which case the Euler shock branch takes its place. 
This branch is extended to the regime of $M_0 > \hat{c}_{f0}$ and connected smoothly to the fast-shock counterpart as mentioned earlier.

\subsection{Shock waves in vanishing normal magnetic field}
Without normal magnetic field, the structure of the shock solutions becomes much simpler because the slow shock loci disappear. 
In fact, (\ref{fast branch}) becomes a quadratic equation with the assumption that $\vh \ne 0$ in this case and,
discarding the solutions that satisfy $\vh \leq 0$, one obtains a unique solution:
\begin{equation}
\vh = \frac{(\gamma -1)M_0^2+A^2+2}{2(\gamma +1)M_0^2} +\sqrt{\left[ \frac{(\gamma -1)M_0^2+A^2+2}{2(\gamma +1)M_0^2} \right] ^2+\frac{(2-\gamma )A^{\color{red}2}}{\gamma (\gamma +1)M_0^2}}.
\end{equation}
The solution belongs to the fast shock because the transverse magnetic field behind the shock is amplified, following from (\ref{Euler_trans_hat}).
Otherwise the shock becomes the Euler shock provided there is no transverse magnetic field. Note here that switch-on shocks are never realized without a normal magnetic field.

\section{How to solve the MHD Riemann problems: in the case of regular solutions without switch-off rarefactions} \label{sec.method}
\subsection{Structure of the regular solutions without switch-off rarefactions}\label{Struct. Reg.}
We review here the way to find the regular solutions, in which no non-regular shock exists, in the case that neither normal nor transverse magnetic field vanishes 
in order to show the basic idea of solving the Riemann problems. 
Then, in the next sub-section, we propose the new strategy to obtain the non-regular solutions, which can also handle vanishing magnetic fields.
\begin{figure}
 \begin{center}
 \includegraphics[scale=0.4]{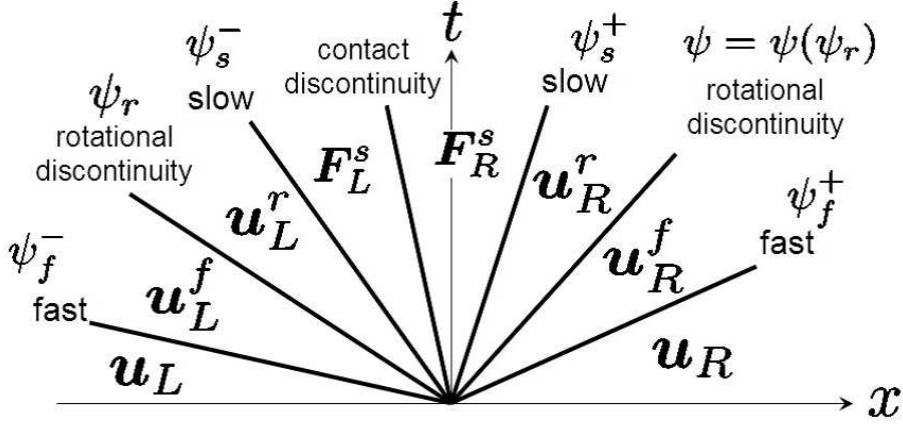}
 \caption{A schematic picture of the regular solution of the MHD Riemann problem in $(x,t)$-plane.
 There are generally seven waves, i.e., a contact discontinuity and fast waves, rotational discontinuities and slow waves running into both sides,
 provided the solution is restricted to the regular one.
 The letters with suffices, $\psi $s, attached to each wave stand for the parameters of each wave.
 The other characters with suffices, $\bm{u}$s and $\bm{F}$s, represent vectors of the conserved quantities in each states and 
 vectors of the quantities which should be continuous across the contact discontinuity respectively.}
 \label{solving_RP}
 \end{center}
\end{figure}

Assuming that both normal and transverse magnetic fields have finite value initially on both sides and ignoring intermediate shocks and switch-off waves, 
the structure of the solution is known \textit{a priori}: fast, \Alfven,\ slow waves fanning out in this order on both sides of a contact discontinuity. 
Since each wave forms a one-parameter family and seven waves exist in the solution, the structure of the solution is determined by fixing the seven parameters. 
One of degrees of freedom is the magnitude of a jump of density at the contact discontinuity, where other six quantities, i.e., pressure, three components of the velocity field and two components
of the transverse magnetic field, should be continuous. Therefore solving the MHD Riemann problems reduces to finding the six parameters that satisfy a requirement that 
the six quantities other than density are continuous across the contact discontinuity. Then remaining parameter associated with a contact discontinuity is necessarily fixed.
Furthermore, we can omit another degree of freedom which parameterizes a rotational discontinuity in either side as pointed out by \citet{T02}.
Thanks to the fact that only the rotational discontinuities rotate the magnetic field, if the angle of the rotation is fixed on either side, 
then another on the other side is necessarily determined to adjust the angle of the transverse magnetic field.
Eventually, there remain five parameters that should be determined to satisfy the conditions of continuity at a contact discontinuity. 

The five parameters can be found by the Newton-Raphson method and, schematically, the system of equations to be solved is 
\begin{equation}
\label{System of Riemann problem}
\bm{F}_L^s( \psi _s^- ;\bm{u}_L^r(\psi _r ;\bm{u}_L^f(\psi _f^-;\bm{u}_L) ) ) -\bm{F}_R^s( \psi _s^+ ;\bm{u}_R^r(\psi _r ;\bm{u}_R^f(\psi _f^+;\bm{u}_R) ) ) = \bm{0} ,
\end{equation}
where $\psi _f^\mp$, $\psi _s^\mp$ and $\psi_r$ are the parameters of the left/right fast wave, left/right slow wave and rotational discontinuity respectively.
$\bm{u}_{L,R}$ is a given initial state on left/right side, i.e., $\bm{u}_{L,R} = {}^t(\rho _{L,R}, p_{L,R}, \bm{v}_{L,R}, \bm{B}_{t L,R})$.
$\bm{u}_{L,R}^f$ is the left/right fast-wave function, which represents the downstream state of the fast wave
and is a function of the upstream variables, $\bm{u}_{L,R}$, and parameter of the fast wave, $\psi _f^\mp$.
Similarly, $\bm{u}_{L,R}^r$ is the left/right rotational discontinuity function and $\bm{F}_{L,R}^s$ is the left/right slow-wave function.
Here, slow-wave functions represent the five downstream variables of the slow waves which have to be continuous across the contact discontinuity, i.e., $p$, $\bm{v}$ and $|\bm{B}_t|$.
See also a schematic picture of the regular solution presented in Fig.~\ref{solving_RP}.

\subsection{Parameterization of the regular waves}
Although each wave forms a one-parameter family as mentioned earlier, finding the variable convenient to control the wave remains as another task.
For instance, slow shock family seems to have no convenient variable to parameterize the Hugoniot locus including the non-regular branch. 
In this sub-section we review the parameterizations to handle the regular waves, ignoring the intermediate shocks and switch-on/off waves \citep[see also][]{T02}.

The problem is how to map the parameters of the waves, which are directly improved by the Newton-Raphson method, to physical quantities, which have appropriate ranges. 
For example, a fast wave including the shock- and rarefaction-wave branches forms a one-parameter family and let $\psi _f$ denote the parameter of the fast wave. $\psi _f$ is improved 
by the Newton-Raphson method as well as the other parameters, $\psi$s, that are associated with other waves (we here omit the plus and minus signs for simplicity) and
the domain of the parameters is $\mathbb{R}$. On the other hand, the fast shocks can be parameterized by its Mach number, $M_0 \in [\hat{c}_{f0}, \infty)$, as mentioned earlier 
and the fast rarefactions can be parameterized by the length of the fast rarefaction locus in phase space, $s$.\footnote{The rarefaction locus in phase space connects the points corresponding to 
a front state and behind state respectively and is formed by integrating the eigenvector, $\bm{r}_{f}$. Hence, the behind state is uniquely given by the length of the locus, $s$, 
provided the front state is also given.}
\textcolor{black}{
That is, the state behind the rarefaction waves, $\bm{r}_\mathrm{behind}$, can be described as
\begin{equation}
\bm{r}_\mathrm{behind} = \int _0^s \bm{r}_f(s') \diff s'.
\end{equation}
}
Note that $s \in [0, s_\mathrm{max})$, where $s_\mathrm{max}$ corresponds to the maximum strength of the fast rarefaction, i.e., the strength of a switch-off rarefaction. 
Now, our concern is to construct a function that maps $\psi _f $ into $M_0$ or $s$.

Ignoring the switch-off rarefactions and switch-on shocks, one can define the fast-wave function as \citep{T02}
\begin{equation} \left\{
\label{para_fast}
\begin{array}{lll}
M_0 = \hat{c}_{f0} + \psi _f &(\psi _f > 0) &\mathrm{(Shock)} , \\
s = s_\mathrm{max} \tanh (-\psi _f ) &(\psi _f \leq 0) & \mathrm{(Rarefaction)} .
\end{array}\right. 
\end{equation}
That is, when $\psi _f$ is positive, the fast wave is the fast shock whose strength is determined by the upper part of (\ref{para_fast}). Otherwise, the fast wave 
is the fast rarefaction whose strength is given by the lower part of (\ref{para_fast}) and the wave never becomes switch-off rarefactions. We omitted the case of $M_0 = \hat{c}_{f0}$ because both 
$s=0$ and $M_0 = \hat{c}_{f0}$ mean that there is no fast wave. 
\textcolor{black}{The fast-shock solution is obtained from the cubic equation (\ref{fast branch}) by substituting this $M_0$. Although the roots have non-trivial structure in general as discussed in \citet{DK}, we can easily pick up the correct root that corresponds to a fast shock as described in Appendix \ref{appB}.
}
Note also that (\ref{para_fast}) is applicable to both left and right fast waves while we omitted plus and minus signs from the variables, 
$M_0, \hat{c}_{f0}, \psi _f, s $ and $s_\mathrm{max}$, for simplicity.

Similarly, discarding the non-regular branch, the slow family can be constructed as
\begin{eqnarray}\left\{
\label{para_slow}
\begin{array}{lll}
\Bth =A (1 -\tanh (\psi _s ))  & (\psi _s > 0) &\mathrm{(Shock)} , \\
s = -\psi _s  & (\psi _s \le 0) &\mathrm{(Rarefaction)} .
\end{array}\right. 
\end{eqnarray}
Here, $s$ is the length of the slow rarefaction locus in phase space. The ranges by this transformation are $\Bth \in (0,A)$ and $s \in [0, \infty)$. Therefore the shock includes only
regular slow shocks and never becomes switch-off nor intermediate shocks. We omitted the case of $\Bth = A$ here in order to prevent from doubly counting 
the situation that there is no slow wave, which is also described as $s = 0$. Note that (\ref{para_slow}) is applicable to both left and right slow waves as same as (\ref{para_fast}).

With respect to rotational discontinuities, the degree of freedom is the rotational angle of the transverse magnetic field. Therefore the parameter, $\psi _r$, can be transformed into 
the rotational angle, $\varphi$, as
\begin{equation}
\label{para_rot}
\varphi \equiv \psi _r \ (\mathrm{mod}\ 2\upi ) .
\end{equation}
This relation is used only for either side because the other rotational angle is necessarily fixed as mentioned in Sec. \ref{Struct. Reg.}; 
The rotational angle on the other side, say right, is automatically adjusted to $\theta _{L} -\theta _{R} +\varphi$, where $\theta _{L,R} $ are the initial angle of the magnetic field 
on left and right side respectively. If $\psi _r$ is the parameter of the right rotational discontinuity, then the rotational angle on left side should be $\theta _R -\theta _L +\varphi$.


\section{How to solve the MHD Riemann problems: in the case including intermediate shocks, switch-on/off shocks and switch-\textcolor{black}{on/}off rarefactions} \label{sec.method1.5}
In this section, we discuss the solutions of MHD Riemann problems, including intermediate shocks, switch-on/off shocks and switch-\textcolor{black}{on/}off rarefactions. 
Furthermore, we take the initial conditions with vanishing magnetic field into account.
There are mainly two differences from the previous section. Firstly, the parameterizations of the waves should be modified to cover all the branches. 
Secondly, the structure of the non-regular solutions cannot be known \textit{a priori} because some waves prohibit the emergence of other waves.
If a left-going $2 \rightarrow 4$ intermediate shock exists, for example, then the left-going rotational discontinuity and slow wave do not appear in the solution because 
$2 \rightarrow 4$ intermediate shocks skip the \Alfven \ and slow speeds. Therefore all the possible combinations of waves should be tried to find the solution and, moreover, the procedure 
to construct the solution should arrange the waves in appropriate order.
We discuss the parameterizations of the waves at first. Then, in the latter part, we discuss the arrangement of the waves, which is considerably associated with the parameterization.


\subsection{Parameterization of the non-regular shocks and switch-\textcolor{black}{on/}off rarefactions}\label{sec.parameter_non_regular}
Including switch-off rarefactions, we modify the fast-wave function (\ref{para_fast}) as follows.
\begin{equation} 
\label{fast_map}
\left\{
\begin{array}{lll}
M_0  = \ \hat{c}_{f0} + \psi _f \quad ( 0 \le \psi _f )  \qquad \qquad \quad \mathrm{(Shock)} , \\
s \ \ = \left\{  
\begin{array}{lll}
-\psi_f &\quad (-s_\mathrm{max} <\psi _f < 0) & \mathrm{(Rarefaction)} , \\
s_\mathrm{max} &\quad (\psi _f \leq -s_\mathrm{max} ) & \mathrm{(Switch\mathchar`- off\ rarefaction)},
\end{array} \right. 
\end{array} \right. 
\end{equation}
where we omitted plus and minus signs for simplicity as the previous section.
This function gives a switch-off rarefaction if $\psi _f \leq -s_\mathrm{max}$.
Note, however, that (\ref{fast_map}) is no longer injective since $\psi _f\leq -s_\mathrm{max}$ is always map into 
a particular value, $s_\mathrm{max}$. This property may cause trouble in the Newton-Raphson iteration, where the derivative of the function is needed, and we discuss the issue in Sec. \ref{sec.NR}.
Note also that the shock's part of (\ref{fast_map}) may not determine the downstream state uniquely because the two branches, the switch-on branch and ordinary Euler one, exist for a given Mach number
provided the upstream transverse magnetic field vanishes and the upstream Mach number satisfies the inequality (\ref{sw-on_ineq}). 
Therefore we need other rules to choose either branch for determining the downstream state uniquely. In our code, this degree of freedom remains as a setting parameter, i.e.,
we select either branch before running the program. 
If one chooses the switch-on shock branch, one should set the direction of the downstream transverse magnetic field as well because the shock can produce the field in arbitrary direction. 
\textcolor{black}{However, as mentioned later there is a good way to adjust the direction automatically
for the initial conditions where the transverse magnetic filed is absent on only one side. For such 
conditions, one does not have to mind the direction beforehand.}


With respect to slow waves, the intermediate shock branches \textcolor{black}{and switch-on rarefaction} should be included. \textcolor{black}{No special modification is necessary in the rarefactions while the parameterization of slow shocks becomes rather complicated, which are given as}
\begin{equation}
\label{slow_map}
\left\{
\begin{array}{lll}
\Bth = g(A, \hat{B}_{t,\mathrm{min}}  , \psi _s) &(\psi _s >0) & \mathrm{(Discontinuity)} , \\
s  = -\psi _s  & (\psi _s \le 0) & \mathrm{(Rarefaction)},
\end{array} \right.  
\end{equation}
where $\hat{B}_{t,\mathrm{min}} $ is the minimum value of the transverse magnetic field in the slow Hugoniot locus given by (\ref{Bthmin}). The function $g$ is defined for $A>0$ by
\begin{equation}
\label{g}
g(A, \hat{B}_{t,\mathrm{min}}, \psi _s ) =  \left\{
\begin{array}{ll}
 A -\psi _s   &(0< \psi _s \le A  +|\hat{B}_{t,\mathrm{min}} | ) , \\
 \psi _s -A -2|\hat{B}_{t,\mathrm{min}}|  & (A  +|\hat{B}_{t,\mathrm{min}} | < \psi _s \le 2|\hat{B}_{t,\mathrm{min}} |) , \\
 -A & (2|\hat{B}_{t,\mathrm{min}} | < \psi _s ).
\end{array} \right. 
\end{equation}
The range of the top equation in (\ref{g}) is $[\hat{B}_{t,\mathrm{min}} , A) \ni \Bth$ while the correspondence to the middle one is $(\hat{B}_{t,\mathrm{min}},-A] \ni \Bth$.
Note that the post-shock state may not be determined uniquely by (\ref{g}) because two branches can exist for a given $\Bth$ as mentioned in Sec. \ref{sec.Fast and slow loci}. 
Therefore we divide the slow Hugoniot loci into 'plus-branches' and 'minus-branches' as mentioned earlier. Then we take the quantities from the plus-branch, which includes
the regular slow shock, the switch-off shock and a part of the intermediate shocks, for $\psi _s\in(0,A  +|\hat{B}_{t,\mathrm{min}} | ]$, i.e., corresponding to the top equation in (\ref{g}). 
Otherwise, we use the minus-branch, which includes the intermediate shocks and rotational discontinuity located at the end point, 
for $\psi _s \in (A  +|\hat{B}_{t,\mathrm{min}} |, \infty )$, corresponding to the middle and bottom equations. 
In case there is no minus-branch, i.e., $\hat{B}_{t,\mathrm{min}} = -A$, the domain for the middle equation becomes the empty set.
Another noteworthy property is that the function gives $180^\circ$ rotational discontinuities, i.e., $\Bth = -A$ and $\vh = 1$, provided $\psi _s \in (2|\hat{B}_{t,\mathrm{min}} |, \infty )$,
reflecting the fact that the terminating points of slow Hugoniot loci give $180^\circ$ rotational discontinuities.  
The advantage that stems from this property is discussed in Sec. \ref{adv}.
For $A=0$, i.e., when the upstream transverse magnetic field is absent, the Euler shock branch may traverse the slow and fast Hugoniot locus plane.
The Euler shocks that belong to the slow branch can be parameterized as follows:
\begin{equation}
\label{Euler_map}
M_0 = 1 + (\hat{c}_{f0} -1) \tanh (\psi _s) \quad (\psi _s > 0).
\end{equation}
This function maps $\psi _s$ into the upstream Mach number of the Euler shock, $M_0 \in (1, \hat{c}_{f0})$. And we use this function instead of $g$ in order to control the slow shock.

With respect to rotational discontinuities, we use the same function (\ref{para_rot}) for determining a rotational angle provided no intermediate shock nor switch-off wave exists in the solution. 
Once intermediate shocks or switch-off waves emerge, however, the degree of freedom associated with a rotational discontinuity disappears as discussed below.
We begin from the case that an intermediate shock or a switch-on/off wave exists only in either side: (i) If an intermediate shock exists, the rotational discontinuity is skipped by the shock 
and the transverse magnetic field is reversed in the side. Therefore the rotational angle on the other side is necessarily fixed to $\theta _{L,R} -\theta _{R,L} +\upi$ for the right/left side to adjust 
the direction of the field.
(ii) If a switch-off shock ($2,3 \rightarrow 4$ shock) or switch-off rarefaction exists, the wave quenches the transverse magnetic field and, as consequences, the rotational discontinuity vanishes.
And no wave produce the magnetic field in the side since only an ordinary rarefaction or ordinary Euler shock can exist behind the switch-off waves. Therefore any switch-off wave 
should appear and hence the rotational discontinuity disappears on the other side. When such waves that prohibit the rotational discontinuity emerge in both sides, no rotational discontinuity 
exists in the solution, of course. In this way, if non-regular shocks or switch-off rarefactions emerge in either side, we do not need consider the rotational angle of the rotational discontinuity in the other side, if any. The problem associated with the disappearance of the degree of freedom is discussed in Sec. \ref{sec.NR}.

\subsection{Structure of the non-regular solutions and how to arrange the waves}
We here discuss the structure of non-regular solutions and propose a process to arrange the waves in appropriate order, which are associated with the parameterization 
discussed in the previous sub-section. As mentioned repeatedly, the structure of the solution is not known \textit{a priori} when the non-regular shocks and switch-\textcolor{black}{on/}off rarefactions are included and
therefore all the possible combinations of the waves should be tried to find the solution. Our method realizes this requirement;
It searches all the patterns and finds the solution automatically in the Newton-Raphson iterations.

Outline of the arranging process is as follows. At first, we arrange the waves on either side, say right, of the contact/tangential discontinuity, based on a given initial guess that controls the right-going waves.
After arranging the right-going waves, we obtain the right state of the contact/tangential discontinuity. Then we take up the other side similarly and we will obtain the left state of the discontinuity, where
the left and right states should satisfy the Rankine-Hugoniot conditions. Until the conditions are satisfied, the process is iterated by the Newton-Raphson method. 
Hereafter, we explain the process to arrange the waves on the left side and the procedure on the other side is almost same.

\subsubsection{In the case that both the transverse and normal magnetic field exist in the left initial condition} \label{roots_1}
We discuss the way to arrange the left-going waves in the case both the transverse and normal magnetic fields exist in the left initial condition. 
In this case, the first wave running to the left can be a fast-family wave, i.e., a fast rarefaction or fast shock,
or an intermediate shock whose upstream speed in rest frame is super-fast or equal to the fast speed, i.e., a '$1 \rightarrow 3$', '$1 \rightarrow 4$', '$1,2 \rightarrow 3$', '$1,2 \rightarrow 4$', 
'$1 \rightarrow 3,4$' or '$1,2 \rightarrow 3,4$' shock.\footnote{It is shown that the shocks designated as '$1,2\rightarrow 2,3$' do not exist.} 
If the parameter that controls the left-going slow wave is positive, i.e., $\psi _s^->0$, we firstly consider the pattern that includes such an intermediate shock and we try inserting an intermediate shock 
which is given by $\psi _s^-$ and its upstream state through the slow-shock function (\ref{slow_map}). 
Note that the 'trial intermediate shock' may not satisfy the assumption that the 
upstream flow speed is not slower than the fast speed, $M_0 \ge \hat{c}_{f0}$, or the trial shock may not be even an intermediate shock since the function (\ref{slow_map}) includes the regular slow shock and 
other intermediate shocks whose upstream flow speed is sub-fast as well. 
Moreover, it can also occur that the slow shock branch does not include the intermediate shock we assumed here, as seen in the left panels in Fig. \ref{fig.locus2}. 
In such a case, we reject the assumption that the leftmost wave is an intermediate shock and alternatively insert 
a fast wave as the leftmost wave, whose strength is given by $\psi_f^-$ through the fast-wave function (\ref{fast_map}). 
Only if the shock satisfies $M_0 \ge \hat{c}_{f0}$, we accept the trial intermediate shock.
If $\psi _s^- \le 0$, on the other hand, we need not consider any intermediate shock because $\psi _s^- \le 0$ gives only slow rarefactions and hence the first wave is necessarily a fast wave 
that is given by $\psi _f^-$ through (\ref{fast_map}). We discuss below the case that the leftmost wave is (a) a fast wave or (b) an intermediate shock.\footnote{We should note that 
the intermediate shocks whose upstream flow speed is super-fast need to be discarded in some cases; otherwise a class of solutions is missed as mentioned in Sec. \ref{elude}.
Therefore we designed our code so that we can choose whether we neglect such intermediate shocks or not before running the program. In the case we discard such shocks, the path starts always at (a).}

(a): We further divide the situation into the four cases as follows. 
(a-i) $\psi _s^- \le 0$ and the fast wave is not a switch-off rarefaction. (a-ii) $\psi _s ^-\le 0$ and the fast wave is a switch-off rarefaction. 
(a-iii) $\psi _s^->0$ and the fast wave is not a switch-off rarefaction. (a-iv) $\psi _s^- >0$ and the fast wave is a switch-off rarefaction. Note that (a-i) and (a-ii) are the simplest cases, 
where we can ignore the possibility of the intermediate shocks. 

(a-i): In this case, the fast wave is followed by a rotational discontinuity and a slow rarefaction in this order. 
The slow rarefaction is controlled by $\psi _s^-$ through the slow-wave function (\ref{slow_map}).
With respect to the rotational discontinuity, its treatment is different between the right side and left one.
On the right side, where we assumed that the waves are arranged before the left side, the rotation angle is 
given by $\psi _r$ through (\ref{para_rot}) while the counterpart on the left side is necessarily determined by the waves on the right side as mentioned in the previous sections.

(a-ii): Since the transverse magnetic field vanishes behind the switch-off rarefactions, the rotational discontinuity disappears and only an ordinary rarefaction or ordinary Euler shock follows. 
We note here that behind switch-off rarefactions, $a _1< c_{A_1}$ is always satisfied, i.e., the fast speed and \Alfven \ speed always degenerate and the acoustic speed and slow 
speed coincide with each other.
Therefore the ordinary wave belongs to slow family and, hence, we here control it by $\psi _s^- \le 0$ thorough the slow-wave function (\ref{slow_map}).
Since $\psi_s^- \le 0$ gives rarefactions, the third wave is necessarily an ordinary rarefaction. 

(a-iii): When $\psi_s^- >0$, we need consider the possibility of the intermediate shocks for the second wave. Since the downstream state of the fast wave 
is designated as '$2$' or '$1,2$' for the shocks and rarefactions respectively, 
the fast shock can be followed by a '$2\rightarrow \star$' or switch-off ($2,3 \rightarrow 4$) shock, where '$\star$' stands for '$3$', '$4$' or '$3,4$', 
and the fast rarefaction can be also followed by one of these non-regular shocks or a '$1,2 \rightarrow \star$' shock.\footnote{It is shown that the shocks 
designated as '$2,3 \rightarrow 3,4$' do not exist.}
Therefore a trial intermediate shock is given again by $\psi _s^-$ through (\ref{slow_map}). Note that the upstream state values
are now given by the downstream state of the fast wave. As before, the trial intermediate shock may not be an intermediate shock desired here and we should confirm that the shock does not
overtake the preceding fast wave.\footnote{Note that the slow-shock function (\ref{slow_map}) contains the rotational discontinuity as mentioned earlier and the rotational discontinuities
can also be inserted behind the fast waves. Therefore we accept the rotational discontinuity that emerges as a trial 'shock' in this moment 
even though the discontinuity is not a shock. The advantage of this treatment is discussed in Sec. \ref{sec.method2}.}
If the trial shock does not satisfy such conditions, we discard it and, as consequences, the following waves are a rotational discontinuity and a regular slow shock, 
which are handled in the same manner as (a-i). 
If the trial shock is acceptable as the second wave, on the other hand, then the downstream state is designated as '$3$', '$4$' or '$3,4$', i.e., $\hat{c}_{s1} <M_1<\hat{c}_{A1}$, $M_1 < \hat{c}_{s1}$ or
$M_1 = \hat{c}_{s1}$ respectively.
For the second case, including the case of switch-off shocks,
no wave follows and the non-regular shock is the last wave. For the third case, only a slow rarefaction can follow while the possible wave for the first case is either a slow rarefaction or regular 
slow shock whose shock speed is smaller than the preceding intermediate shock. 
Note, however, that there remains no parameter associated with the left slow wave because we have already used $\psi _s^-$ for the intermediate shock. Therefore we control the slow wave behind 
the intermediate shock by introducing an extra parameter, $\psi _{ex}^-$, which gives the strength of the slow wave through the regular-slow-wave function (\ref{para_slow}). 
This parameter is also iteratively improved by the Newton-Raphson method, which is discussed in Sec. \ref{sec.NR}.

(a-iv): As mentioned in (a-ii), the switch-off rarefaction is followed only by an ordinary wave. 
Since $\psi _s^- >0$, the following wave is an ordinary Euler shocks whose strength is given by $\psi _s^-$ through the Euler-shock function (\ref{Euler_map}).

(b): If the first wave is an intermediate shock, which is controlled by the slow function (\ref{slow_map}), then the downstream state is either '$3$', '$4$' or '$3,4$'. 
Therefore the patterns of the following waves are the same as that discussed in the latter part of (a-iii); That is, no wave or an extra slow wave follows.
Here, we also introduce the extra parameter $\psi _{ex}^-$ for the slow wave.

\begin{figure}
\begin{center}
\includegraphics[scale=0.2]{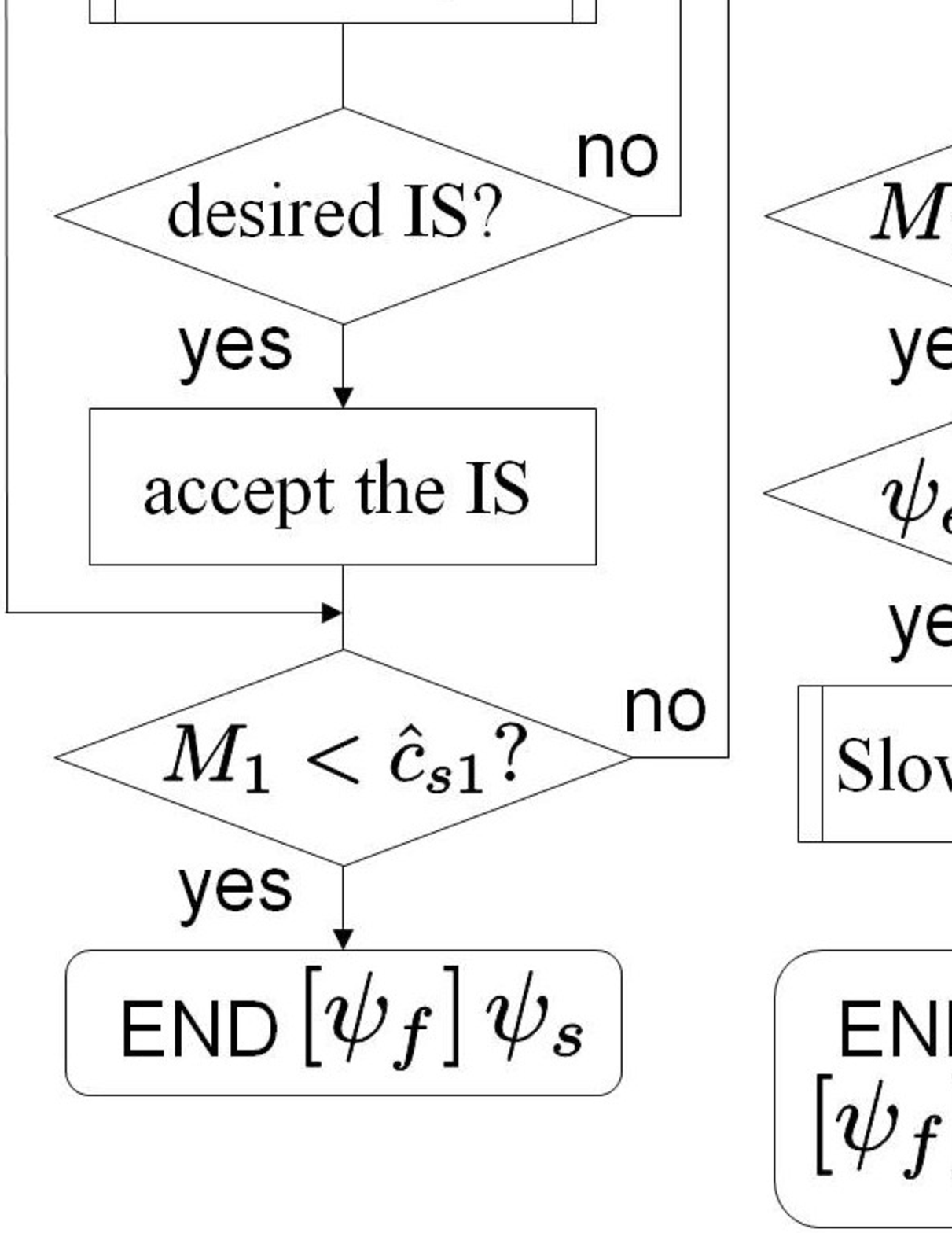}
\caption{The flow chart for arranging the waves in one side where there are both the transverse and normal magnetic fields. The squares with double lines on each side represent the subroutines 
that insert a wave with the use of the corresponding function and parameter. The letters, $\psi$s, attached to these functions are the parameters which the function uses. More specifically,
the functions entitled 'trial IS' give the trial intermediate shock whose strength is determined by $\psi _s$ through (\ref{slow_map}). The 'Fast' functions give the fast wave, controlled
by $\psi _f$ through (\ref{fast_map}), including the switch-off rarefactions. The 'Euler' function gives the ordinary Euler shock, controlled by $\psi _s$ through (\ref{Euler_map}). 
The 'Ordinary RW' function gives the ordinary rarefaction, controlled by $\psi _s$ through (\ref{slow_map}). The 'Rot.' functions give the rotational discontinuity 
whose rotational angle is determined by $\psi _r$ unless the other rules determine the angle due to the waves on the other side.
The 'Regular Slow' function returns the regular slow wave, controlled by $\psi _{ex}$ through (\ref{para_slow}). 
The 'Slow RW' function gives the slow rarefaction, controlled by $\psi _s$ or $\psi _{ex}$ through (\ref{slow_map}). The 'Slow SW' function gives the regular slow shock, 
controlled by $\psi _s$ through (\ref{para_slow}). The designations, (a-i)-(a-iv) and (b), attached to the branches correspond to the paths that are mentioned in Sec. \ref{roots_1}. The 
letters, $\psi$s, described in the terminals are the free parameters that are used in the path and improved by the Newton-Raphson method in the succeeding process. Note that the parameters 
given in square brackets mean that the parameter is not a free parameter if the path passes through (b) as discussed in Sec. \ref{sec.NR}. Similarly, the letter in curly parentheses means that 
the parameter is not a free parameter only if the accepted trial 'shock' is a rotational discontinuity. The parameters given in round brackets mean that the parameter is not always used in the path. 
}
\label{flow_chart_1}
\end{center}
\end{figure}

The procedure to arrange the waves discussed above are summarized in Fig. \ref{flow_chart_1} as a flow chart.
Note also that the number of the waves and, as consequences, the number of the free parameters are different for each case.
The issue is associated with the way to treat the parameters in the Newton-Raphson method, which is discussed in Sec. \ref{sec.NR}.

\subsubsection{In the case that the initial transverse magnetic field is absent in the left side}
\textcolor{black}{
Firstly, note that if the transverse magnetic field exists on the other side, then we arrange the 
waves on that side. This is because some switch-on wave may emerge in this side. Because 
the magnetic field should coincide at the contact discontinuity, the direction of the transverse magnetic field that is produced by the switch-on wave is necessary determined by arranging the waves on the other side beforehand. In this sub-section, we suppose that we already arranged the waves on the right side.}

Without the transverse magnetic field, the fast and slow speed degenerates into the \Alfven \ or acoustic speed, depending on the magnitudes of those speeds. 
Accordingly, we separately discuss the two situations: (p) $\hat{c}_{A0} \le 1$ and (q) $\hat{c}_{A0} > 1$, where $\hat{c}_{A0}$ is the \Alfven \ speed of the given initial condition normalized 
by the acoustic speed. 

(p): As shown in Appendix \ref{appA}, there is no chance for the switch-on shocks in this case and, hence, only the Euler shocks are allowed if any shock runs. Recalling 
$\hat{c}_{f0} = 1 \ge \hat{c}_{A0} = \hat{c}_{s0} $ and a fact that the flow speed changes from super-acoustic to sub-acoustic across Euler shocks, the flow speed in front of the Euler shock is super-fast. 
Therefore the Euler shock belongs to the fast branch in this case and is controlled by $\psi _f^- \ge 0$ through (\ref{fast_map}). Similarly, the ordinary rarefaction is controlled by $\psi _f^- < 0$ since the 
ordinary rarefactions flow with acoustic speed. 
\textcolor{black}{ 
After the ordinary wave, a switch-on slow rarefaction can follow if $\hat{c}_{A1} \le 1$, where 
$\hat{c}_{A1}$ stands for the \Alfven \ speed behind the first wave normalized by the acoustic speed.
Since we control the switch-on rarefactions through (\ref{slow_map}), it follows only if $\psi _s^-$ is negative. Otherwise, no wave follows and only the ordinary wave runs on the side.
}

\begin{figure}
\begin{center}
\includegraphics[scale=0.2]{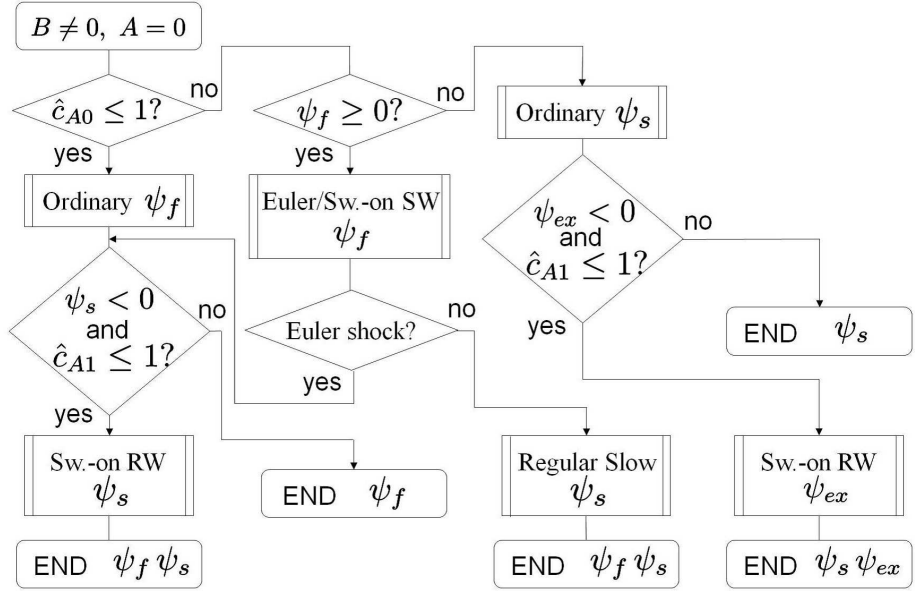}
\caption{The flow chart for arranging the waves when there is no transverse magnetic field initially while the normal magnetic field exists. 
\textcolor{black}{
The function entitled 'Euler/Sw.-on SW' is controlled by $\psi _f$ through (\ref{fast_map}) and gives an ordinary rarefaction wave if  $\psi _f < 0$ or Euler or Switch-on shock otherwise. Note here that the switch-on branch is used only when the branch is selected beforehand. The 'Regular Slow' function gives a regular slow shock, controlled by $\psi _s$ through (\ref{para_slow}). The 'Sw.-on RW' one handles a switch-on slow rarefaction which is given through (\ref{slow_map}).
The 'Ordinary' one gives an ordinary rarefaction or Euler shock. Here, one with $\psi _f$ is controlled through (\ref{fast_map}) while one with $\psi _s$ is handled through (\ref{slow_map}) and (\ref{Euler_map}) for rarefactions and shocks respectively.}
}
\label{flow_chart_2}
\end{center}
\end{figure}

\textcolor{black}{
(q): In this case both the Euler shocks and switch-on shocks are allowed.
Recalling $\hat{c}_{f0} = \hat{c}_{A0} > 1 = \hat{c}_{s0} $, the Euler shock lies in both the fast and slow branches. Note here that we need select either of switch-on shock branch or Euler shock one that is used for the Mach numbers in the overlap region before running the program as mentioned in Sec. \ref{sec.parameter_non_regular}.
And two wave-patterns are possible as explained below: 
(q-i) an Euler shock or ordinary rarefaction which is possibly followed by a switch-on rarefaction and 
(q-ii) a switch-on shock followed by a slow shock or rarefaction. The wave-pattern is determined as follows. If $\psi _f^- > 0$, the first wave is either the switch-on shock or Euler one, which are given by $\psi _f^-$ through (\ref{fast_map}). In the case of the Euler shock, there is a chance for a following switch-on rarefaction like the case of (p). Hence, the pattern (q-i) where the flow speed is super-fast in front of the shock is realized. In the case of the switch-on shocks, on the other hand, the pattern (q-ii) is realized. The following slow wave is given by $\psi _s^-$ through (\ref{slow_map}).
If $\psi _f^- \le 0$, which has no corresponding rarefaction wave now, then the first wave is controlled by $\psi _s^-$ through (\ref{slow_map}). $\psi _s^- \le 0$ gives an ordinary rarefaction and the wave pattern comes to (q-i). Note that switch-on rarefactions never follow in this case because $\hat{c}_A$ increases across rarefaction waves and $\hat{c}_{A1}$ is necessarily larger than unity.
On the other hand, $\psi _s^- > 0$ gives an Euler shock whose flow speed is super-slow and sub-fast.
The shock is followed by a switch-on rarefaction if $\hat{c}_{A1} \le 1$. Since we already used 
$\psi _s^-$, we control the switch-on rarefaction by introducing $\psi _{ex}^-$. Then the pattern (q-i) where the flow speed is sub-fast and super-slow in front of the shock is realized.
Note that the number of the waves is one or two in these cases. See also the flow chart presented in Fig. \ref{flow_chart_2}.
}

\subsubsection{The case without normal magnetic field}
In this case, the structure of solutions is \textit{a priori} known: two fast waves fanning out on both sides of a tangential discontinuity. The fast waves are controlled by $\psi _f^\mp$
thorough (\ref{fast_map}) on the left and right sides respectively. Note that switch-off rarefaction waves do not exist and $s_\mathrm{max} = \infty$ in this case.
Different from the contact discontinuities, the Rankine-Hugoniot conditions of the tangential discontinuities require the continuity of the total pressure and normal velocity.
Hence, there are always two fast waves and two matching conditions.

\section{Other technical details}\label{sec.method2}
\subsection{Modified Newton-Raphson method}  \label{sec.NR}
We discuss how to modify the Newton-Raphson method in order to handle the case that the number of the parameters changes in the iteration process. This modification is necessary because 
the number of the waves in the solution changes due to the intermediate shocks, switch-on/off waves or extra waves as mentioned in the preceding section. Furthermore, since the mappings (\ref{fast_map}) 
and (\ref{slow_map}) are not injective for the switch-off rarefaction and the rotational discontinuity respectively, 
the differentials of the quantities at the contact discontinuity with respect to $\psi _f$ or $\psi _s$ in such cases are zero and, as a result, the Jacobi matrix becomes singular. 
Therefore, avoiding the singularity, the parameters associated with those waves should be omitted in the Newton-Raphson procedure, i.e., the parameters are not
improved but hold their values.

On the other hand, there are always five conditions that should be satisfied at the contact discontinuity unless the normal magnetic field vanishes.
Therefore we need ignore some equations and find the solution of the reduced system and then, as a post process, we check whether the other conditions are satisfied.
There seems to be no special strategy for selecting the equations that are omitted although 
the code is designed so that the equations for the magnitude of the transverse magnetic field, $z,y,x$-components of the velocity and pressure are ignored in this order 
when the number of equations is adjusted.

We also note that the number of equations may be reduced although the Jacobi matrix is regular;
This occurs when the initial condition is confined in the $(x,y)$-plane due to the non-existence of the $z$-components of the velocity and magnetic fields. 
If none of the waves arranged on both sides cannot produce the $z$-components, for example such case that two fast waves and two slow waves 
fan out on both of the contact discontinuity and a $2\rightarrow 3$ intermediate shock runs into the right side, then the differentials of the differences of $v_z$ and $B_z$ at the contact discontinuity 
with respect to any parameter become zero since $v_z$ and $B_z$ are absent throughout the space for any combination of the parameters unless the wave-pattern changes and the rotational 
discontinuities appear. Therefore these equations associated with $z$-component should be neglected, otherwise the corresponding rows of the Jacobi matrix lead the singularity. 
Accordingly, we also reduce the number of the parameters to three, corresponding to that of the equations, provided there are more than three parameters. 
Since there seems to be no general prescription for choosing the parameters that are discarded, 
we remain the issue as a setting parameter, i.e., we plan which parameters are neglected before running the program.

If the normal magnetic field vanishes, on the other hand, these modifications are never necessary because there are always two fast waves, which are not switch-off rarefactions,
running into both sides of a tangential discontinuity and there are just two matching conditions at the tangential discontinuities: the continuity of the total pressure and normal velocity.
That is, the Jacobian is always a $2\times 2$ matrix and does not become singular.


\subsection{How to obtain the maximum strength of the fast rarefactions}
As mentioned repeatedly, the fast rarefaction branches terminate at the switch-off rarefactions and therefore there is the maximum strength of the fast rarefaction for the given upstream state
provided the normal magnetic field has any finite value.
Once the initial condition is given, the maximum strength is known since the state in front of the switch-off rarefaction is the given initial state. Principally, we obtain the strength, 
$s_\mathrm{max}$, by solving the equation below.
\begin{equation}
\bm{B}_{t,\mathrm{behind}}(s_\mathrm{max}) = \int _0 ^{s_\mathrm{max}} \frac{\bm{B}_t(s')}{(c_A(s')/c_{f}(s'))^2 -1} \diff s' = \bm{0},
\end{equation}
where we integrate the fast eigenfunction, $\bm{r}_f$, and picked up the component of the transverse magnetic field. The equation is solved by numerically integrating the integrand and finding 
the value of such $s$ that the transverse magnetic field is quenched. 
Then the $s_\mathrm{max}$ is used throughout the calculation. Note that values of $s_\mathrm{max}$ on the right and left sides are generally different.

\subsection{Remark on the class of solutions that eludes the search} \label{elude}
As mentioned in the footnote in Sec. \ref{roots_1}, there is the class of solutions that the algorithm cannot find unless we discard the intermediate shocks whose upstream flow speed is greater than 
fast speed. More specifically, we may miss the solutions that include $2 \rightarrow  \star$ intermediate shocks, where '$\star$' stands for '$3$', '$4$' or '$3,4$', provided the initial condition allows
the $1 \rightarrow \star$ intermediate shocks. For example, suppose that an initial condition whose left state allows the emergence of the $1 \rightarrow \star$ intermediate shocks
has a solution that includes a left-going $2 \rightarrow 4$ intermediate shock whose strength is given by $\psi _s^- = \psi _0 >0$ through (\ref{slow_map}). 
And suppose that we also know all the other value of the $\psi$s that parameterize each wave in the solution. Then, if we give the $\psi $s as the initial guess and follow naively 
the flow chart (Fig. \ref{flow_chart_1}), will the solution reproduce? The answer might be no because, following the flow chart, the first step is inserting a trial intermediate shock with hope that
the fastest wave is a $1 \rightarrow \star$ intermediate shock. If the $\psi _0$ gives a $1 \rightarrow \star$ shock in this first step, then we follow the path (b) and there is no chance 
for the $2\rightarrow 4$ intermediate shock to be considered. To produce the solution that includes the $2\rightarrow 4$ shock, we should reject the first trial intermediate shock.
Therefore we designed our code as we can search such hidden solutions by discarding always the first trial intermediate shock; It is a setting parameter whether the first trial intermediate shock is always
neglected or not. Trying both the settings, we can find all the solutions.

\subsection{The advantage of including the rotational discontinuity in the slow-shock function} \label{adv}
The non-regular-slow-shock function (\ref{slow_map}) includes the rotational discontinuity, as mentioned earlier, which lies at the end point of the slow Hugoniot locus. 
Owing to this feature, the rotational discontinuity followed by a slow wave can be realized with two ways: 
the combination of $\psi _r$ and $\psi _s$, where the parameters give the rotational discontinuity and slow wave through the function of rotational discontinuities (\ref{para_rot}) and 
the slow-wave function (\ref{para_slow}) respectively, 
or the pair of $\psi _s$ and $\psi _{ex}$, where the $\psi _s$ now gives the rotational discontinuity through (\ref{slow_map}) while the $\psi _{ex}$ gives the slow wave through (\ref{para_slow}).
Since the degeneracy violates the one-to-one correspondence of the wave parameters and the structure of the solution, the parameterization may seem to be awkward.
However, thanks to this parameterization, the non-regular solutions and regular solution can form a one-parameter family
as the $2 \rightarrow 3$ intermediate shocks continuously change toward the rotational discontinuity. In fact, some initial conditions have uncountably infinite solutions that form a one-parameter 
family of $\psi _s$ whose end point is the regular solution and smoothly connected to the non-regular solutions that include a $2 \rightarrow 3$ intermediate shock instead of the rotational discontinuity. 
Such examples are shown in the next section and our previous paper \citep{TY12a}.

\section{Examples of the exact solutions}\label{sec.results}
In this section, we show some examples of the exact solutions of MHD Riemann problems in order to demonstrate our code.
See also \citet{TY12a}, which presents many examples of the exact solutions that are obtained by our code.

\begin{figure}
\begin{tabular}{cc}
\begin{minipage}{0.45\hsize}
\begin{center}
\includegraphics[scale=0.26]{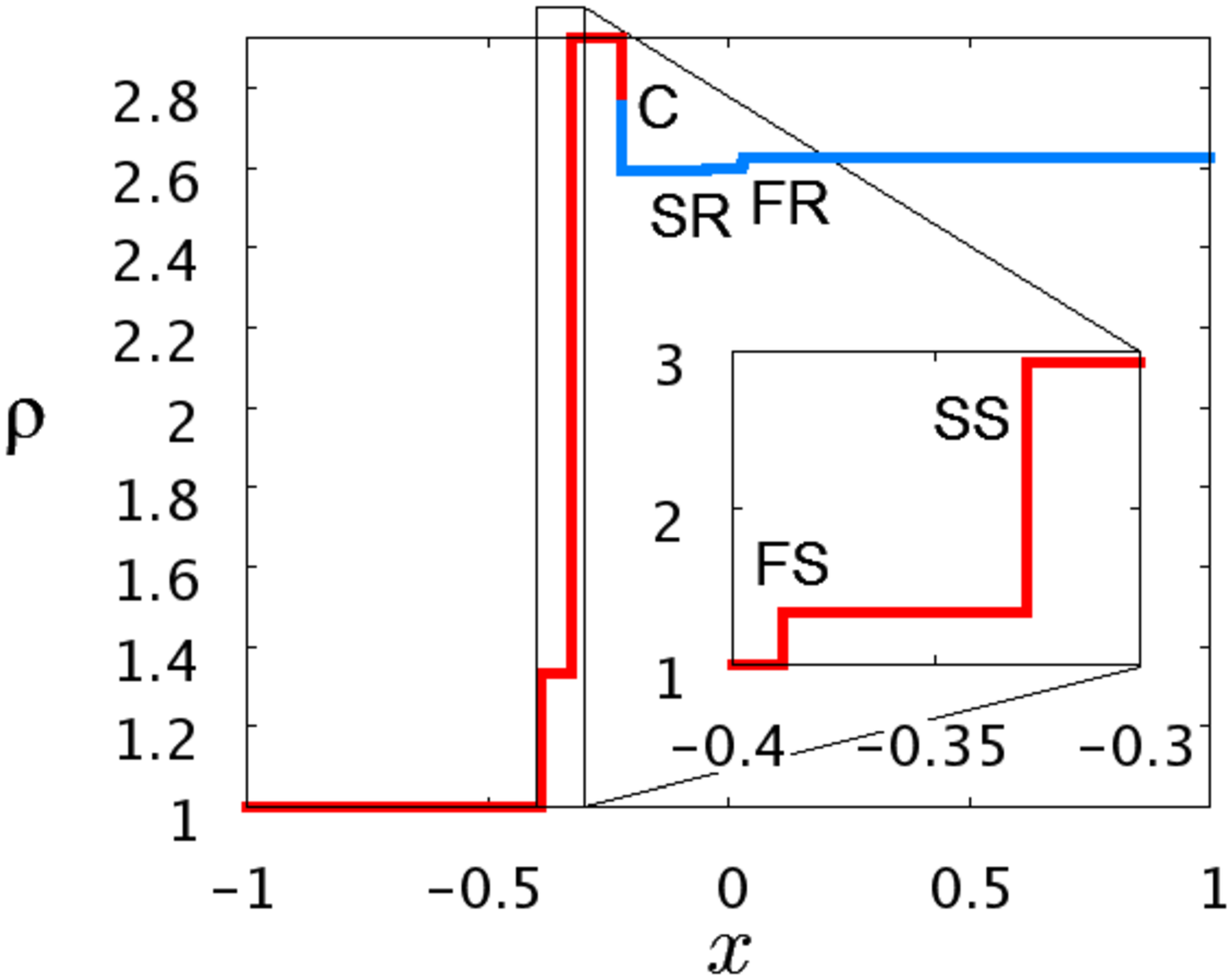}
\end{center}
\end{minipage} &
\begin{minipage}{0.45\hsize}
\begin{center}
\includegraphics[scale=0.26]{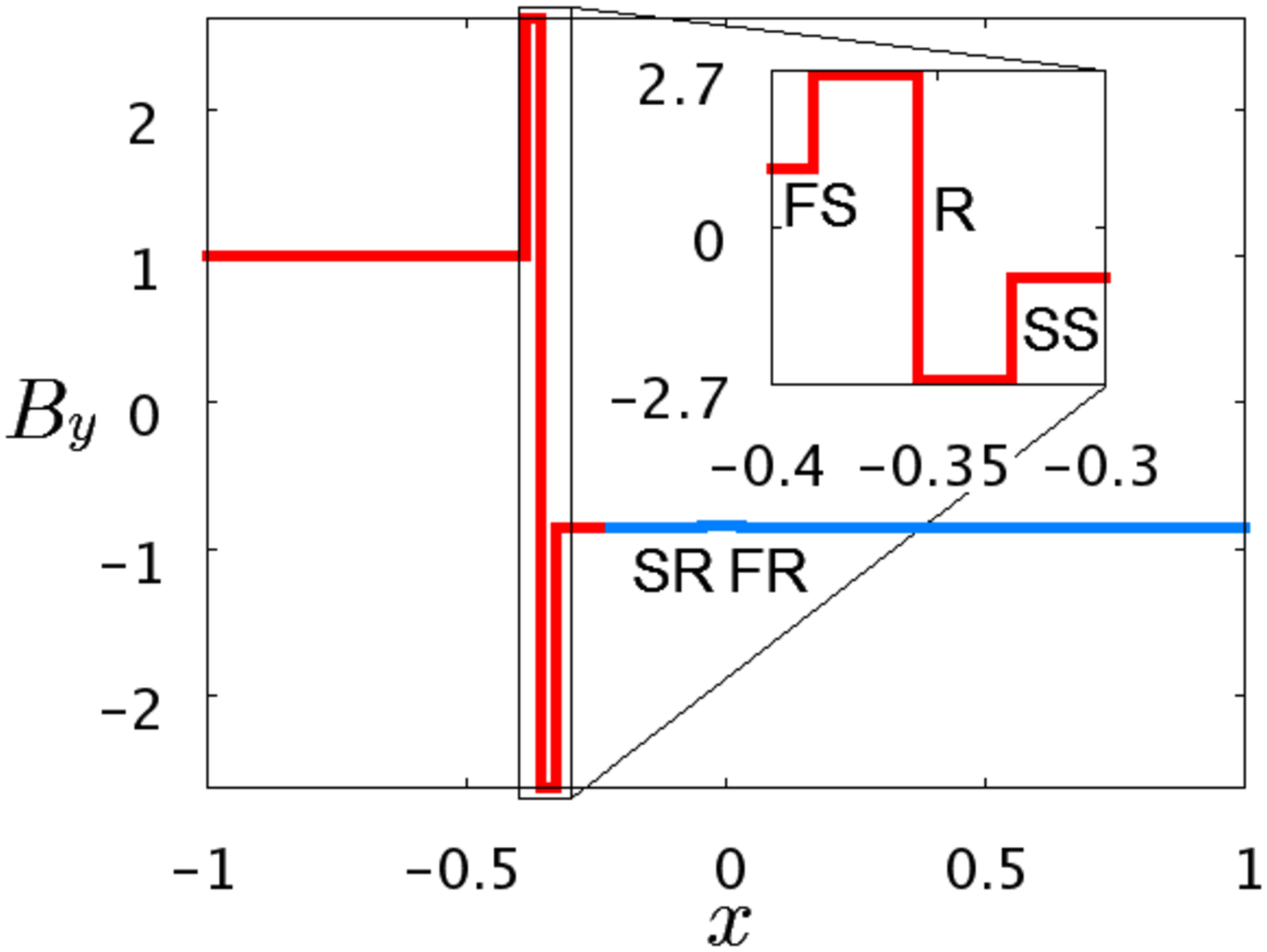}
\end{center}
\end{minipage} \\
\begin{minipage}{0.45\hsize}
\begin{center}
\includegraphics[scale=0.26]{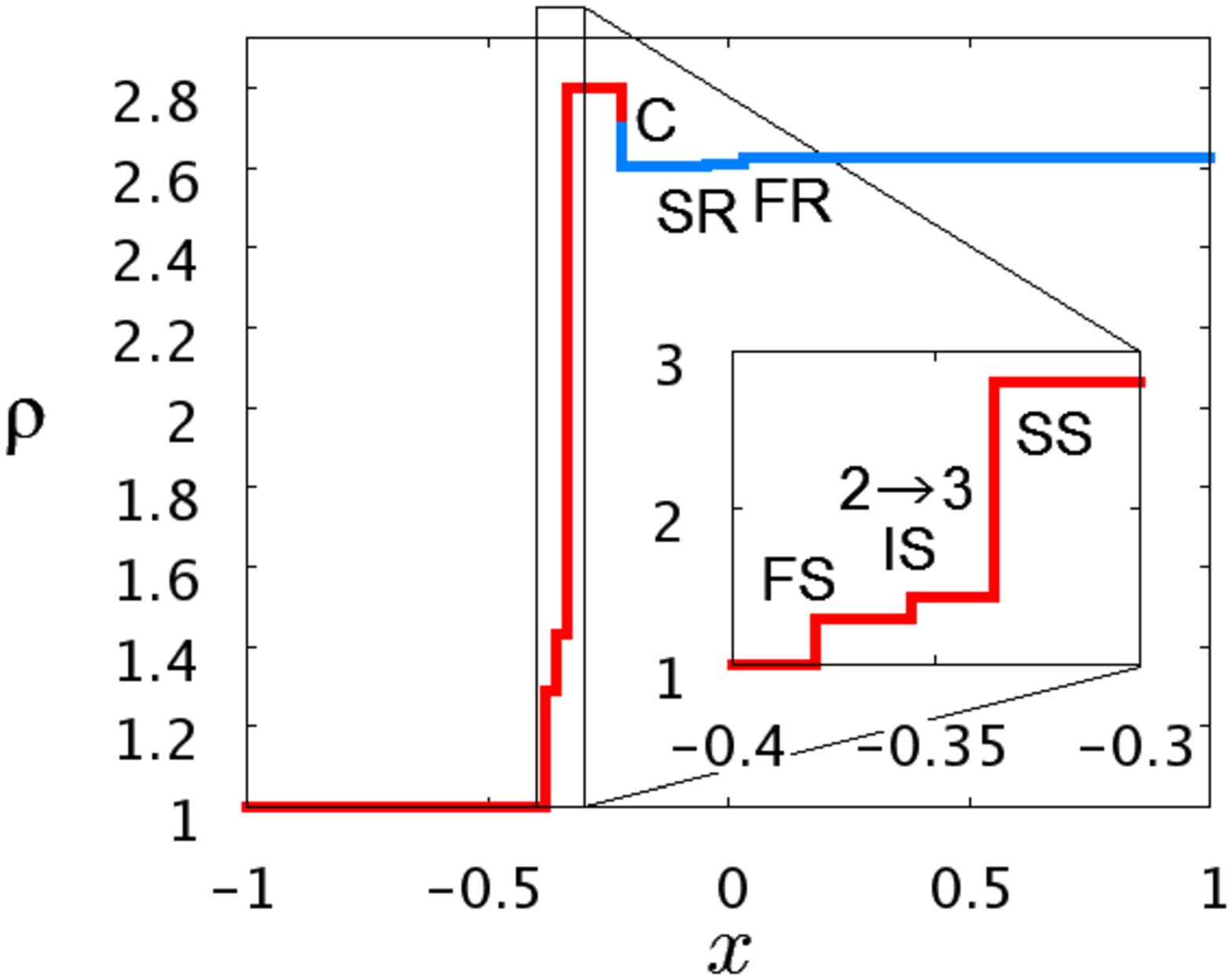}
\end{center}
\end{minipage} &
\begin{minipage}{0.45\hsize}
\begin{center}
\includegraphics[scale=0.26]{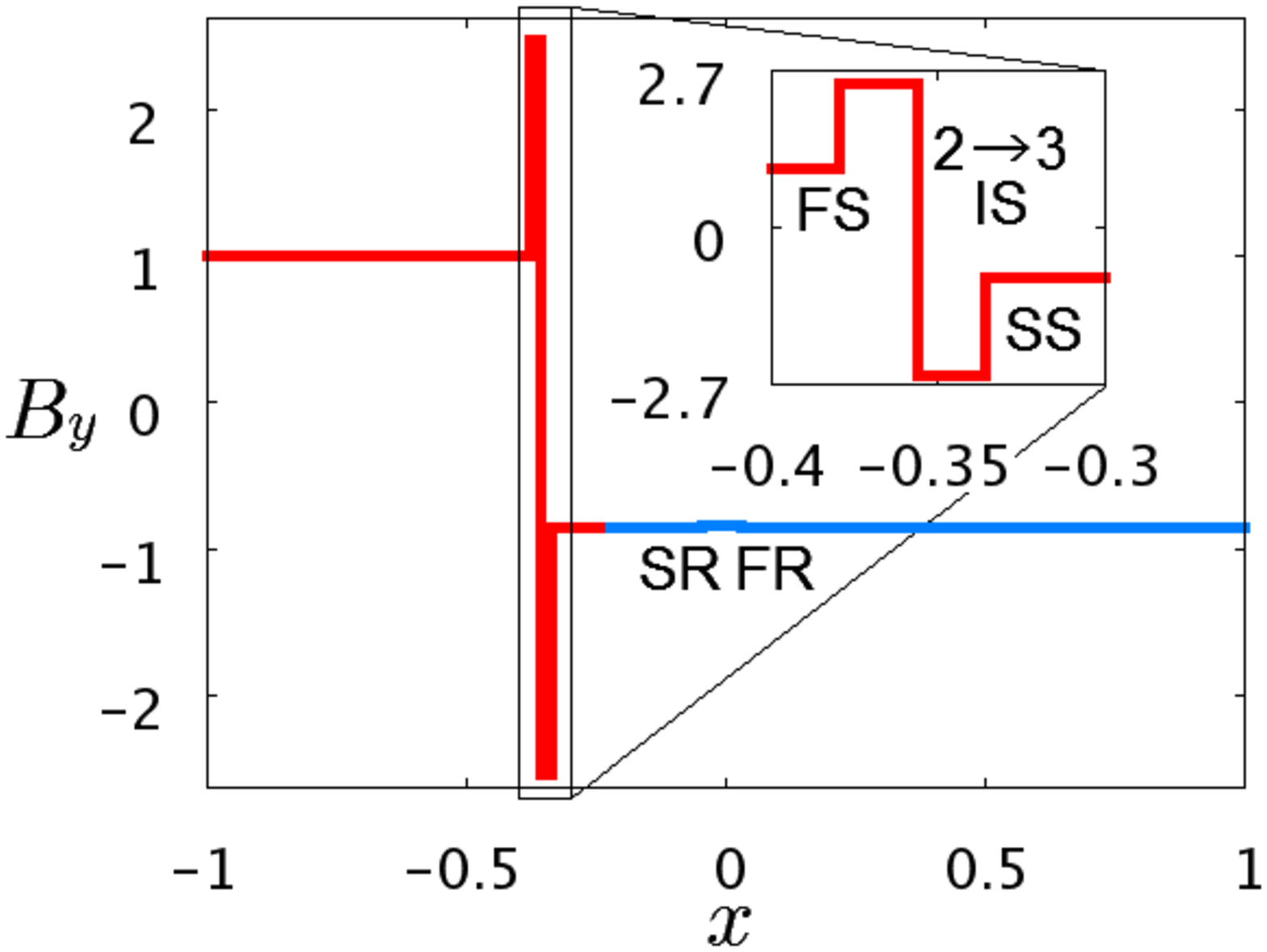}
\end{center}
\end{minipage} \\
\begin{minipage}{0.45\hsize}
\begin{center}
\includegraphics[scale=0.26]{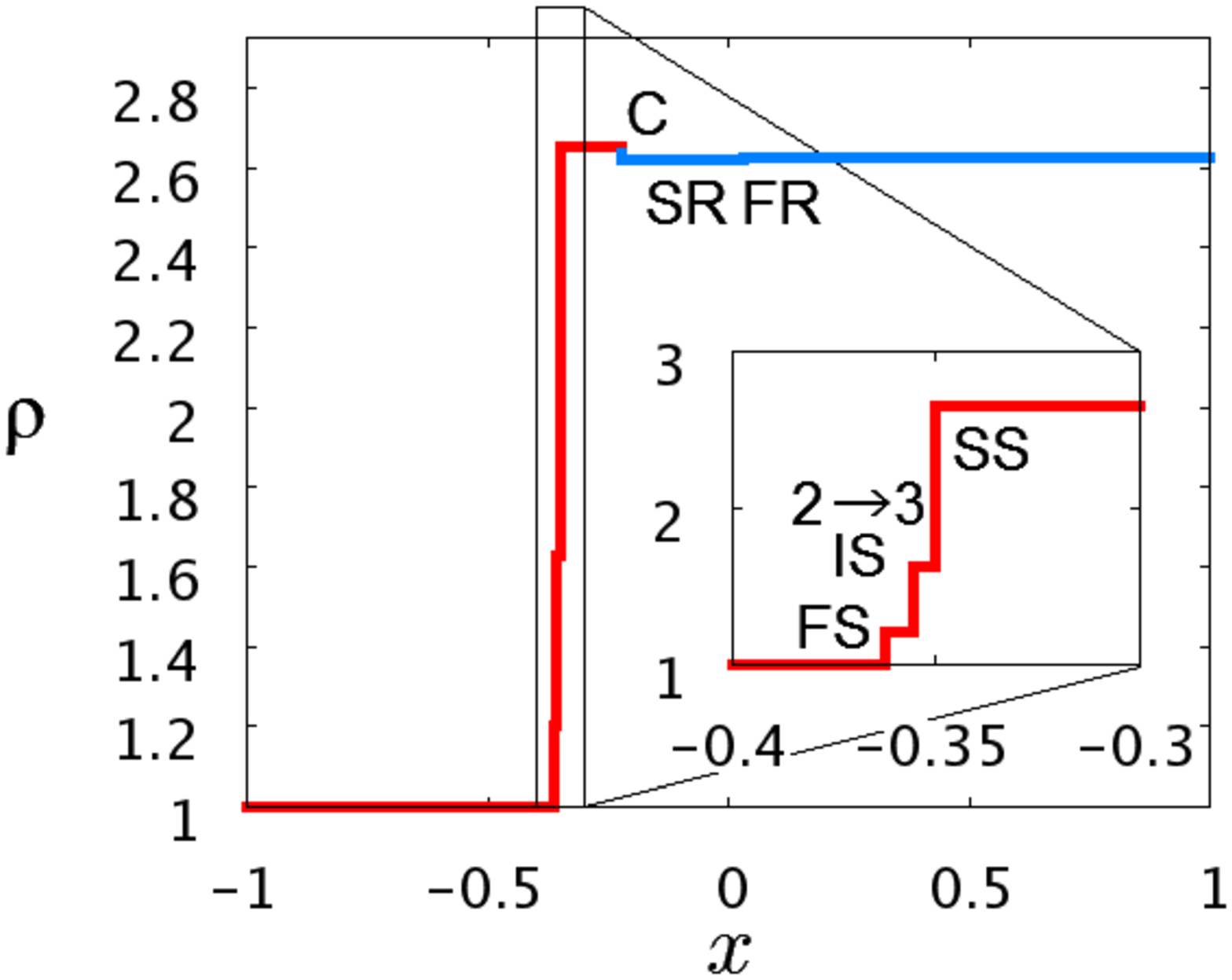}
\end{center}
\end{minipage} &
\begin{minipage}{0.45\hsize}
\begin{center}
\includegraphics[scale=0.26]{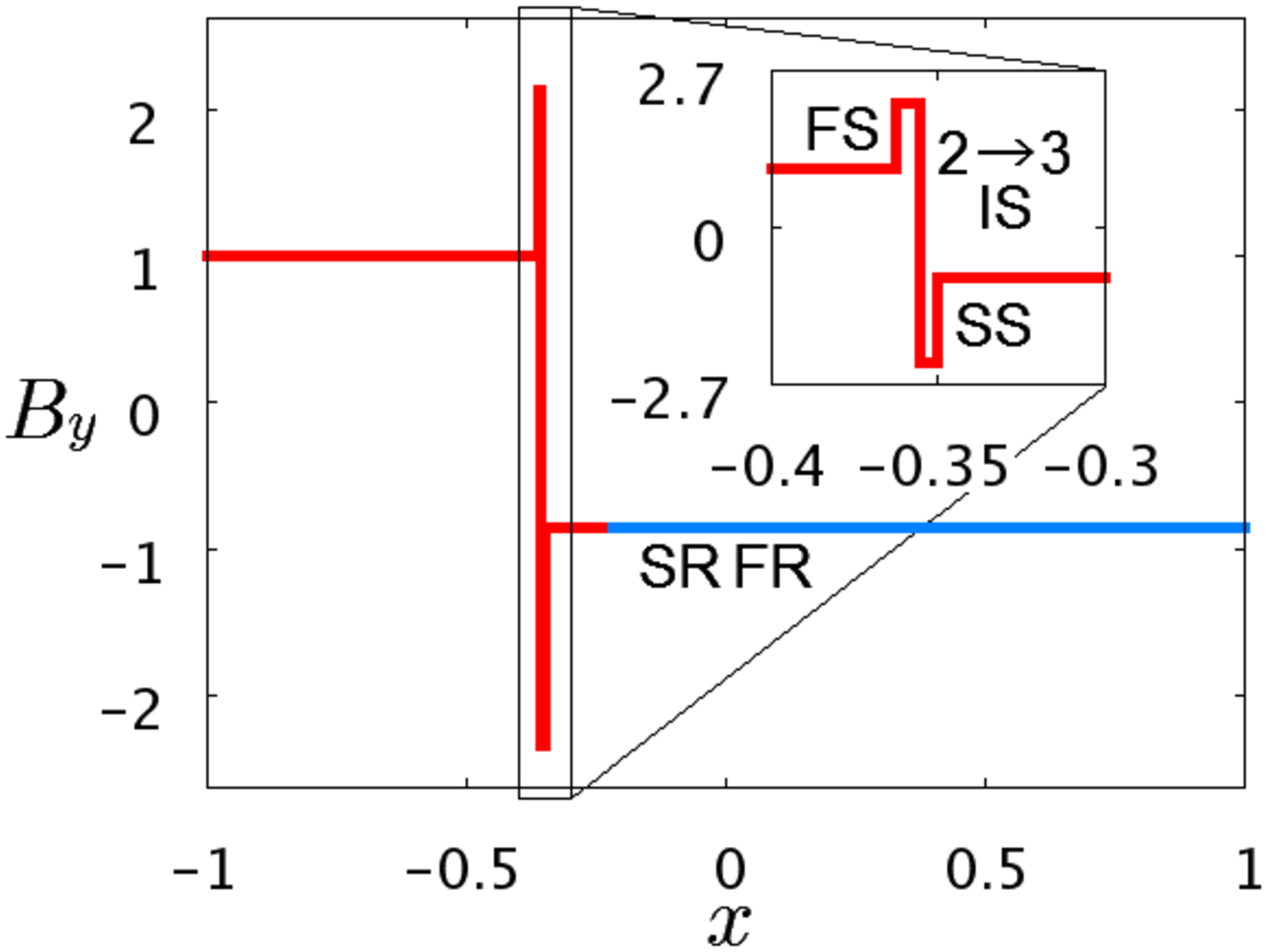}
\end{center}
\end{minipage} \\
\begin{minipage}{0.45\hsize}
\begin{center}
\includegraphics[scale=0.26]{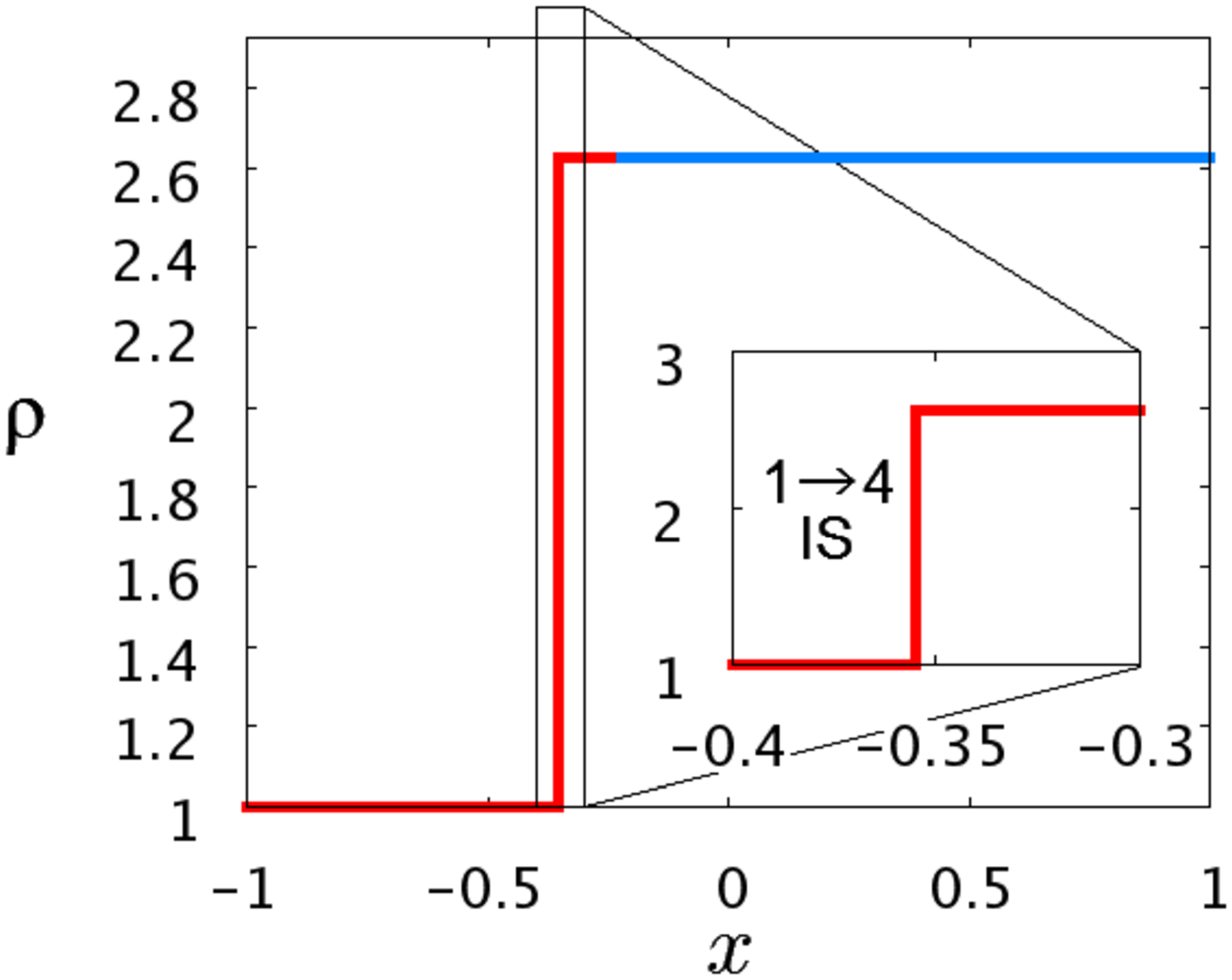}
\end{center}
\end{minipage} &
\begin{minipage}{0.45\hsize}
\begin{center}
\includegraphics[scale=0.26]{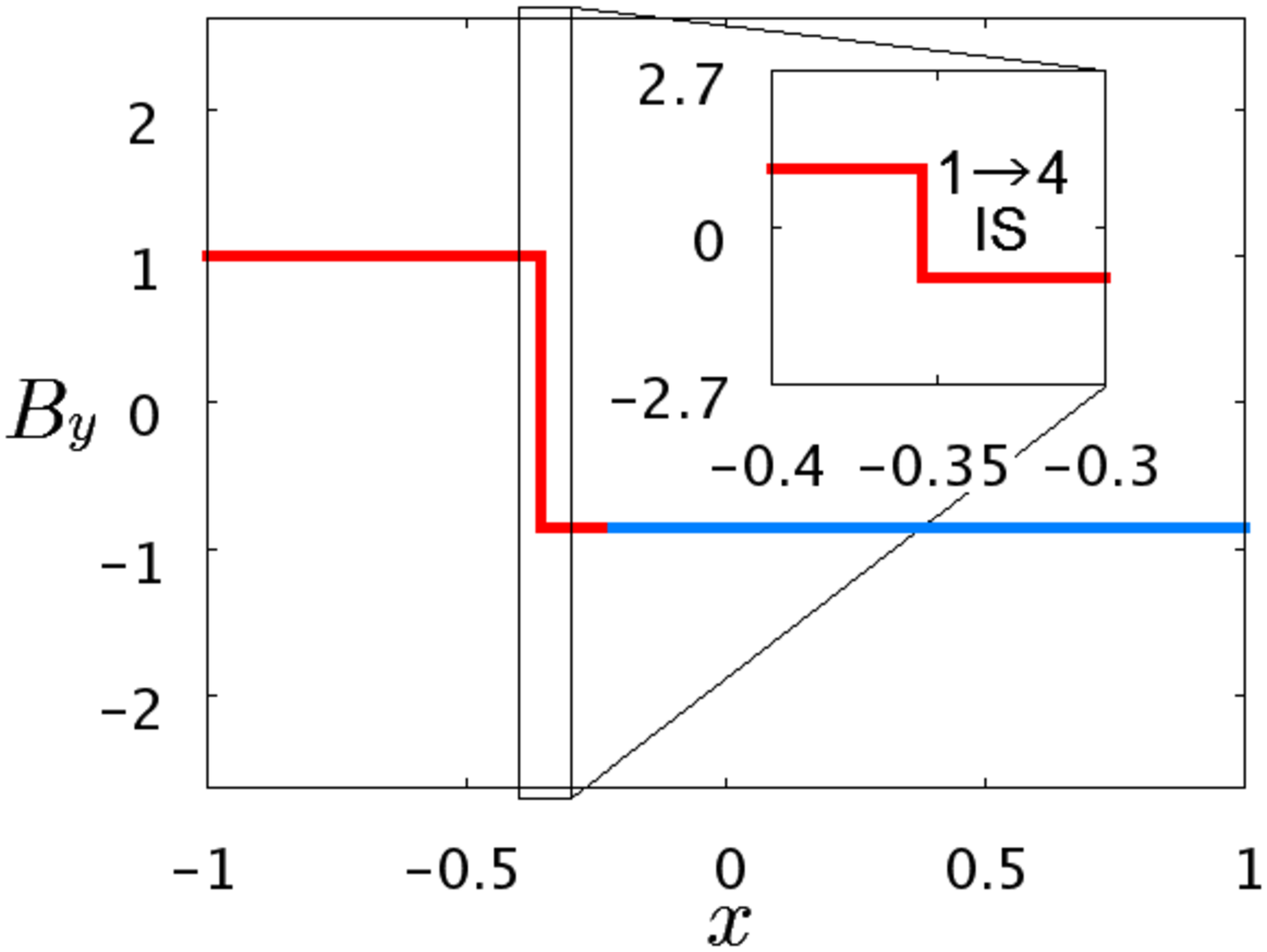}
\end{center}
\end{minipage} 
\end{tabular}
\caption{The regular solution and some non-regular solutions for an initial condition which can be connected by a $1 \rightarrow 4$ intermediate shock. 
The waves in the red and blue portions that are separated by the contact discontinuity are left- and right-going respectively. 
The designations FS, SS, R, FR, SR, C and IS represent the fast shock, slow shock, rotational discontinuity, fast rarefaction, slow rarefaction, contact discontinuity and intermediate shock respectively. 
The insets are the close-ups of indicated regions.}
\label{1-4}
\end{figure}

\begin{figure}
\begin{tabular}{cc}
\begin{minipage}{0.45\hsize}
\begin{center}
\includegraphics[scale=0.26]{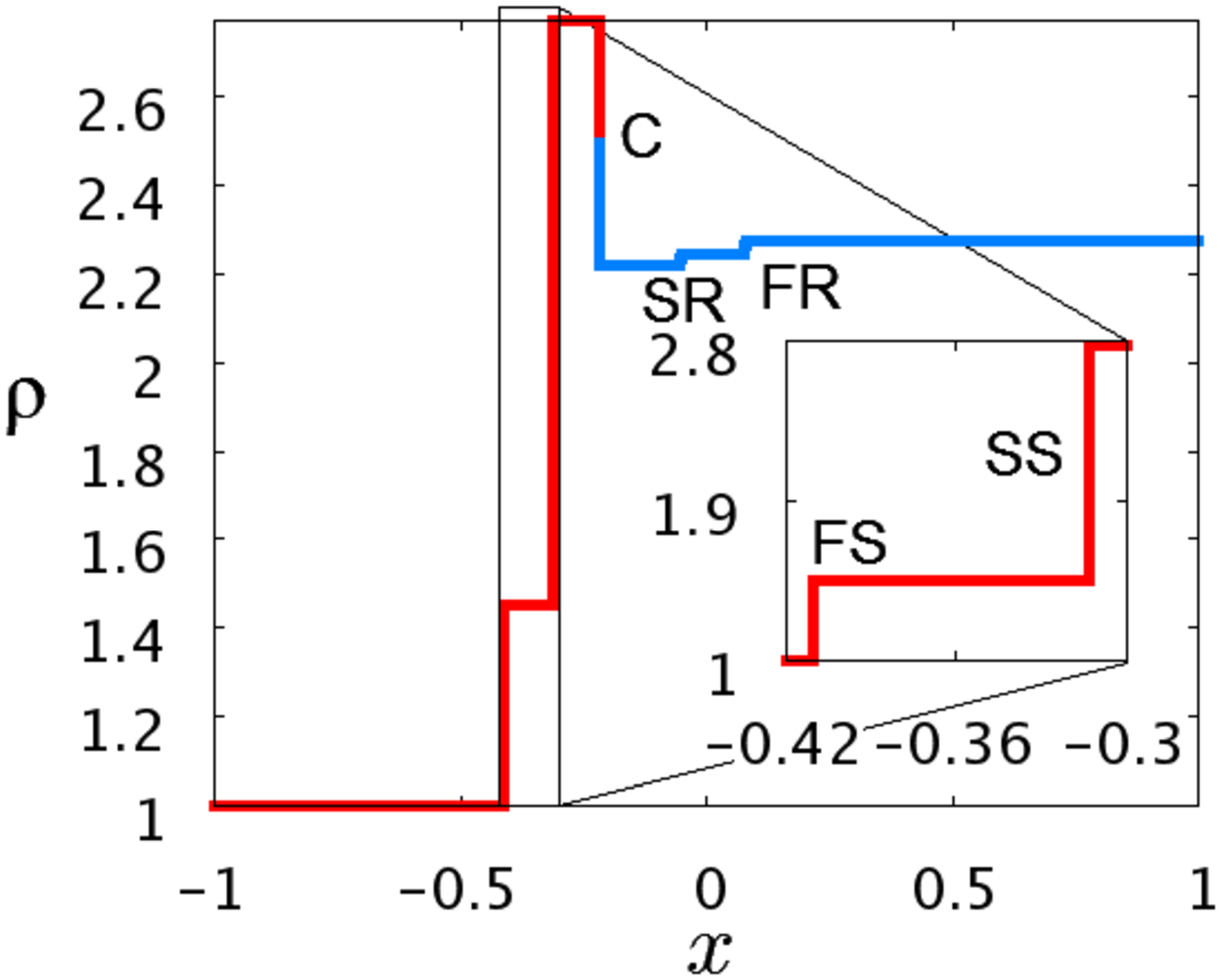}
\end{center}
\end{minipage} &
\begin{minipage}{0.45\hsize}
\begin{center}
\includegraphics[scale=0.26]{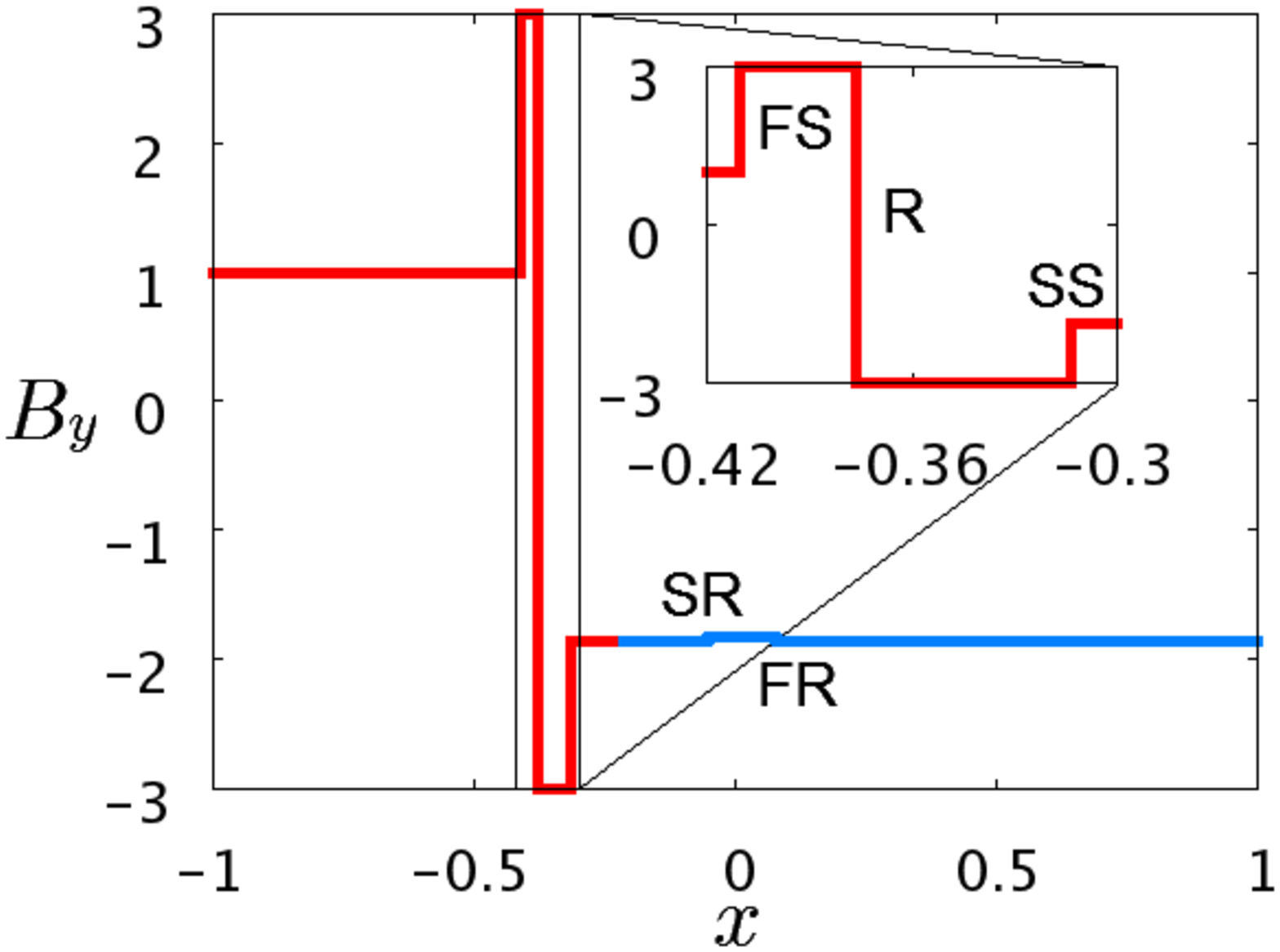}
\end{center}
\end{minipage} \\
\begin{minipage}{0.45\hsize}
\begin{center}
\includegraphics[scale=0.26]{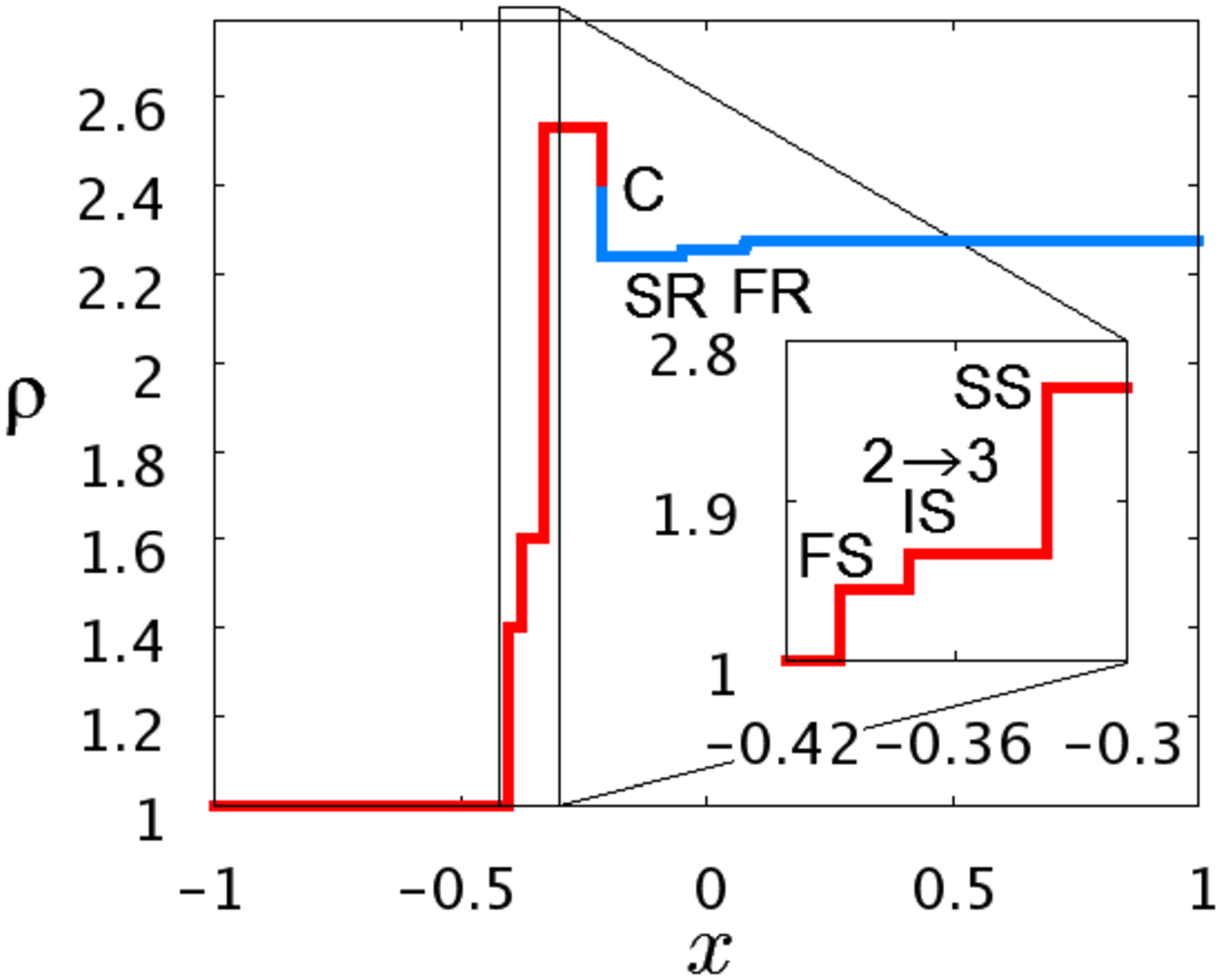}
\end{center}
\end{minipage} &
\begin{minipage}{0.45\hsize}
\begin{center}
\includegraphics[scale=0.26]{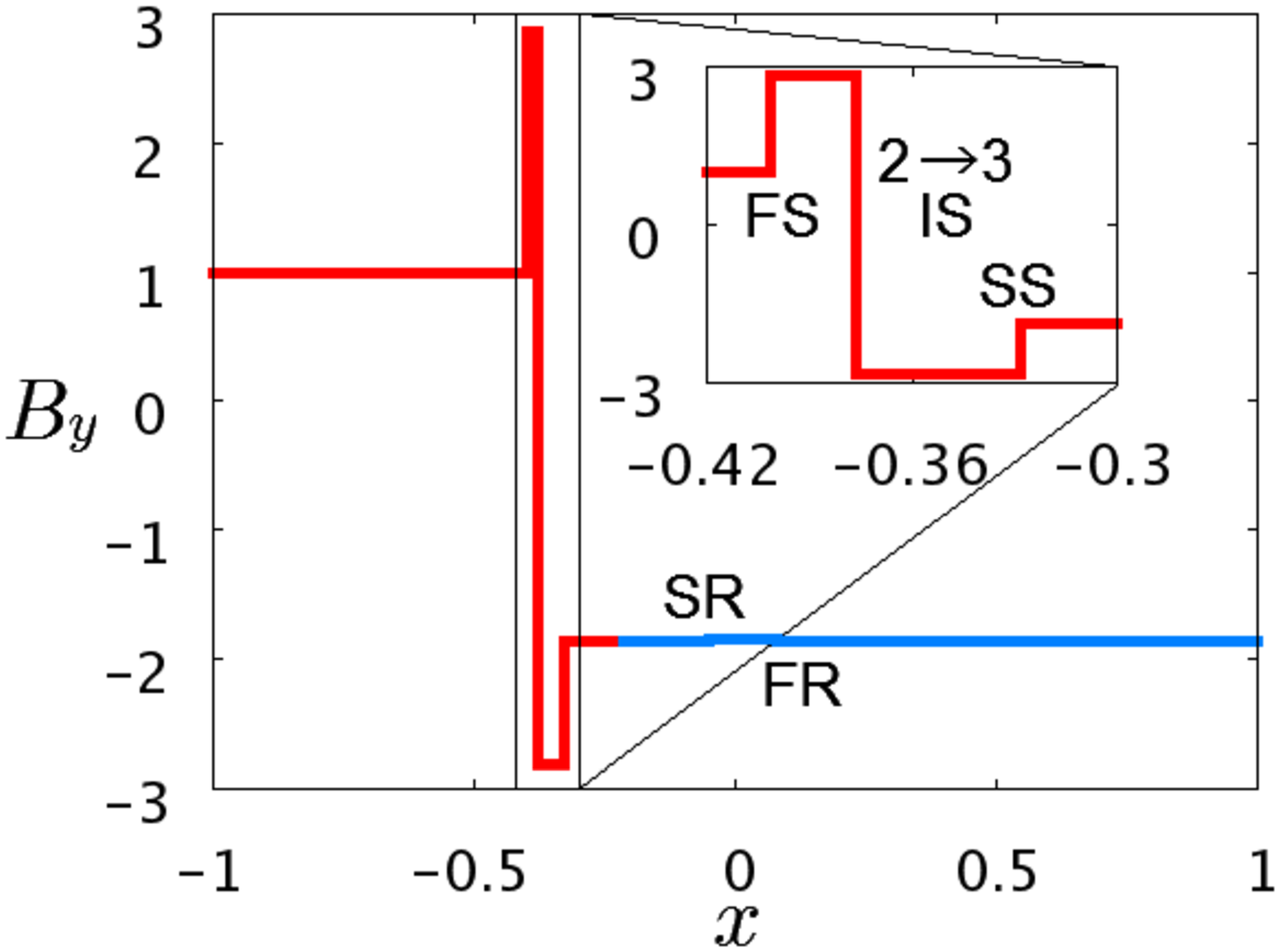}
\end{center}
\end{minipage} \\
\begin{minipage}{0.45\hsize}
\begin{center}
\includegraphics[scale=0.26]{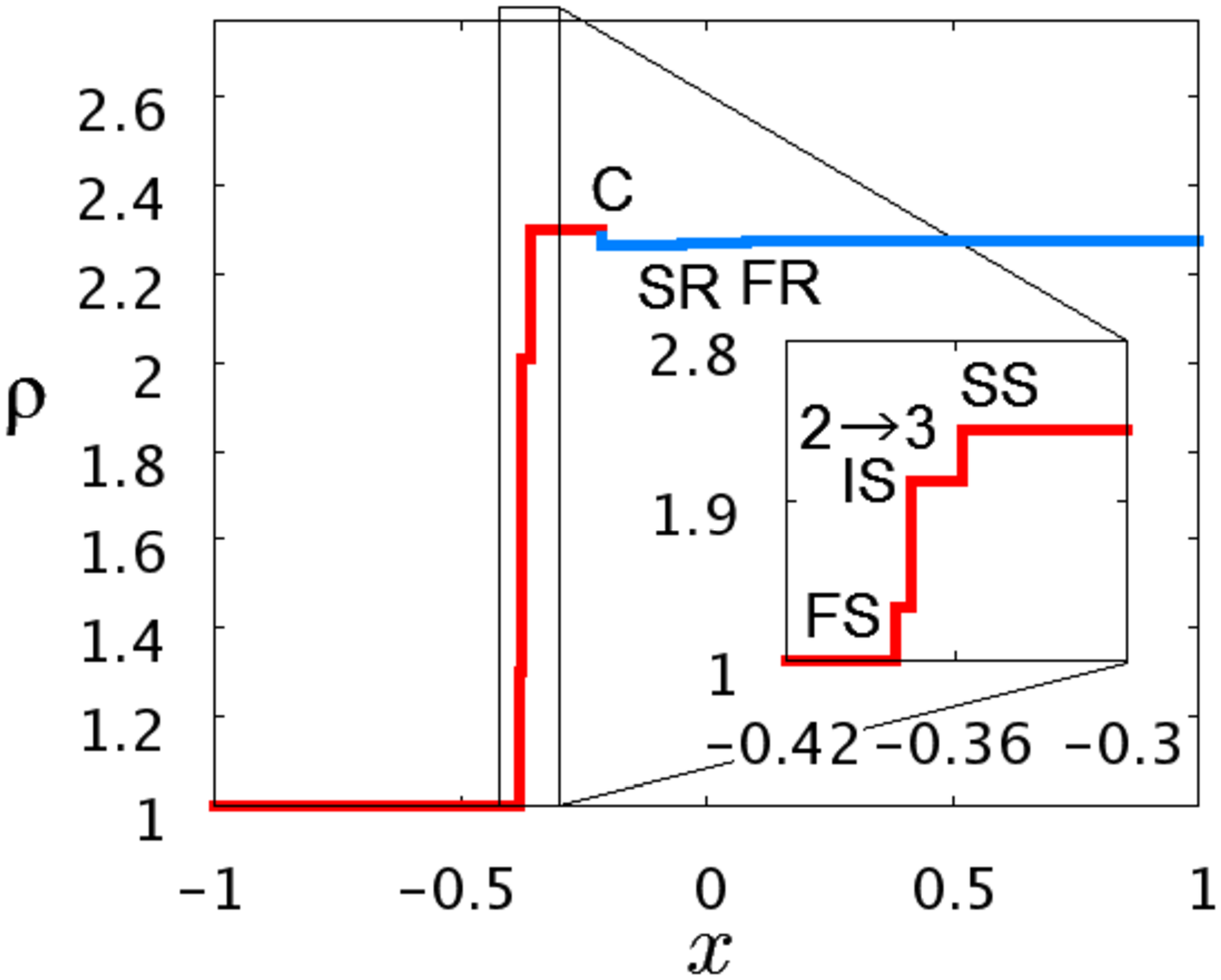}
\end{center}
\end{minipage} &
\begin{minipage}{0.45\hsize}
\begin{center}
\includegraphics[scale=0.26]{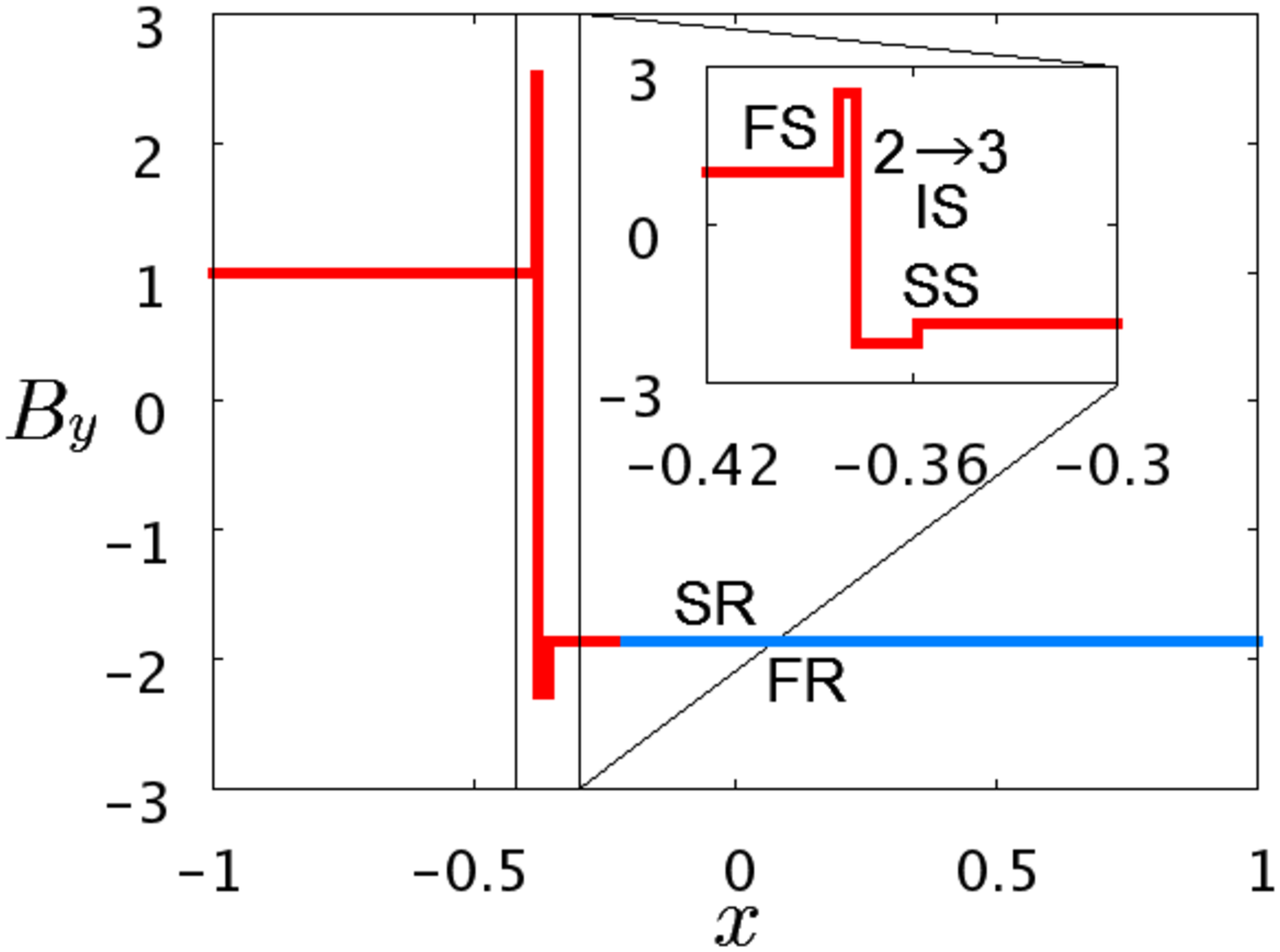}
\end{center}
\end{minipage} \\
\begin{minipage}{0.45\hsize}
\begin{center}
\includegraphics[scale=0.26]{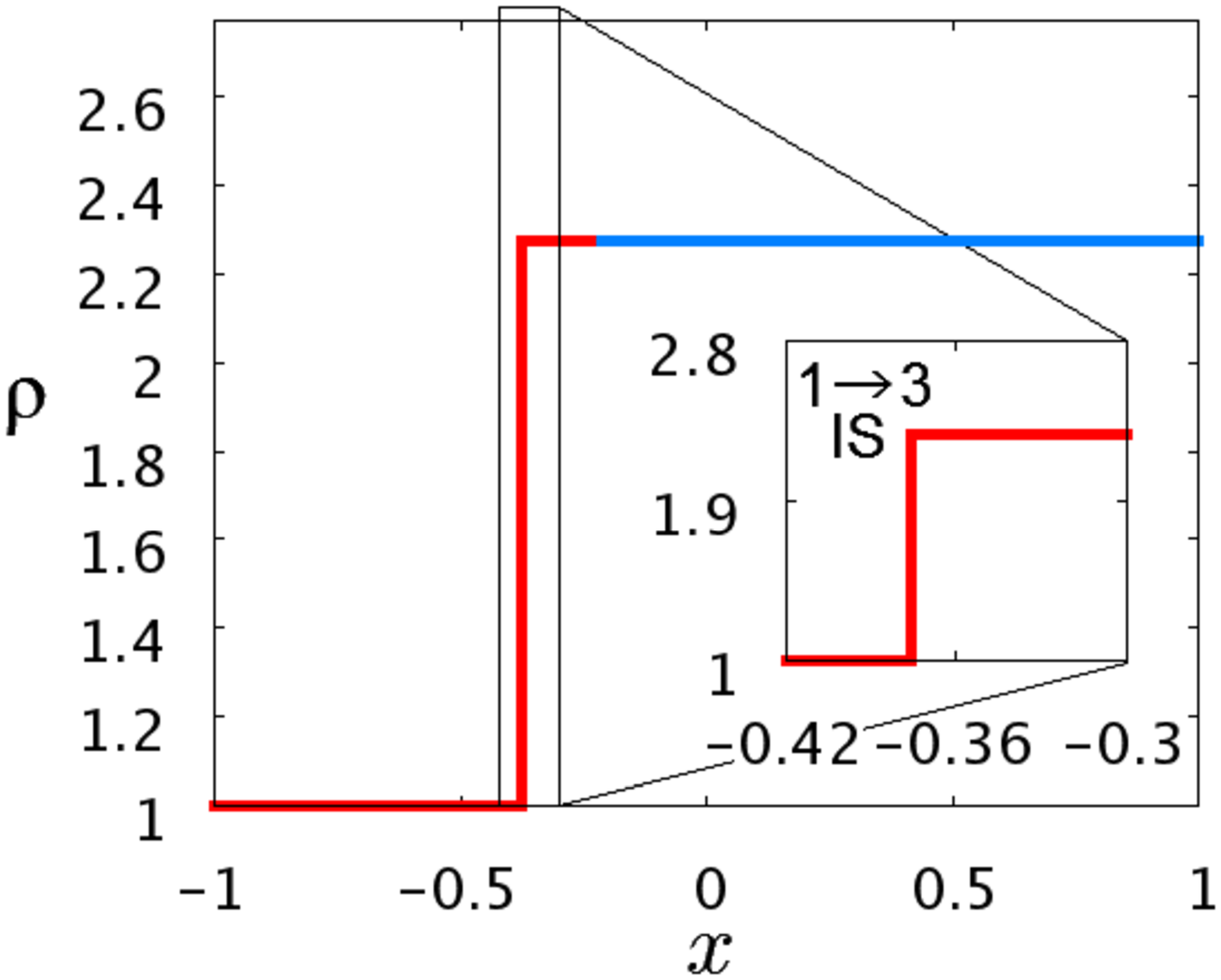}
\end{center}
\end{minipage} &
\begin{minipage}{0.45\hsize}
\begin{center}
\includegraphics[scale=0.26]{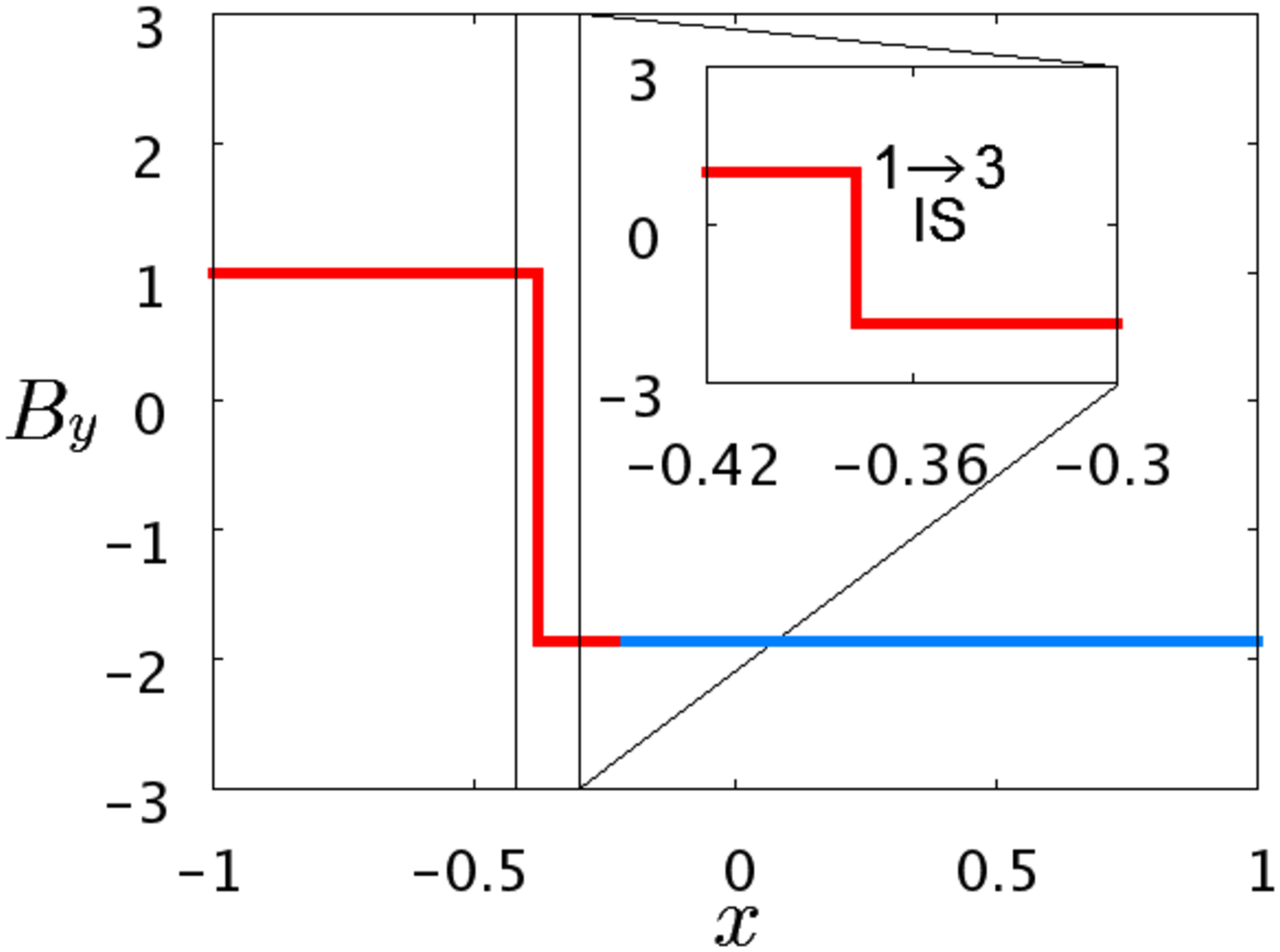}
\end{center}
\end{minipage} 
\end{tabular}
\caption{The regular solution and some non-regular solutions for an initial condition which can be connected by a $1 \rightarrow 3$ intermediate shock. 
The notations are the same as in Fig. \ref{1-4}.}
\label{1-3}
\end{figure}

\begin{figure}
\begin{tabular}{cc}
\begin{minipage}{0.45\hsize}
\begin{center}
\includegraphics[scale=0.26]{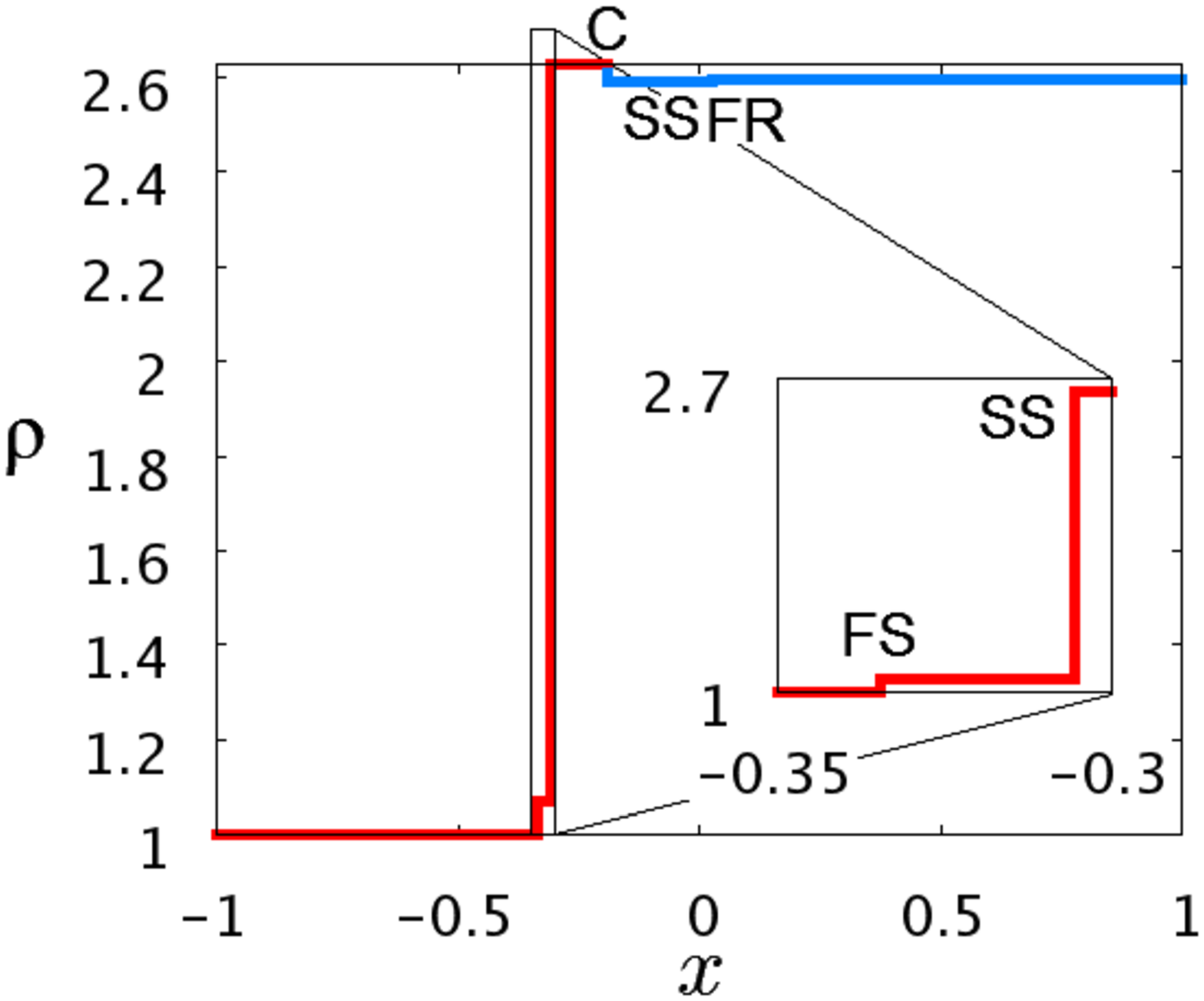}
\end{center}
\end{minipage} &
\begin{minipage}{0.45\hsize}
\begin{center}
\includegraphics[scale=0.26]{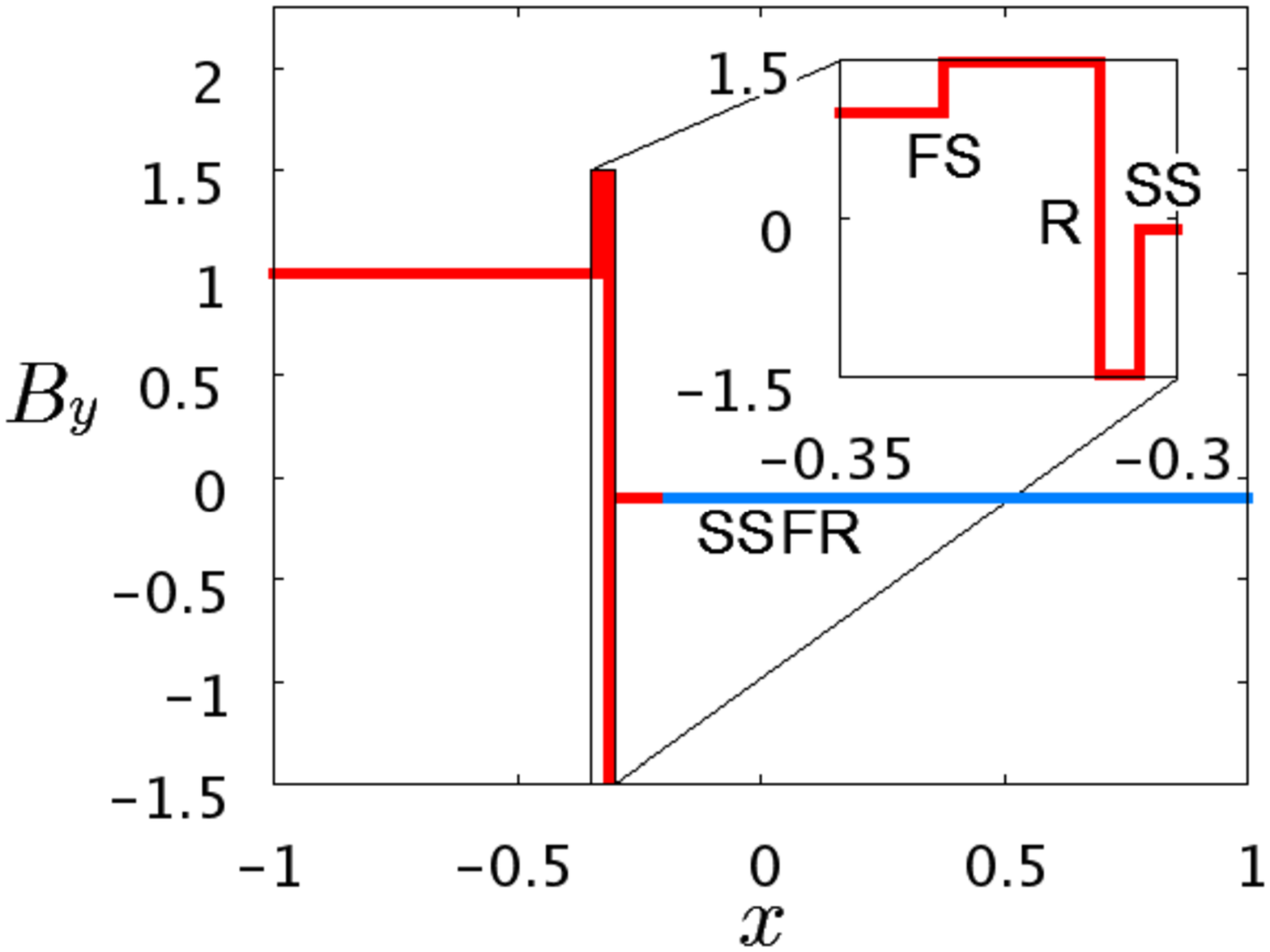}
\end{center}
\end{minipage} \\
\begin{minipage}{0.45\hsize}
\begin{center}
\includegraphics[scale=0.26]{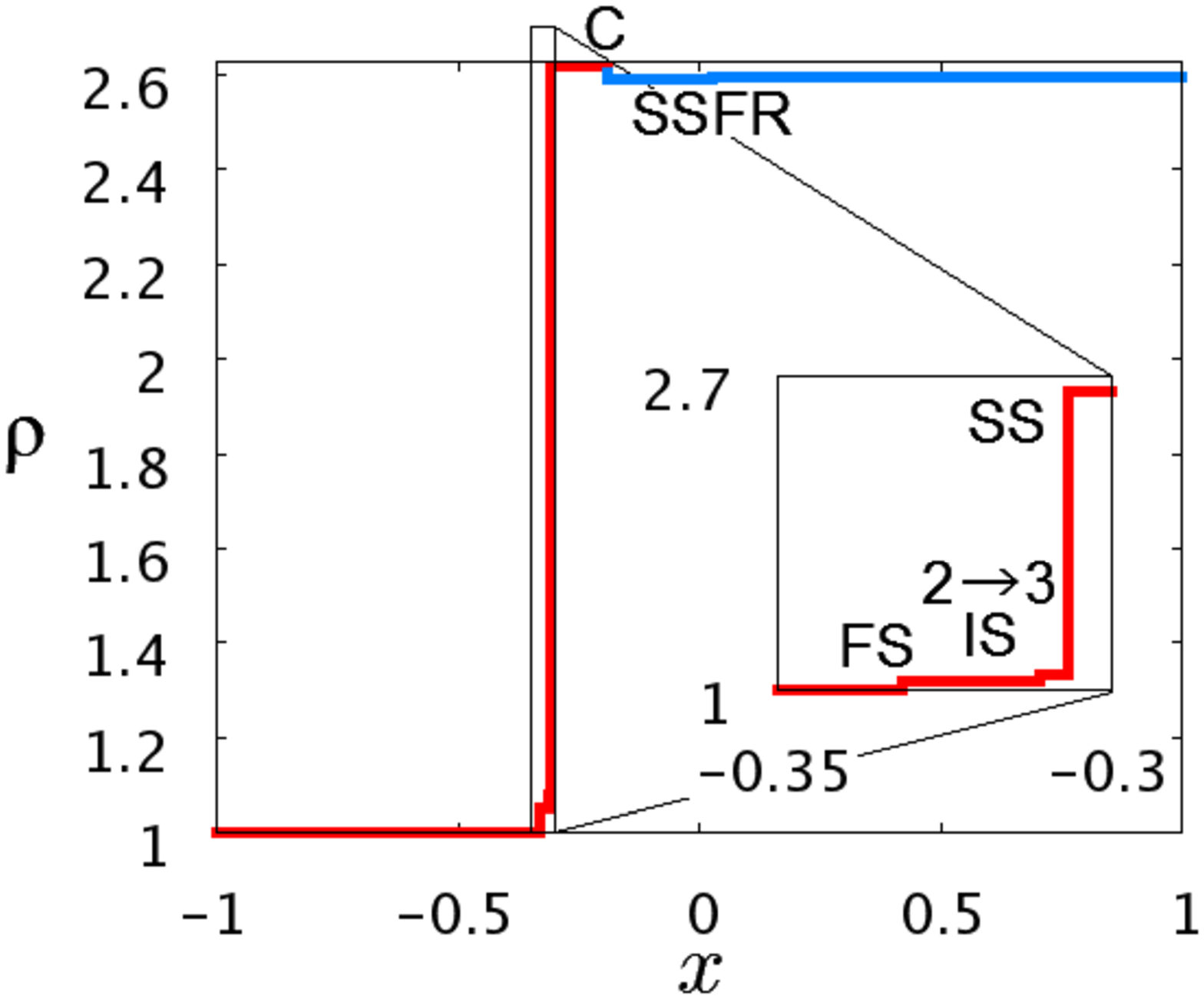}
\end{center}
\end{minipage} &
\begin{minipage}{0.45\hsize}
\begin{center}
\includegraphics[scale=0.26]{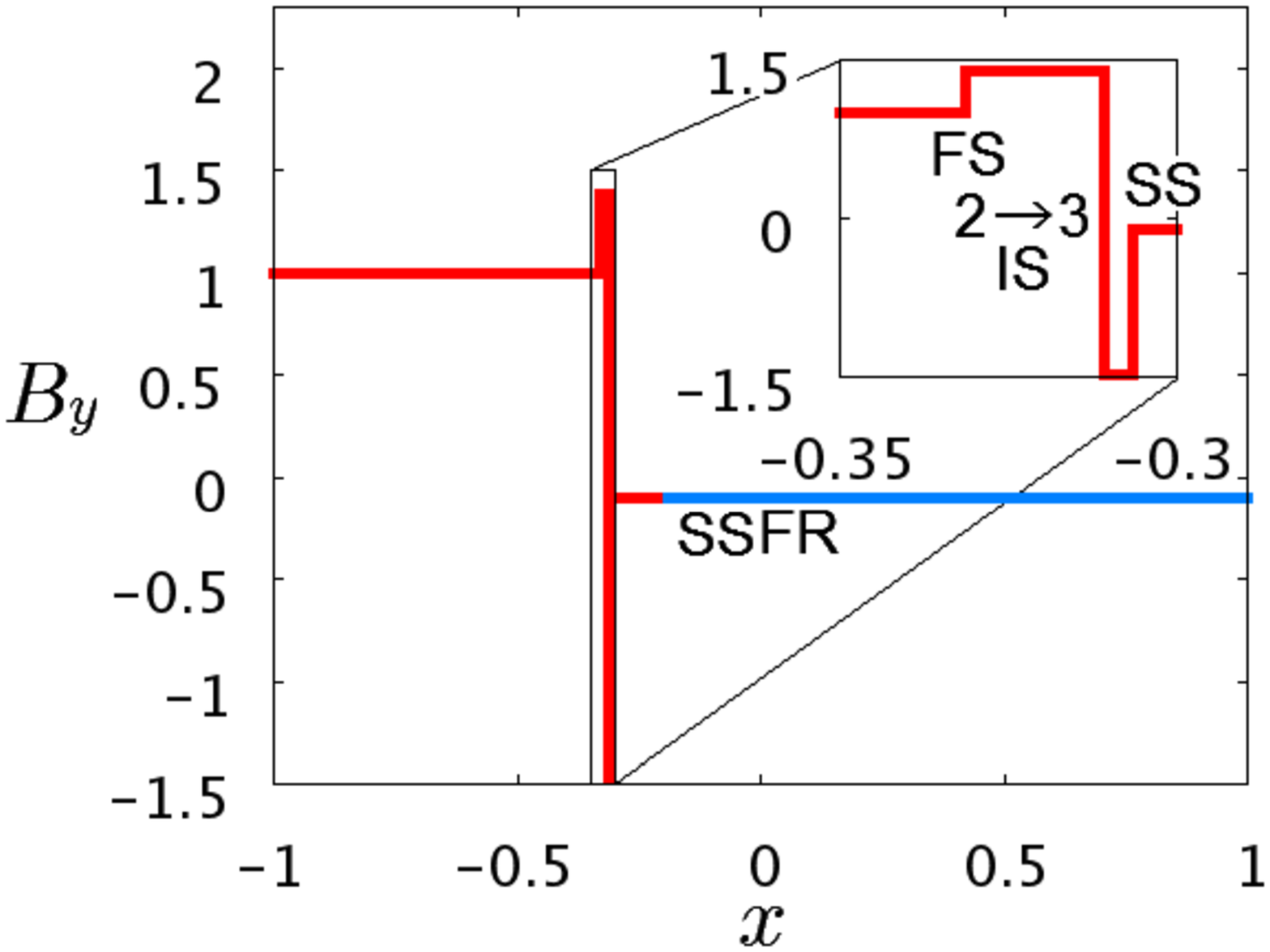}
\end{center}
\end{minipage} \\
\begin{minipage}{0.45\hsize}
\begin{center}
\includegraphics[scale=0.26]{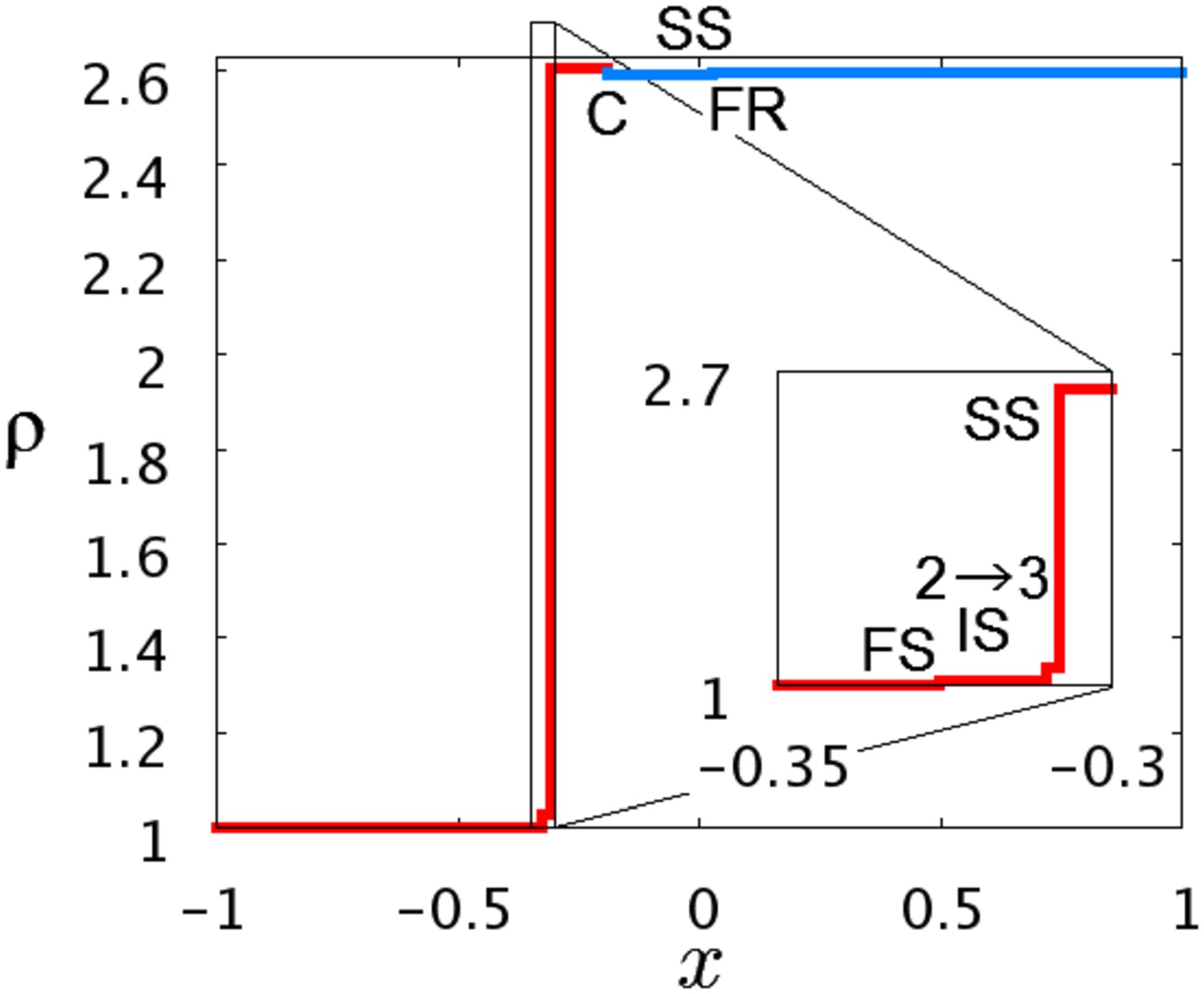}
\end{center}
\end{minipage} &
\begin{minipage}{0.45\hsize}
\begin{center}
\includegraphics[scale=0.26]{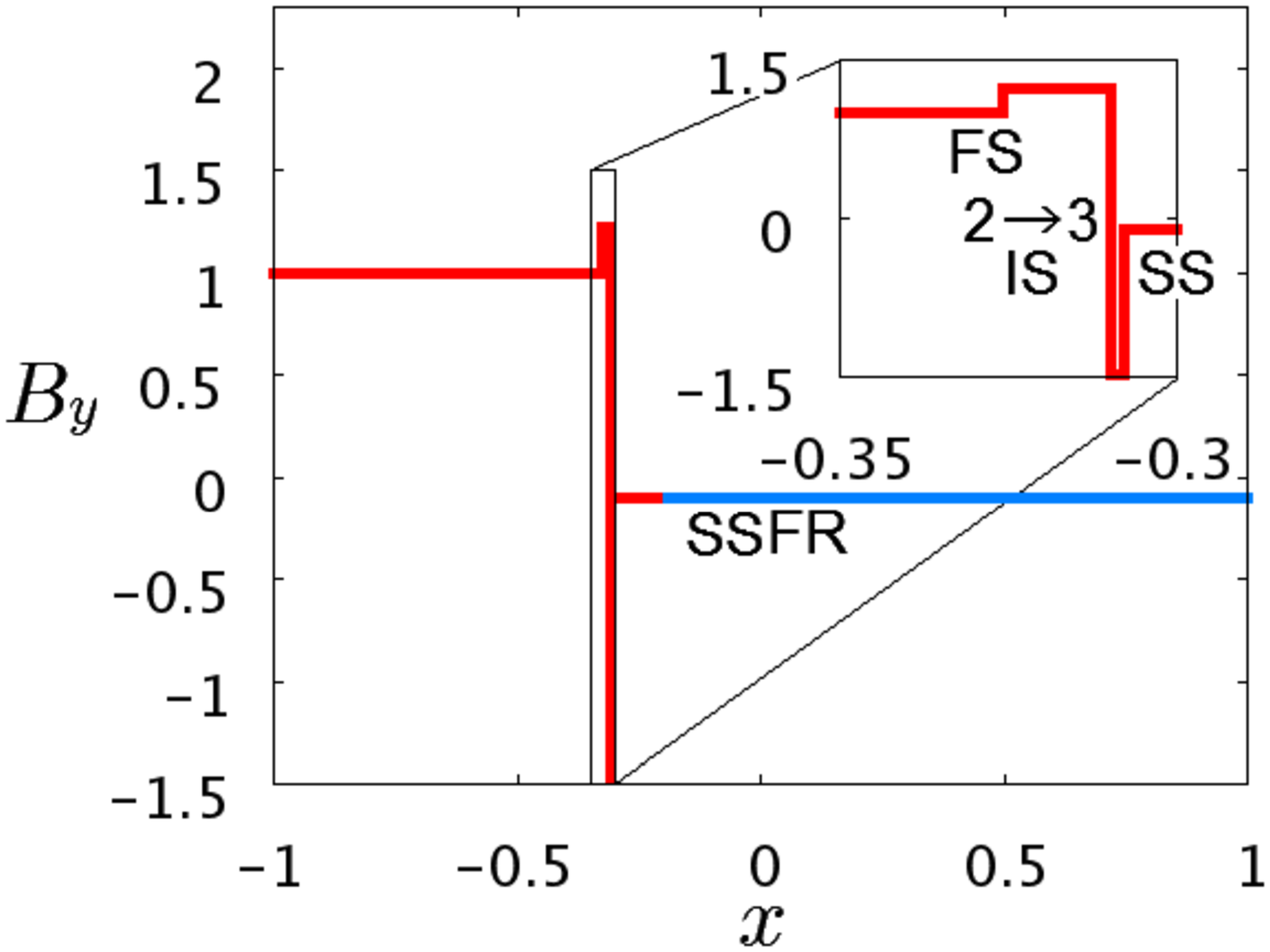}
\end{center}
\end{minipage} \\
\begin{minipage}{0.45\hsize}
\begin{center}
\includegraphics[scale=0.26]{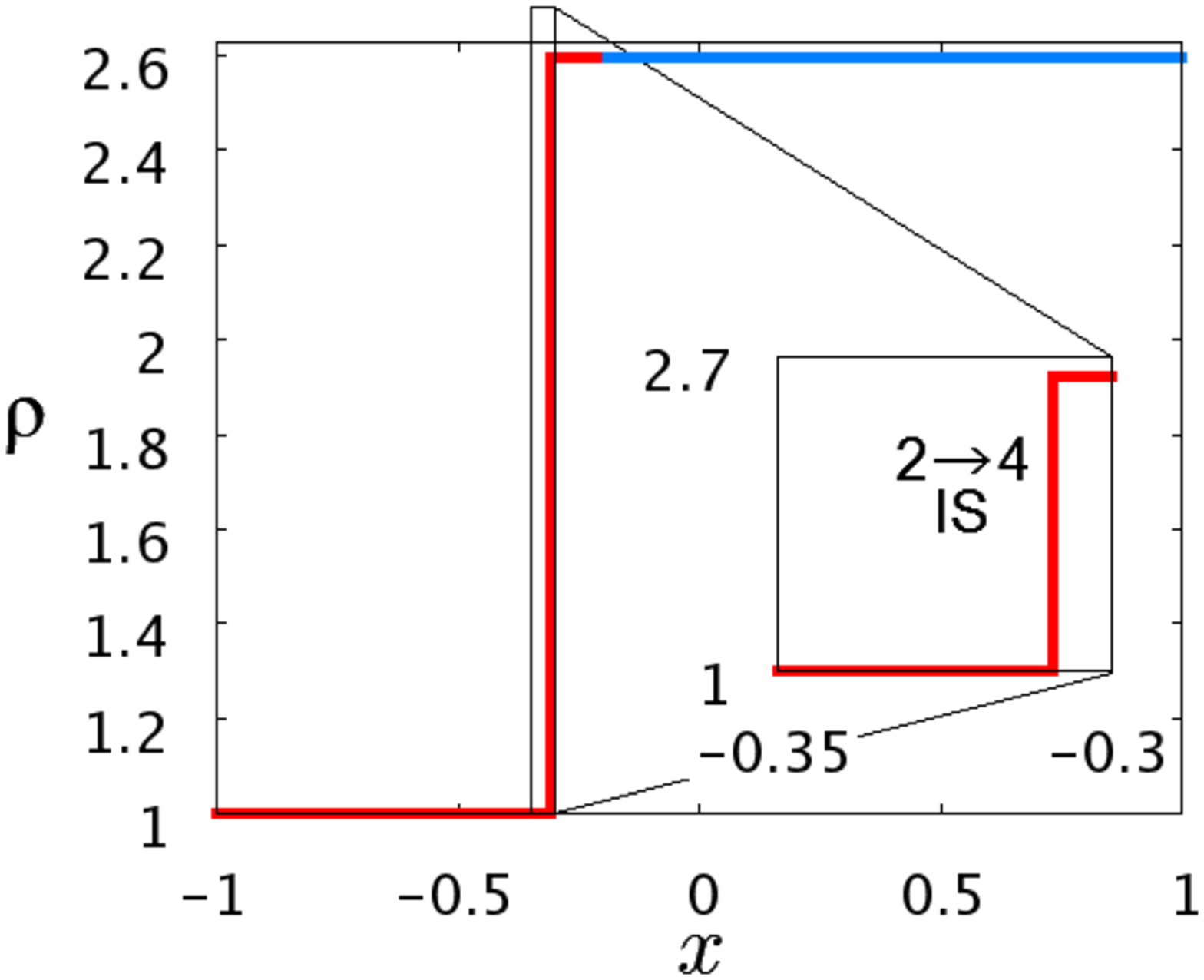}
\end{center}
\end{minipage} &
\begin{minipage}{0.45\hsize}
\begin{center}
\includegraphics[scale=0.26]{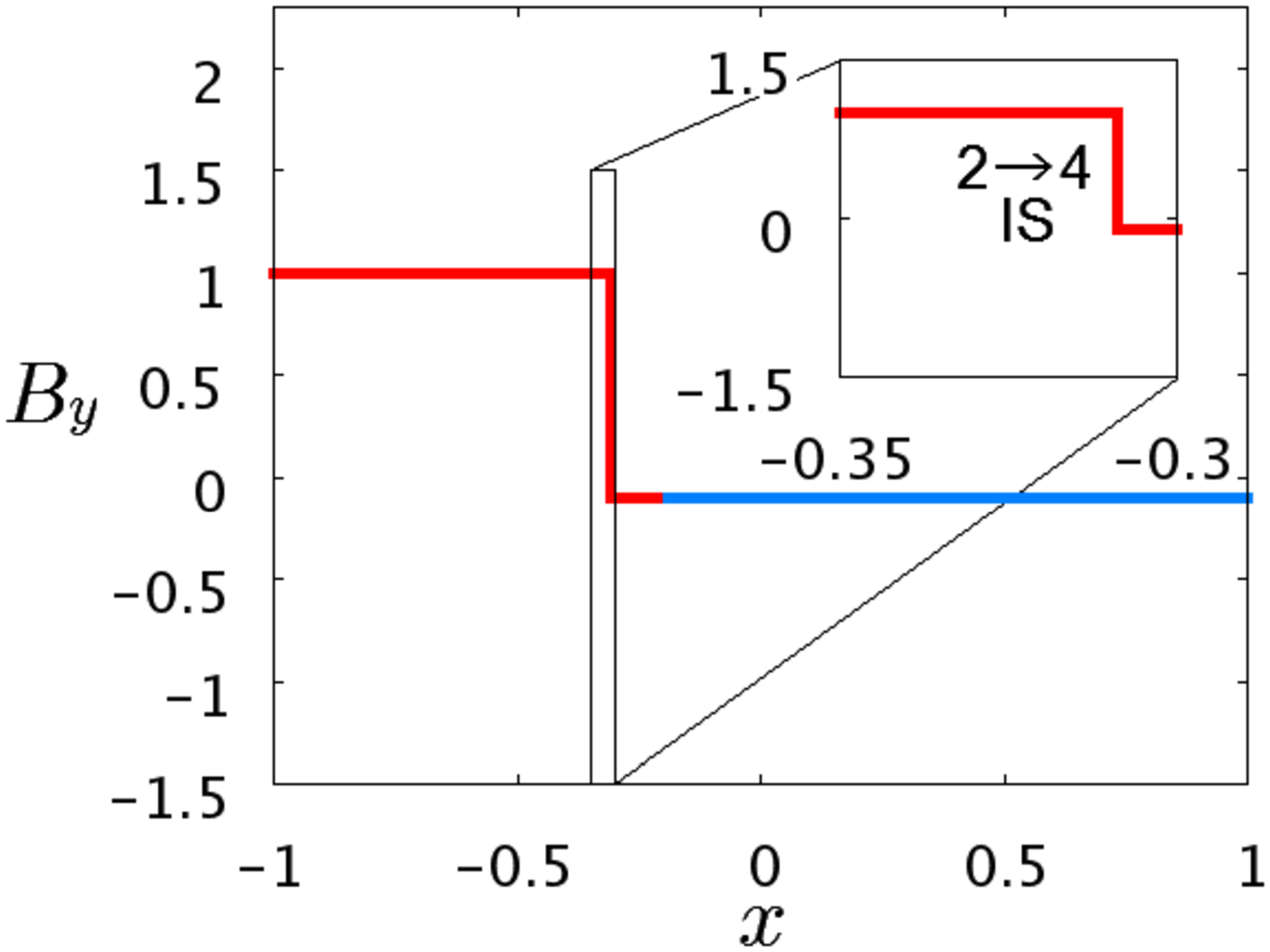}
\end{center}
\end{minipage} 
\end{tabular}
\caption{The regular solution and some non-regular solutions for an initial condition which can be connected by a $2 \rightarrow 4$ intermediate shock. 
The notations are the same as in Fig. \ref{1-4}.}
\label{2-4}
\end{figure}

\begin{figure}
\begin{tabular}{cc}
\begin{minipage}{0.45\hsize}
\begin{center}
\includegraphics[scale=0.26]{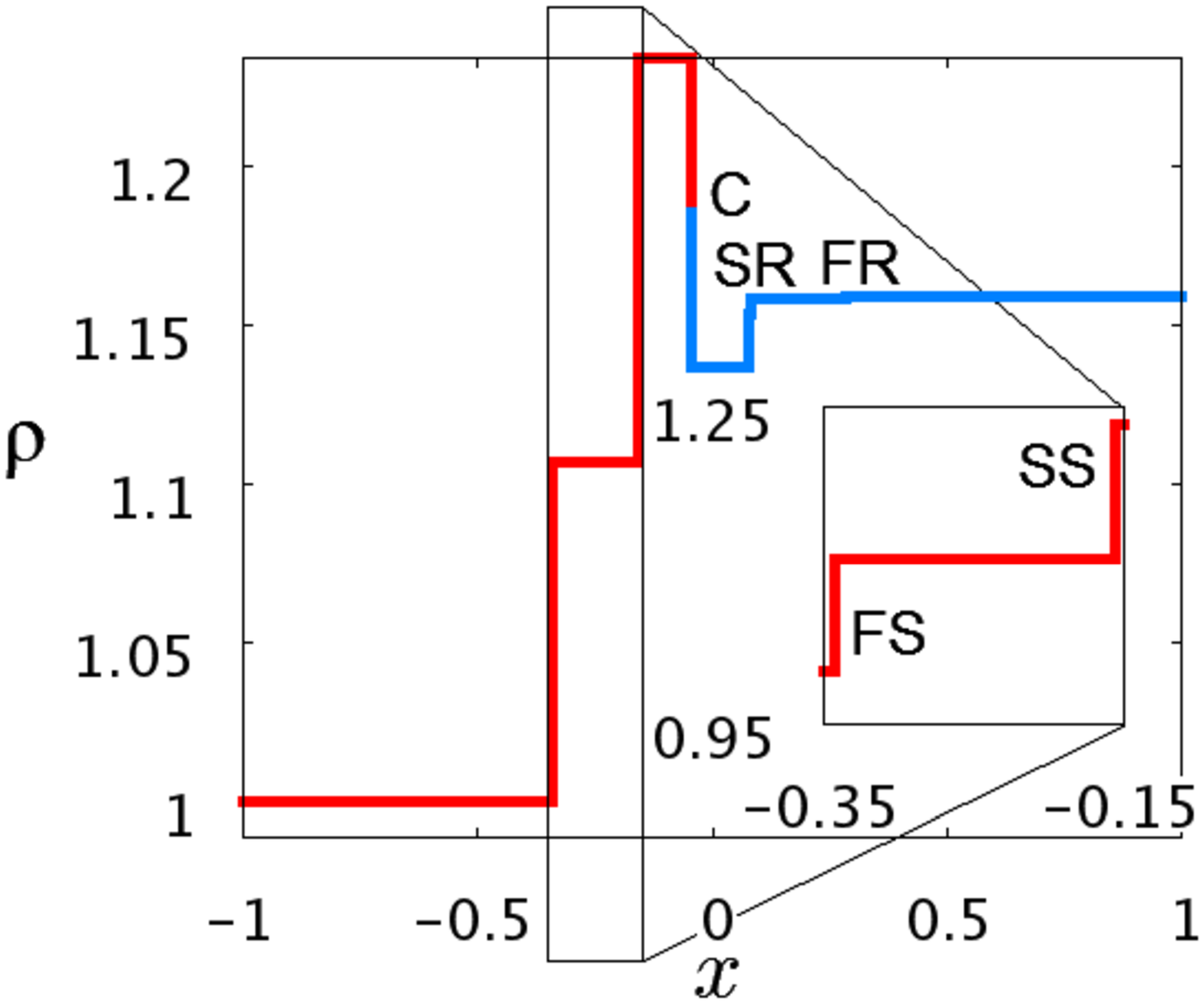}
\end{center}
\end{minipage} &
\begin{minipage}{0.45\hsize}
\begin{center}
\includegraphics[scale=0.26]{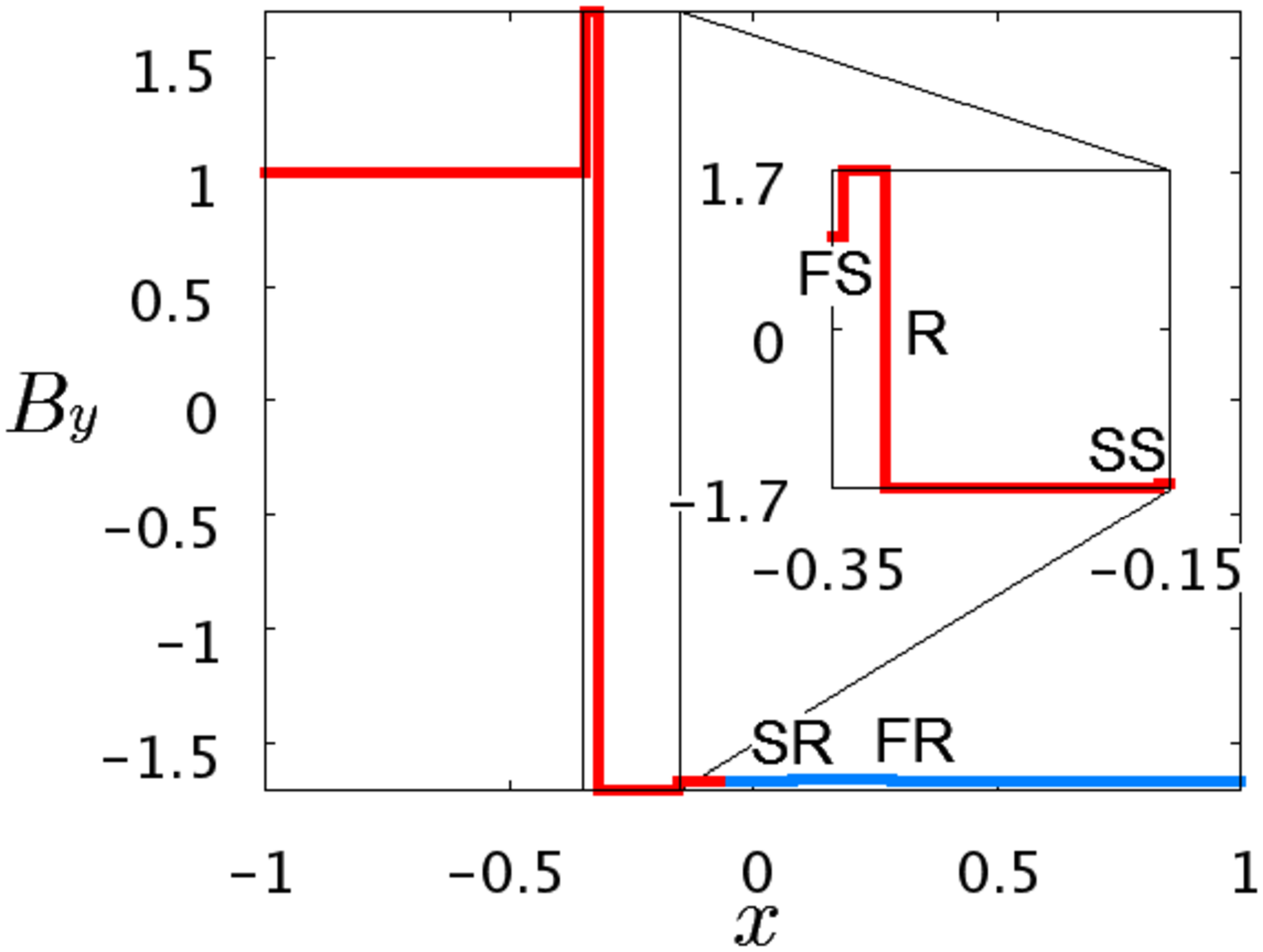}
\end{center}
\end{minipage} \\
\begin{minipage}{0.45\hsize}
\begin{center}
\includegraphics[scale=0.26]{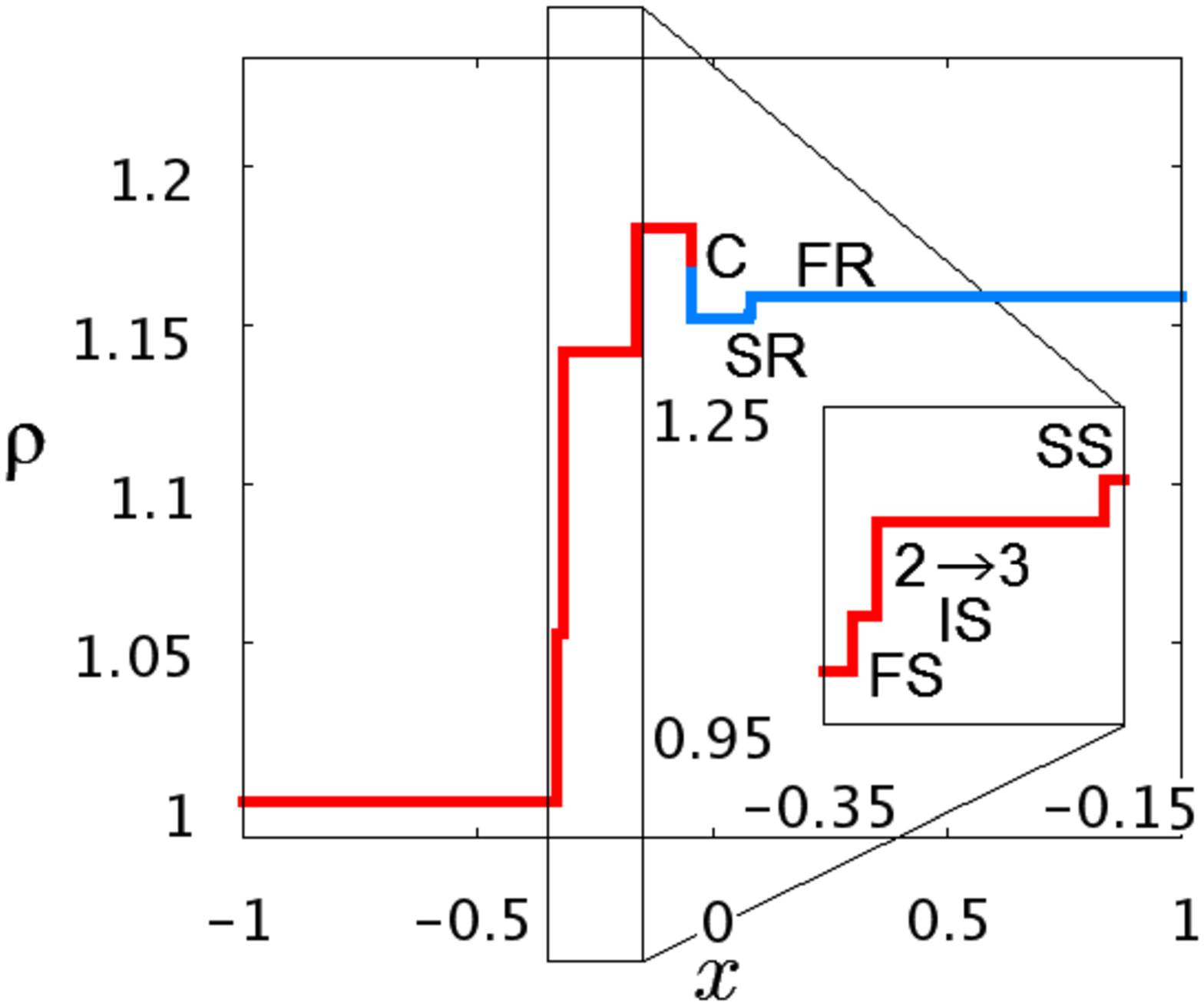}
\end{center}
\end{minipage} &
\begin{minipage}{0.45\hsize}
\begin{center}
\includegraphics[scale=0.26]{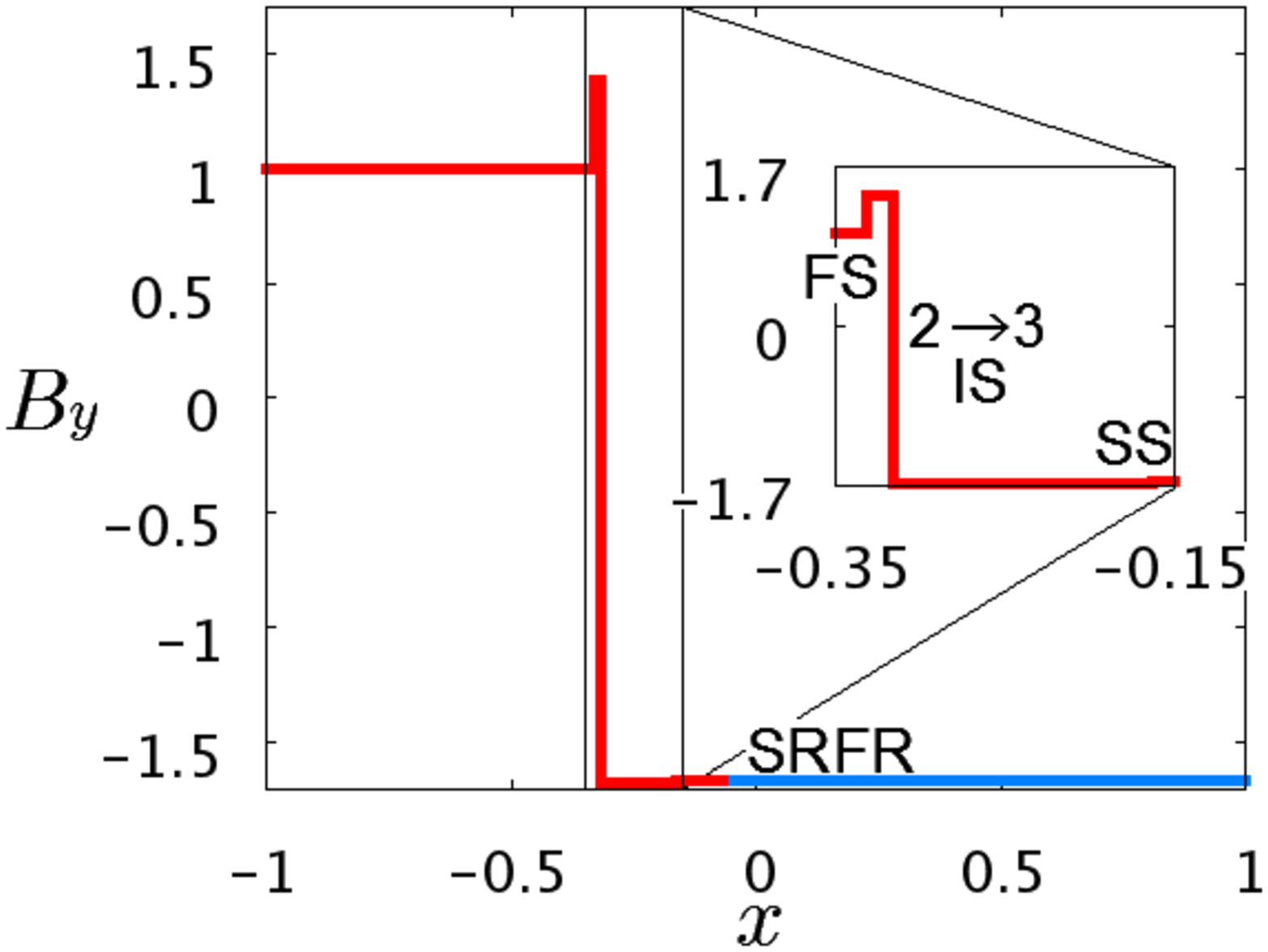}
\end{center}
\end{minipage} \\
\begin{minipage}{0.45\hsize}
\begin{center}
\includegraphics[scale=0.26]{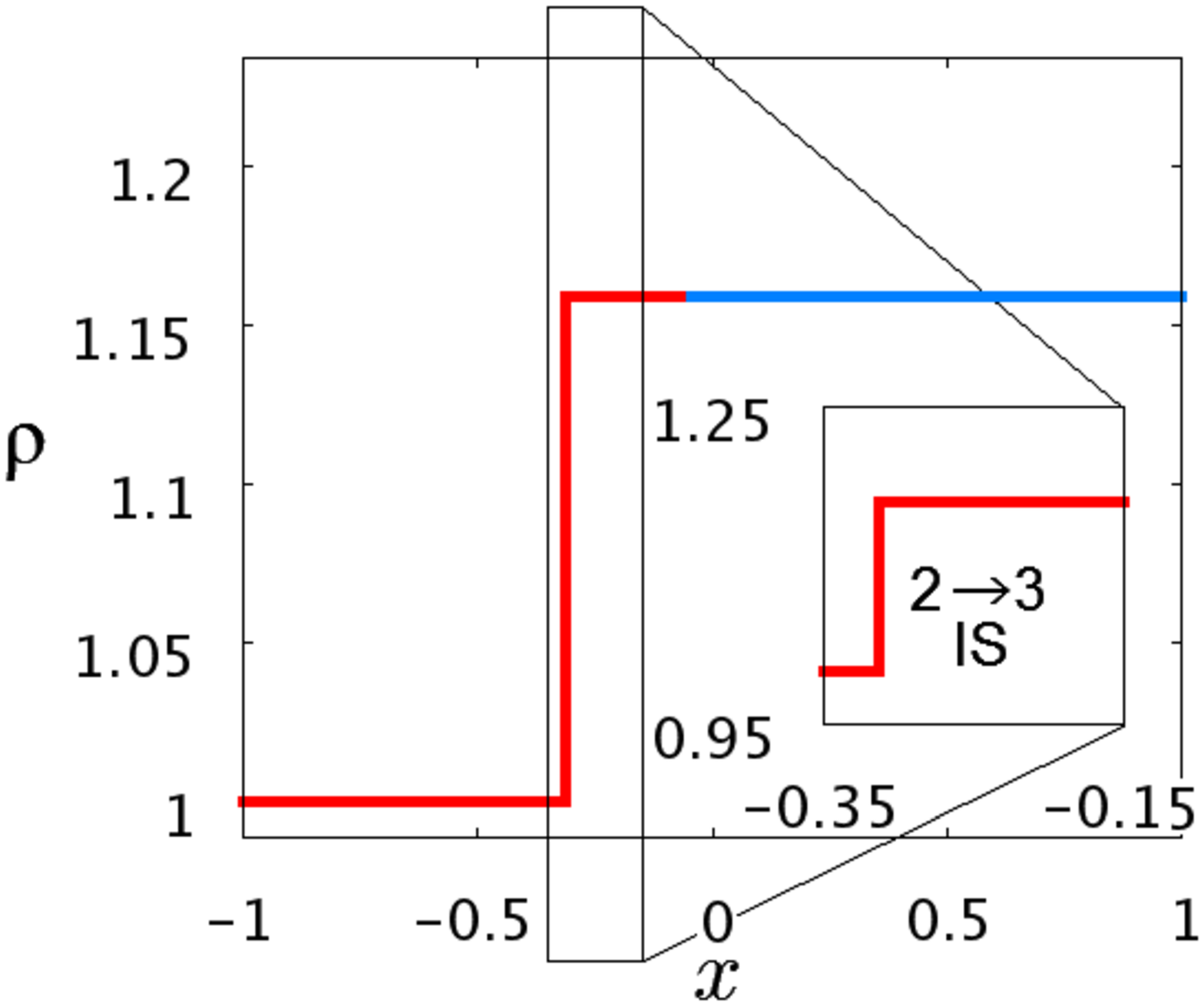}
\end{center}
\end{minipage} &
\begin{minipage}{0.45\hsize}
\begin{center}
\includegraphics[scale=0.26]{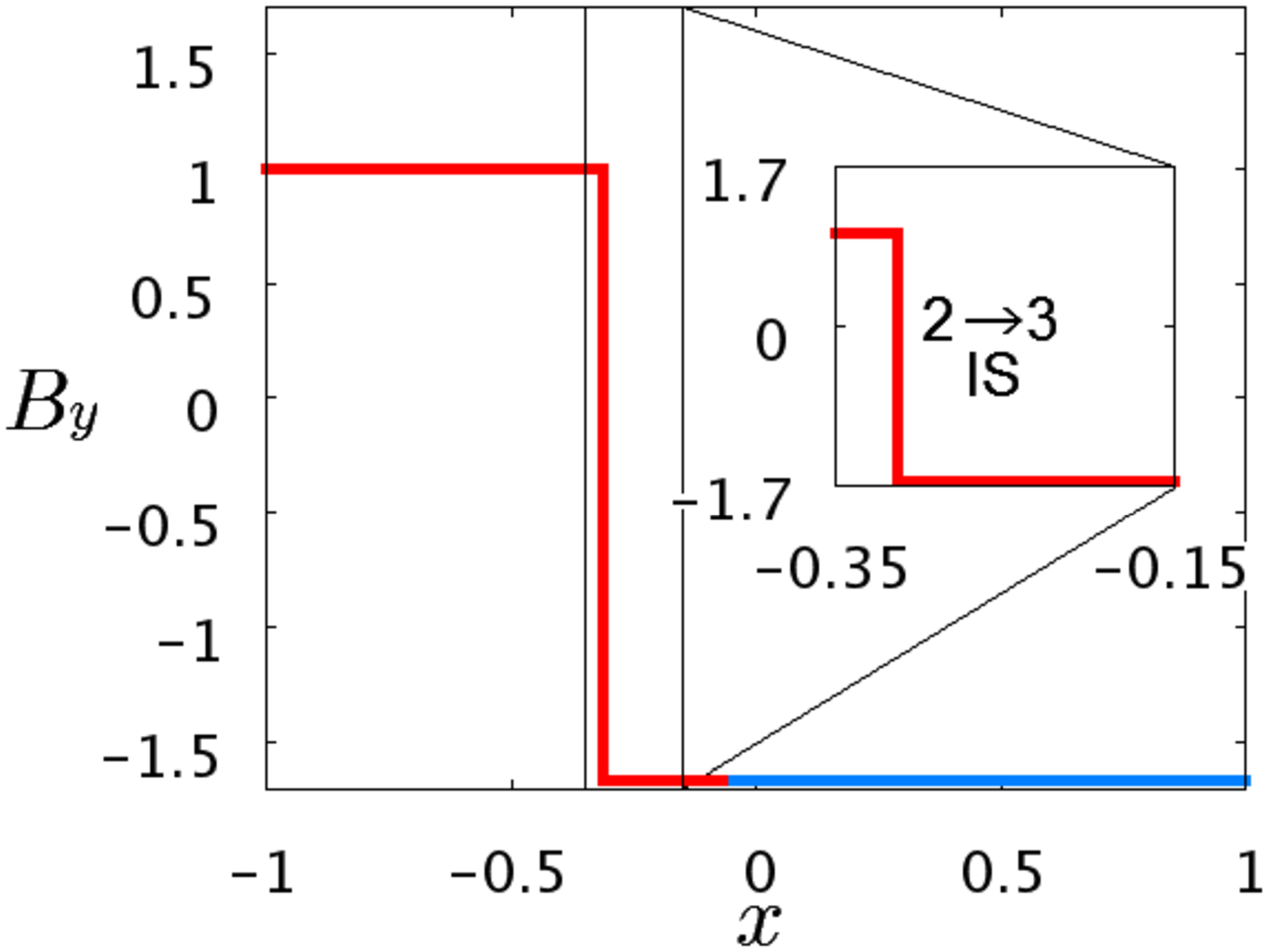}
\end{center}
\end{minipage} \\
\begin{minipage}{0.45\hsize}
\begin{center}
\includegraphics[scale=0.26]{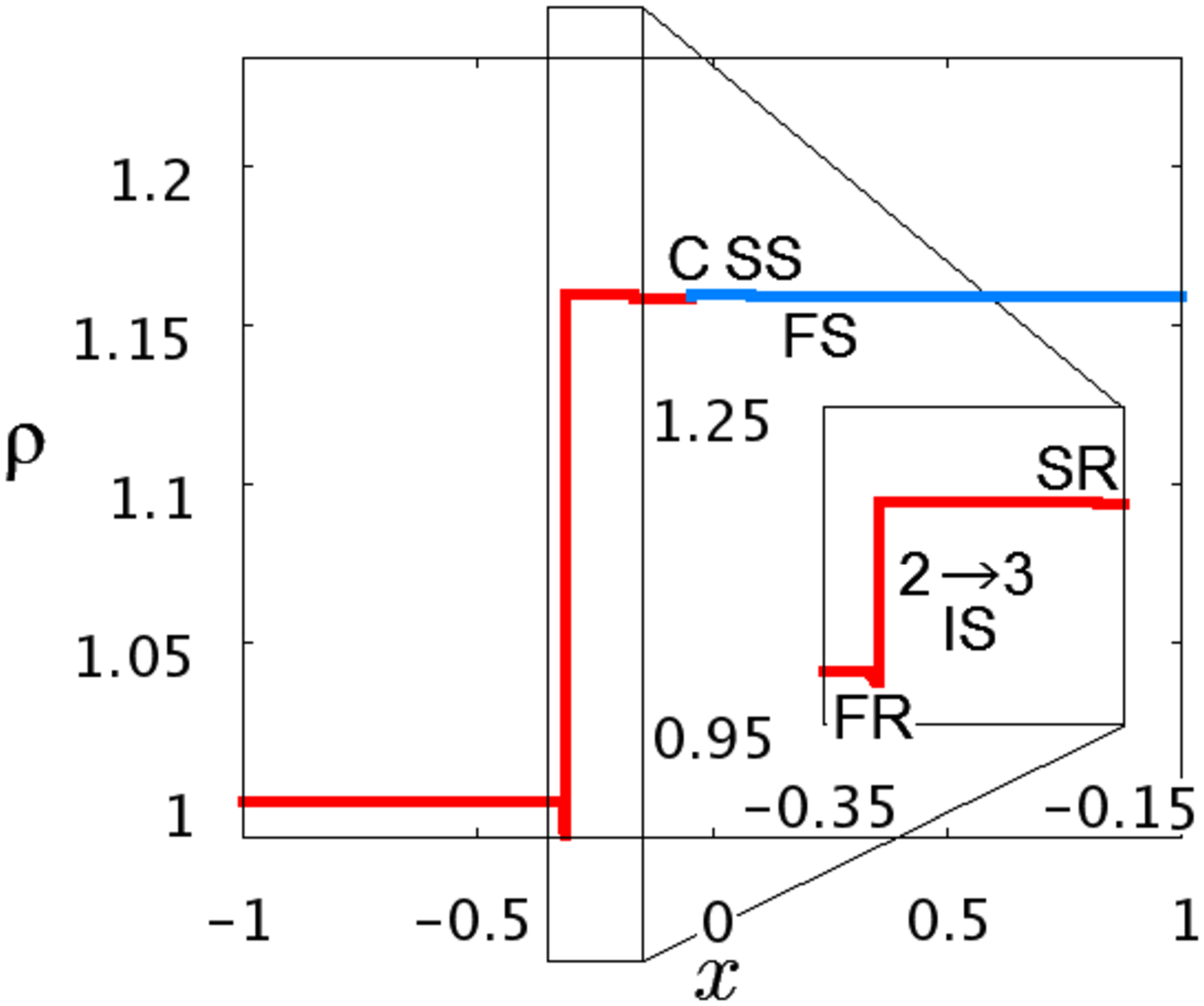}
\end{center}
\end{minipage} &
\begin{minipage}{0.45\hsize}
\begin{center}
\includegraphics[scale=0.26]{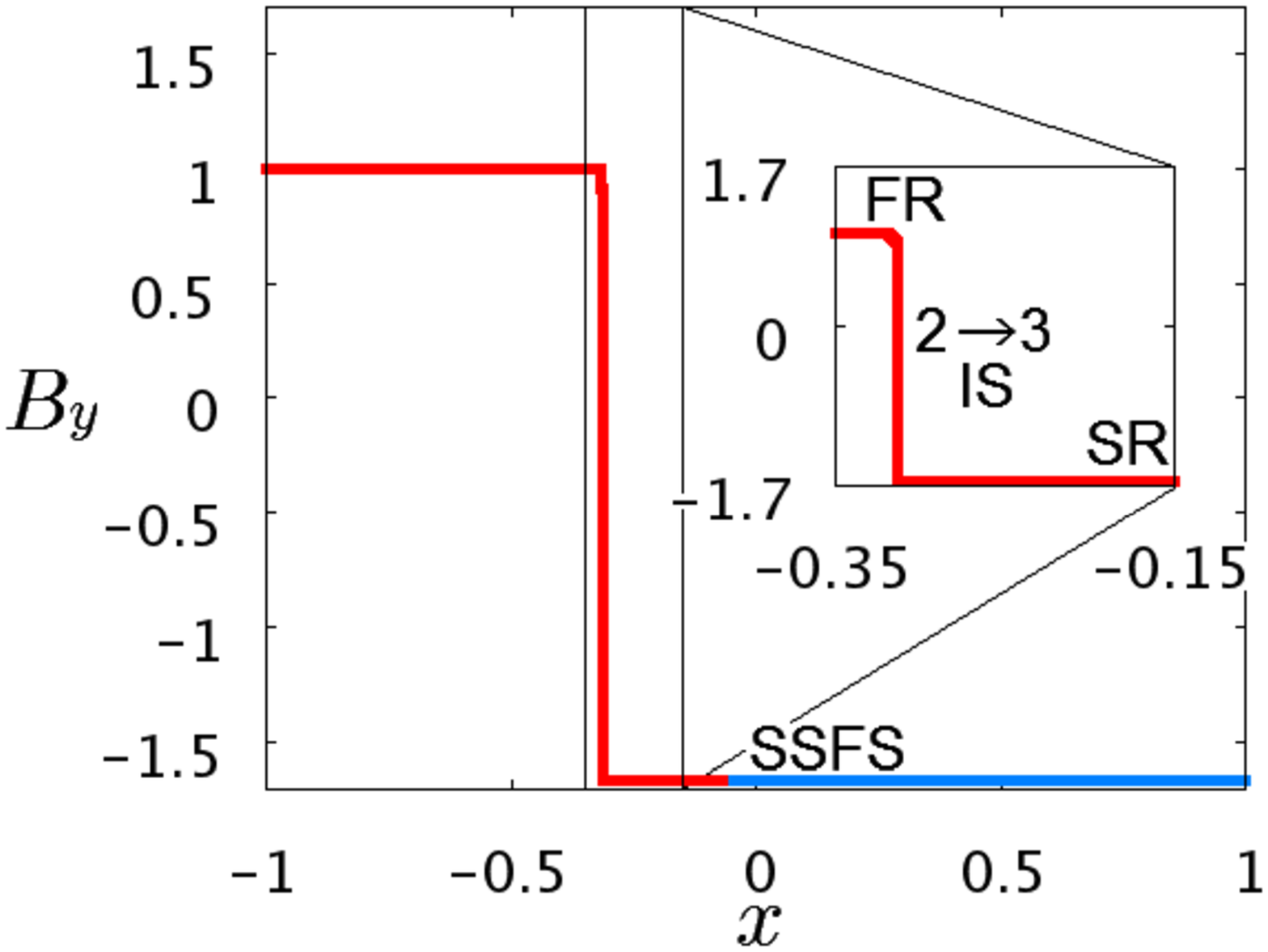}
\end{center}
\end{minipage} 
\end{tabular}
\caption{The regular solution and some non-regular solutions for an initial condition which can be connected by a $2 \rightarrow 3$ intermediate shock. 
The notations are the same as in Fig. \ref{1-4}. In this sequence, the left-going fast shock becomes rarefaction across the solution that consists of only a $2\rightarrow 3$ intermediate shock.
The tail of the left-going fast rarefaction and $2\rightarrow 3$ intermediate shock are gradually coming closer,
inferring that the end point of this sequence is the solution that includes a left-going $1,2\rightarrow 3$ intermediate shock attached to a fast rarefaction.}
\label{2-3}
\end{figure}

Firstly, we present the solutions for an Riemann problem whose initial condition satisfies the Rankine-Hugoniot conditions of a $1\rightarrow 4$ intermediate shock: 
\begin{eqnarray}
(\rho_L,\ p_L,\ v_{xL},\ v_{yL},\ v_{zL},\ B_{yL},\ B_{zL}) =  (1,\ 1,\ 0,\ 0,\ 0,\ 1,\ 0), \qquad \qquad \qquad \\
(\rho_R,\ p_R,\ v_{xR},\ v_{yR},\ v_{zR},\ B_{yR},\ B_{zR})  \qquad \qquad \qquad \qquad \qquad \qquad \qquad \qquad \ \nonumber \\
 = (2.622826,\ 8.930218,\ -2.196843,\ -1.571584,\ 0,\ -0.8600000,\ 0),
\end{eqnarray}
with $B_n = 3$ and $\gamma = 5/3$. The discontinuity is initially located at $x=0$.
Some of the solutions at $t=0.1$ are shown in Fig. \ref{1-4}, where we show the profiles of the density and transverse magnetic field. Note that the transverse magnetic fields 
are confined in $(x,y)$-plane in these solutions. 
As shown in the figure, the initial condition can be connected not only by a $1\rightarrow 4$ shock (the bottom panels) but also by other wave-patterns.
The top panels are the regular solution, which consists of a fast shock, $180^\circ$-rotational discontinuity and slow shock that run into the left side of a contact discontinuity and a fast and slow rarefaction
waves that run on the other side. The second and third rows show some non-regular solutions that contain a $2\rightarrow 3$ intermediate shock instead of the rotational discontinuity, which 
is responsible for reversing the transverse magnetic field. Although these two solutions resemble each other, the close-ups reveal the difference that the three shock waves change their strength
as well as the fast and slow rarefactions. In fact, we discovered the uncountably infinite solutions which contain a $2 \rightarrow 3$ intermediate shock whose strength is different from each other.
The sequence is parameterized by the strength of the left-going $2\rightarrow 3$ intermediate shock, i.e., $\psi _s^-$, and is obtained by gradually altering $\psi _s^-$ that is fixed in the modified 
Newton-Raphson method. Note that the rotational discontinuity in the regular solution is represented by the terminal point of the slow Hugoniot locus as mentioned in Sec. \ref{adv}. 
As $\psi _s^-$ approaches a finite value, the speeds of the left-going fast, slow and $2\rightarrow 3$ shocks come closer to each other while the right-going fast and slow
rarefactions weaken their strength. The solution that includes only a $1\rightarrow 4$ intermediate shock corresponds to the limit of the coincidence of the three shock speeds.
The reason why there are uncountably infinite solutions is explained as follows; Since the fields are confined in $(x,y)$-plane, there are only four non-trivial Rankine-Hugoniot conditions,
continuity of $p, v_x, v_y, B_y$; On the other hand, there are five waves in the solutions as long as a rotational discontinuity or $2\rightarrow 3$ intermediate shock exist; That is, the system is 
under-determined and hence there remains an extra degree of freedom, which brings the existence of the uncountably infinite solutions.

As the second example, we present the solutions for an initial condition that is connected by a $1 \rightarrow 3$ intermediate shock. The initial discontinuity located at $x=0$ is given as
\begin{eqnarray}
(\rho_L,\ p_L,\ v_{xL},\ v_{yL},\ v_{zL},\ B_{yL},\ B_{zL}) =  (1,\ 1,\ 0,\ 0,\ 0,\ 1,\ 0), \qquad \qquad \qquad \\
(\rho_R,\ p_R,\ v_{xR},\ v_{yR},\ v_{zR},\ B_{yR},\ B_{zR})  \qquad \qquad \qquad \qquad \qquad \qquad \qquad \qquad \ \nonumber \\
 = (2.272607,\ 7.696652,\ -2.106806,\ -2.280515,\ 0,\ -1.8600000,\ 0),
\end{eqnarray}
with $B_n = 3$ and $\gamma = 5/3$. Some of the solutions at $t=0.1$ are shown in Fig. \ref{1-3}, displaying the profiles of the density and transverse magnetic field.
We note that the transverse magnetic fields are confined in $(x,y)$-plane in these solutions. 
Like the previous example, we obtained a sequence of the solutions composed by various waves. The top panels show the regular solution that consists of a $180^\circ$ rotational discontinuity and 
fast and slow shocks fanning out on the left side of a contact discontinuity and fast and slow rarefactions on the other side. The second and third rows show some non-regular solutions 
that include a $2\rightarrow 3$ intermediate shock, which reverses the transverse magnetic field instead of the rotational discontinuity. The bottom panels show a non-regular solution that consists
of only a $1\rightarrow 3$ intermediate shock. The close-ups reveal the difference of these solutions while we note that the strengths of the rarefactions also differ from each other.
Like the previous example, these solutions form a one-parameter family that is parameterized by $\psi _s^-$, which controls the left-going slow-family wave.
Asymptotically, the fast shock and $2\rightarrow 3$ intermediate shock appear to merge at first while all the three shock speeds are coming closer to each other as $\psi _s^-$ reduces.
Although this asymptotic behavior infers the existence of the solutions that include a left-going $1\rightarrow 3$ intermediate shock and slow shock, such a solution is not found. 
Hence, the solution including three shocks jumps to one that is composed of only a $1 \rightarrow 3$ shock before the two shocks merge. 
Considering the reason why there are uncountably infinite solutions, this feature may be natural. Once a $1\rightarrow 3$ intermediate shock is formed, the under-determination of the system
is lost and the system becomes determined one. Therefore there is only one solution that includes a $1\rightarrow 3$ intermediate shock (the bottom panels).

As the third example, we pick up an initial condition that is connected by a $2 \rightarrow 4$ intermediate shock:
\begin{eqnarray}
(\rho_L,\ p_L,\ v_{xL},\ v_{yL},\ v_{zL},\ B_{yL},\ B_{zL}) =  (1,\ 1,\ 0,\ 0,\ 0,\ 1,\ 0), \qquad \qquad \qquad \\
(\rho_R,\ p_R,\ v_{xR},\ v_{yR},\ v_{zR},\ B_{yR},\ B_{zR})  \qquad \qquad \qquad \qquad \qquad \qquad \qquad \qquad \ \nonumber \\
 = (2.593746,\ 7.352303,\ -1.897120,\ -1.068836,\ 0,\ -0.1000000,\ 0),
\end{eqnarray}
with $B_n = 3$ and $\gamma = 5/3$. Some of the solutions at $t=0$ for the initial discontinuity located at $x=0$ are presented in Fig. \ref{2-4}. Note that the transverse magnetic fields
are confined in $(x,y)$-plane in these solutions. The top panels show the regular solution that consists of a fast shock, 
$180^\circ$ rotational discontinuity and slow shock running into the left side of a contact discontinuity and fast rarefaction and slow shock running on the other side. The second and third ones present some
non-regular solutions that include a $2\rightarrow 3$ intermediate shock, which is responsible for reversing the transverse magnetic field. The solution including $2\rightarrow 4$ intermediate shock is
shown in the bottom panels. There are also uncountably infinite solutions like the previous examples since these solutions form a one-parameter family parameterized by $\psi _s^-$. 
Although all the left-going shocks come closer asymptotically, the $2\rightarrow 3$ shock and slow shock appear to merge before the fast shock and $2\rightarrow 3$ shock converge, 
inferring the asymptotic solution that includes a fast shock and $2\rightarrow 4$ intermediate shock. However, like the previous example, such a solution is not found. 
Hence, the solution including three shocks jumps to one that includes only a $2\rightarrow 4$ intermediate shock.

Finally, we give the solutions for an initial condition that is connected by a $2 \rightarrow 3$ intermediate shock:
\begin{eqnarray}
(\rho_L,\ p_L,\ v_{xL},\ v_{yL},\ v_{zL},\ B_{yL},\ B_{zL}) =  (1,\ 1,\ 0,\ 0,\ 0,\ 1,\ 0), \qquad \qquad \qquad \\
(\rho_R,\ p_R,\ v_{xR},\ v_{yR},\ v_{zR},\ B_{yR},\ B_{zR})  \qquad \qquad \qquad \qquad \qquad \qquad \qquad \qquad \ \nonumber \\
 = (1.159467,\ 1.479053,\ -0.4315320,\ -2.541677,\ 0,\ -1.658269 ,\ 0),
\end{eqnarray}
with $B_n = 3$ and $\gamma = 5/3$. Some of the solutions at $t=0.1$ for the discontinuity located at $x=0$ are shown in Fig. \ref{2-3}. Note that the transverse magnetic fields are confined in 
$(x,y)$-plane in these solutions. The top panels are the regular solution that consists of a $180^\circ$ rotational discontinuity and fast and slow shocks on the left side of a contact discontinuity and
fast and slow rarefactions on the other side. The second ones present a non-regular solution that includes a $2\rightarrow 3$ intermediate shock instead of the rotational discontinuity, which reverses
the transverse magnetic field. The third ones is the non-regular solutions that is composed of only a $2\rightarrow 3$ intermediate shock. The bottom ones show a non-regular solutions that
consists of a fast rarefaction, $2\rightarrow 3$ intermediate shock and slow rarefaction fanning out on the left of a contact discontinuity and fast and slow shock on the other side. Alike the previous examples,
these solutions cannot be parameterized by $\psi _s^-$. Instead, they are parameterized by $\psi _s^+$ that controls the right-going slow wave.
As $\psi _s^+$ increases, the left-going fast shock in the solutions becomes weaker and changes into a rarefaction wave across the solution that includes only a $2\rightarrow 3$ intermediate shock.
As $\psi _s^+$ increases further, the tail of the fast rarefaction and the $2\rightarrow 3$ intermediate shock come closer to each other, and the fast rarefaction becomes stronger. 
Therefore we concluded that the solution will reach one that includes a compound wave formed by a left-going $1,2\rightarrow 3$ intermediate shock attached to a fast rarefaction wave.
The solution including a compound wave must be an end point of the sequence because the value of $\psi _s^+$ asymptotically 
approaches a finite value and appears to converge in the limit and we can find no solution for $\psi_s^+$ larger than the asymptotic value.

\section{Summary} \label{sec.summary}
In the paper, we presented an exact Riemann solver that can handle the intermediate shocks and switch-on/off waves. Our solver can handle any initial condition even when 
the normal or transverse magnetic field is absent. These features are realized for the first time; Previous studies discarded these non-regular shocks or initial conditions with vanishing magnetic field. 
Although our method refers one in \citet{T02}, we drastically improved it to handle all types of non-regular shocks and switch-\textcolor{black}{on/}off rarefactions and the details of the techniques are released 
for the first time. Since the types of waves generated and their order are not known \textit{a priori} in MHD Riemann problems once such non-regular waves are considered, we developed the method 
that can arrange the waves in all possible order and search the structure of the solution automatically. Due to the variability of the number of the waves generated, we modified the Newton-Raphson method to 
adjust the number of the independent variables and equations.
Thanks to these techniques, all the solutions are found for a given initial condition, which has never been achieved by other authors \citep[e.g.][]{ATJweb}.
Our method works well indeed as shown in Sec. \ref{sec.results} and our previous paper \citep{TY12a}, where we presented the examples of the exact solutions, which include the regular and non-regular
ones. As demonstrated, our solver can investigate the structure of the solution space in detail. Therefore the solver is a powerful instrument to solve the outstanding problem of the existence and uniqueness
of solutions of MHD Riemann problems.

Since our method is based on the Newton-Raphson method, there might be a solution that exists far away from the sequence of the solutions and hence eludes our search. 
Therefore any strategy that finds such a particular solution should be studied in future research.
Aiming for application to the numerical schemes like the Godunov scheme, any reasonable way to find a good initial guess must be also investigated in future work.

\appendix
\section{The valid range of switch-on shocks} \label{appA}
As mentioned in Sec. \ref{sec.Fast and slow loci}, switch-on shocks are possible only when the Mach number satisfies the inequality (\ref{sw-on_ineq}).
Such Mach numbers exist only if $\hat{c}_{A0} := c_{A0}/a_0 = \sqrt{B^2/\gamma} > 1$ as shown below.

(i) In the case of $\sqrt{B^2/\gamma } > 1$. Recalling the degeneracy (\ref{degenerate_2}), the fast speed equals to the \Alfven \ speed, i.e., $\hat{c}_{f0} = \hat{c}_{A0} = \sqrt{B^2/\gamma}$.
Then,  
\begin{equation}
\left( \frac{\gamma +1}{\gamma -1}\frac{B^2}{\gamma} -\frac{2}{\gamma -1}\right) -\hat{c}_{f0}^2 
= \frac{2}{\gamma -1}\left( \frac{B^2}{\gamma } -1\right)
> 0,
\end{equation}
and, since $\gamma > 1$, the value in the square root is positive.
Therefore there is a finite range in (\ref{sw-on_ineq}) where the switch-on shocks are possible. 

(ii) $\sqrt{B^2/\gamma } \le 1$. In this case, the fast speed degenerates into the acoustic speed, i.e., $\hat{c}_{f0} = 1$. Then
\begin{equation}
\left( \frac{\gamma +1}{\gamma -1}\frac{B^2}{\gamma} -\frac{2}{\gamma -1}\right) -\hat{c}_{f0}^2 
= \frac{\gamma +1}{\gamma -1}\left( \frac{B^2}{\gamma } -1\right)
\le 0.
\end{equation}
Therefore there is no Mach number that satisfies the inequality (\ref{sw-on_ineq}) and the switch-on shocks are never possible.

\section{The correct root in the cubic equation} \label{appB}
\textcolor{black}{
We use the cubic equation for $\hat{v}$ (\ref{fast branch}) to obtain the fast shock solution. This equation has three roots in general and hence we should correctly find the correct root. Here, we discuss how to pick it up, which turns to be easy as shown below.}

\textcolor{black}{
The point is that we use (\ref{fast branch}) only for fast shocks. Then, since the state in front of the shock is super-fast, the parameter $M_0$ in (\ref{fast branch}) is always larger than $\hat{c}_{f0}$ when we solve it. Therefore all the roots correspond to some super-fast solution; One is a fast shock and the others, if any, are intermediate shocks whose upstream flow speed is super-fast. 
We note here that $\hat{v}$ is larger than $B^2/(\gamma M_0^2)$ if we assume that the shock is super-Alfvenic and the transverse magnetic field is not reversed and we can also show that $\hat{v}$ is smaller than $B^2/(\gamma M_0^2)$ for intermediate shocks. Therefore, the correct root is always larger than the others. Then, we can easily get the fast shock solution by giving an initial guess as $\hat{v}$ = 1.0 in Newton-Raphson method. 
}

\bibliographystyle{jpp}

\bibliography{bibliography}

\end{document}